%% file: 000_main.tex
\input{00_preamble}

\begin{document}

\twocolumn[\vspace{-1.5cm}\maketitle\vspace{-1cm}
	\normalsize{\begin{shaded}
		\noindent\abstract
	\end{shaded}
	\vspace{0.4cm} 
	}
]

\begin{figure} [!b]
\begin{minipage}[t]{\columnwidth}{\rule{\columnwidth}{1pt}\footnotesize{\textsf{\affiliation}}}\end{minipage}
\end{figure}

\input{1_Introduction/0_main}
\input{2_Background/0_main}

\input{3_Modeling/0_main}
\input{4_Parameters/0_main}

\input{5_Results/0_main}
\input{6_Conclusion/0_main}

\section*{Acknowledgements}
This work was funded by the German Federal Ministry of Education and Research (BMBF) as part of the \textit{ZABSES - Zinc-air batteries for stationary electricity storage} project (funding reference: 03XP0505A).

\section*{Conflict of Interest}
The authors declare no conflict of interest.

\section*{CRediT authorship contribution statement}
\textbf{Noah Lettner:} Conceptualization, Methodology, Software, Validation, Formal analysis, Investigation, Data curation, Writing - original draft, Visualization.
\textbf{Felix K. Schwab:} Conceptualization, Writing - review \& editing, Supervision, Project administration.
\textbf{Birger Horstmann:} Conceptualization, Methodology, Resources, Writing - review \& editing, Supervision, Project administration, Funding acquisition.

\begin{shaded}
\noindent\textsf{\textbf{Keywords:} \keywords} 
\end{shaded}

\setlength{\bibsep}{0.0cm}
\bibliographystyle{Template/Wiley-chemistry}
\bibliography{Quellen_Paper}

\clearpage

\input{7_Appendix/0_main}

\clearpage

\end{document}

%% file: 00_preamble.tex
\documentclass{Template/WileyChemistry-template}

\title{\raggedright Thermodynamically consistent modeling of ion exchange membranes in multi-ionic environments}

\author{
\begin{minipage}{\textwidth}%
Noah Lettner \textsuperscript{[a,b]}, Felix K. Schwab\textsuperscript{[a,b]}, Birger Horstmann \textsuperscript{[*,a,b,c]} 
\end{minipage}
}

\newcommand{\affiliation}{
\begin{itemize}

\item[{[a]}] German Aerospace Center (DLR), Institute of Engineering Thermodynamics, Wilhelm-Runge-Str. 10, 89081 Ulm, Germany

\item[{[b]}] Helmholtz Institute Ulm (HIU), Helmholtzstr. 11, 89081 Ulm, Germany

\item[{[c]}] Ulm University, Albert-Einstein-Allee 11, 89069 Ulm, Germany

\item[{[*]}] Corresponding author, E-mail: birger.horstmann@dlr.de
\end{itemize}
}

\renewcommand{\dedication}{
	\begin{minipage}{\textwidth}
		Dedication (optional, leave blank if no dedication is required)
	\end{minipage}
}

\renewcommand{\abstract}{ \input{0_Abstract/0_main}}

\input{Template/colors}

\newcommand{\keywords}{
	membrane modeling \textbullet\
	filtration \textbullet\  
	selectivity \textbullet\
	partitioning \textbullet\
	permeability \textbullet\
	Donnan potential \textbullet\
	mean-field \textbullet\
	mass-action \textbullet\
	SDE \textbullet\
	Donnan-Manning \textbullet\
	low-T*
}

\usepackage{etoolbox}
\usepgfplotslibrary{groupplots}
\usepackage{outlines}
\usepackage{xfrac}
\tikzset{
  dashed/.style={dash pattern=on 4pt off 2pt} ,
  dashdotted/.style={dash pattern=on 6pt off 2pt on 2pt off 2pt}
}

\AtBeginEnvironment{figure}{
  \let\oldimport\import
  \renewcommand{\import}[2]{%
    {\fontsize{9}{11}\selectfont\oldimport{#1}{#2}}%
  }
}

\usepackage[framemethod=TikZ]{mdframed} 
\mdfdefinestyle{important}{%
  backgroundcolor=black!5,
  roundcorner=6pt,
  hidealllines=true,
  nobreak,
  innerleftmargin=6pt,
  innerrightmargin=6pt,
  innertopmargin=3pt,
  innerbottommargin=3pt,
  skipabove=\abovedisplayskip,
  skipbelow=\belowdisplayskip,
}
\newmdenv[style=important]{impeq}

\DeclareFontFamily{OT1}{pzc}{}
\DeclareFontShape{OT1}{pzc}{m}{it}{<-> s * [1.10] pzcmi7t}{}
\DeclareMathAlphabet{\mathpzc}{OT1}{pzc}{m}{it}

%% file: 0_Abstract/0_main.tex
Ion exchange membranes are useful for a wide range of applications, including water desalination, fuel cells, and aqueous batteries. Accordingly, a variety of models for ion exchange membranes has been proposed, each emphasizing different aspects that govern their static and dynamic properties. By reviewing these models, we identify key physical contributions and beneficial modeling strategies. Based on these insights, we derive a thermodynamically consistent model by combining mass-action site occupation with mean-field electrostatic interactions along the polymer backbone. In this derivation, we explicitly account for multicomponent electrolytes at elevated concentrations. The parameters of the resulting model relate closely to those of other models, but gain consistency and interpretability through the underlying derivation. A discussion of the model parameters highlights redundancies and linkages between quantities that are commonly treated independently.
Comparison to experimental data shows that both static and dynamic membrane properties are reproduced with good accuracy by the presented model. This makes it a promising basis for theory-driven membrane optimization and supports the tailored design of ion exchange membranes for various technologies.

%% file: Template/colors.tex
\definecolor{KITgreen}{RGB}{0,150,130}
\definecolor{KITred}{RGB}{162,34,35}
\definecolor{KITblue}{RGB}{70,100,170}
\definecolor{KITorange}{RGB}{223,155,27}
\definecolor{KITyellow}{RGB}{252,229,0}
\definecolor{KITpurple}{RGB}{163,16,124}
\definecolor{KITmaygreen}{RGB}{140,182,60}
\definecolor{KITcyan}{RGB}{35,161,224}
\definecolor{KITbrown}{RGB}{167,130,46}
\definecolor{ULMblue}{RGB}{125,154,170}
\definecolor{KITblack}{RGB}{0,0,0}
\definecolor{KITdarkgrey}{RGB}{64,64,64}
\definecolor{KITgrey}{RGB}{128,128,128}
\definecolor{KITlightgrey}{RGB}{192,192,192}

%% file: 1_Introduction/0_main.tex
\section{Introduction}
\label{C:Introduction}
Membrane filtration has been a subject of extensive research for over a century. Consequently, there is a wide range of commonly applied models \cite{Bowen2002, Kamcev2015, Meyer1936, Teorell1935, Morrison1965, Weber2004, Freger2020}, but the scientific discussion is still ongoing \cite{Freger2020, Freger2025, Luo2018, Kitto2022, Yaroshchuk2019, Bannon2024, Crothers2020}. Especially the mechanisms of partitioning \cite{Crothers2020, Freger2023} and ion permeability \cite{Freger2020} remain a matter of debate. \par 
One crucial point is also the applicability of these membrane models for salt mixtures and broadly varying ambient conditions \cite{Santafe-Moros2008, Wang2023}, which is commonly outside the model delimitation. Such conditions are encountered, e.g., in aqueous batteries. Various types of aqueous batteries are researched with increasing extent, mostly for their advantageous properties regarding environmental sustainability and safety aspects such as flammability, as well as their competitiveness in terms of cost; yet they commonly suffer from low cycle or shelf life \cite{Chao2020, Liang2022, Posada2017, Ju2022, Borchers2021}. Previous research has pointed out that aqueous batteries can be improved by selective membranes that limit the crossover of certain species which contribute to degradation \cite{Dembele2023, Tsehaye2020, Tsehaye2021}. In order to assess the potential of this approach, a suitable membrane model for the variety of conditions encountered during cycling and storage of such batteries is required. \par
This work proposes a thermodynamically consistent basis for membrane modeling, incorporating mass-action occupation of active sites as proposed in the low-T* model \cite{Freger2020}, while also including mean-field nearest neighbor interactions in analogy to the Donnan-Manning model \cite{Kamcev2015}. Further, exclusion mechanisms as targeted by the Steric Donnan Dielectric (SDE) model \cite{Bowen2002} are discussed and accounted for in the model. \par 
The proposed framework does not introduce new electrostatic principles; its novelty lies in the consistent coupling of occupation statistics to membrane partitioning and transport, in a formulation that structurally extends to multicomponent electrolytes at elevated concentrations.\par
The following sections are structured as follows: Section \ref{C:Background} introduces the concept of membrane filtration, discusses relevant scales and gives an overview of selected existing models. Section \ref{C:Modeling} derives the proposed modeling framework, while section \ref{C:Parameter_estimation} discusses and estimates its parameters. Section \ref{C:Results} analyses the model behavior and compares its results to experimental data and existing models. Section \ref{C:Multi-ionic} discusses the applicability of the derived model to salt mixtures. 

%% file: 2_Background/0_main.tex
\section{Background}
\label{C:Background}
The topic of membranes is a broad field spanning many purposes, approaches and technical applications. Thus, any model has to be tailored to its intended application to maintain a realistic number of meaningful parameters. Within the scope of this work, it is assumed that there is neither a relevant gas phase, nor a significant difference in pressure across the membrane. Further, the membrane is considered to be in equilibrium with an aqueous salt solution, with swelling and hydration being approximately constant on the relevant timescales.\par
Within this delimitation, membranes are usually characterized by their predominant selectivity mechanism, which is either size or charge filtration. A typical categorization of size filtering membranes ranges from microfiltration (filtering macroscopic particles), through ultrafiltration (large molecules) and nanofiltration (large ions) to reverse osmosis (all ions)\cite{Oatley2004}. \par 
Charge filtering membranes are usually categorized by the charge of the active groups.  
Cation exchange membranes (CEM) bear negative effective charge on the polymer backbone from acidic groups dissociating in the hydrated state, making them more permeable for cations. The opposite goes for anion exchange membranes (AEM). If a membrane contains both acidic and basic functional groups and the membrane charge subsequently depends on the pH of the solution, they are referred to as amphoteric \cite{Luo2018,Sollner1932, Liu2021}. Composite architectures, such as layered or patterned membranes are not considered separately here, because they locally correspond to these categories. To avoid separate terminology for CEMs and AEMs, we denote ionic charge relative to the membrane charge; \textit{coions} have the same charge sign, \textit{counterions} the opposite.\par
The subsequent sections present relevant background information for charge and size filtering membranes.

\subsection{Membranes as multiscale systems}
In general, membranes consist of a solid or near-solid matrix interspersed with regions accessible to solvent and solutes. From a macroscopic perspective, they are usually treated as effective transport media, whose properties are influenced by various statistically averaged subscale effects. \par 
For clarity, we divide the subscale into a \textit{microscale} associated with local thermodynamic and kinetic effects ($\lesssim \mathcal{O}(1\,\text{nm})$) and a \textit{mesoscale} associated with porosity-tortuosity effects ($\lesssim \mathcal{O}(1-100\,\text{nm})$).\par
The following sections show how micro- and mesoscale effects can be modeled for use in a macroscale membrane model.
\subsection{Established microscale models}
\label{C:Microscale_models}
This section briefly discusses selected existing models that introduce or highlight key concepts later used for the model derived in this work; namely the Steric Donnan Dielectric model (SDE) \cite{Bowen1997, Bowen2002}, the Donnan-Manning model (D-M) \cite{Kamcev2015}, and the Low Reduced Temperature model (low-T*) \cite{Freger2020}. In the terminology adopted here, these are considered microscale models, since they describe local partitioning and transport within membrane pores and can technically be combined with any mesoscale model. Many further important developments and membrane models can be found in literature \cite{Meyer1936, Teorell1935, Teorell1953, Donnan1911, Gibbs1878, Morrison1965, Pappenheimer1951, Giddings1968, Deen1987, Yaroshchuk2019}. \par
\subsubsection{Steric Donnan Dielectric model}
\label{C:SDE_model}
\begin{figure}[h]
    \centering
    \def\svgwidth{6.6cm}
    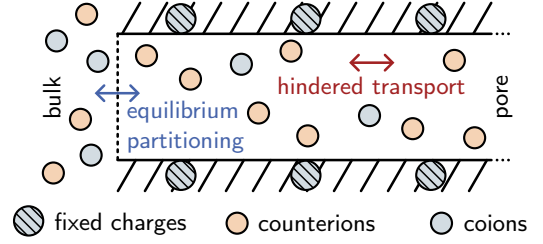
    \caption{Schematic representation of the SDE model. Colors indicate the sign of ion charge, the hatched pattern denotes fixed sites and pore walls.}
    \label{F:SDE_model_schematic}
\end{figure}

The Steric Donnan Dielectric (SDE) model considers narrow pores that partially exclude ions at their entrance and impede the transport of those that enter. Specifically, it applies equilibrium partitioning at the pore entrance and hindered transport theory within the pores. This is depicted in figure \ref{F:SDE_model_schematic}.
\par
Equilibrium partitioning is described by a partitioning factor $S$, which relates pore ($\mathrm{p}$) and bulk ($\mathrm{b}$) concentrations $c_i$. It accounts for electric (Donnan, $\mathrm{do}$), steric ($\mathrm{st}$) and dielectric ($\mathrm{de}$) contributions:
\begin{equation}
    S_i = \frac{c_i^\mathrm{p}}{c_i^\mathrm{b}} = S_i^\mathrm{st} S_i^\mathrm{de} S_i^\mathrm{do} 
    \label{E:Partitioning_SDE}
\end{equation}
Expressions for steric exclusion in various pore geometries as derived by Giddings et al. \cite{Giddings1968} are shown in the supporting information \ref{C:Steric_exclusion_Giddings}. Assuming cylindrical pores, the steric exclusion factor is described by the following function of hydrated particle radius $r_i^\mathrm{h}$ and pore radius $r_\mathrm{p}$:
\begin{equation}
    S_i^\mathrm{st}
    = 
    \left(1 - \frac{r_i^\mathrm{h}}{r_\mathrm{p}}\right)^2
    \label{E:Steric_exclusion}
\end{equation}
Dielectric exclusion is modeled using the Born model for the electrostatic contribution to the solvation energy $W$ of ions  \cite{Born1920}:
\begin{equation}
\begin{split}
    &S_i^\mathrm{de}
    =
    \exp{\left(-\frac{\Delta W_i}{k_\mathrm{B} T}\right)}
    \\
    &\text{with}
    \quad
    \Delta W_i = \frac{{z_i}^2 e^2}{8\pi\varepsilon_0 r_i^\mathrm{h}} \left(\frac{1}{\varepsilon_\mathrm{p}} - \frac{1}{\varepsilon_\mathrm{b}}\right)
    \label{E:Dielectric_exclusion}
\end{split}
\end{equation}
Here, $k_\mathrm{B}$ is the Boltzmann constant, $T$ the absolute temperature, $z$ the charge number, $e$ the elementary charge and $\varepsilon_0$ the vacuum permittivity. Whilst the relative permittivity of the bulk $\varepsilon_\mathrm{b}$ is usually well known, the relative permittivity within the pore $\varepsilon_p$ has to be modeled or fitted to experiments \cite{Wang2021}.\par
Donnan exclusion couples partitioning to charge neutrality through the implicitly determined Donnan potential $\Phi_\mathrm{D}$:
\begin{equation}
    S^\mathrm{do} = \exp{\left(-\frac{z_i e}{k_\mathrm{B} T} \Phi_\mathrm{D}\right)}
    \label{E:Donnan_exclusion}
\end{equation}
Charge neutrality has to account for the effective membrane charge density $q_\mathrm{eff}$ which is usually treated as an adjustable parameter: 
\begin{equation}
    \sum_i z_i c_i^\mathrm{p} = - q_\mathrm{eff} 
    \label{E:Charge_neutrality_SDE}
\end{equation}
Transport within the pores is modeled by a Nernst-Planck equation with modified transport coefficients:
\begin{equation}
    N_i^\mathrm{p} = - k^\mathrm{d}_i \left( D_i^\mathrm{b} \nabla c_i^\mathrm{p} + \frac{z_i c_i^\mathrm{p} D_i^\mathrm{b} F}{RT} \nabla \Phi \right)+ k^\mathrm{c}_i c_i^\mathrm{p} v
    \label{E:Nernst_Planck_SDE}
\end{equation}
Here, $D_i^\mathrm{b}$ is the bulk diffusion coefficient, $F$ the Faraday constant, $R$ the gas constant, $\Phi$ the electric potential and $v$ the flow velocity. Diffusive and a convective hindrance factors are denoted as $k^\mathrm{d}$ and $k^\mathrm{c}$.\par
These hindrance factors depend on pore geometry and on the particle to pore size ratio $\lambda = r_i^\mathrm{h}/r_\mathrm{p}$. Various expressions for these dependencies $k_\mathrm{d} \left( \lambda \right)$ and $k_\mathrm{c} \left( \lambda \right)$ in cylindrical and in slit pores have been reviewed by Dechadilok and Deen \cite{Dechadilok2006, Deen1987} and can be found in the supporting information \ref{C:SDE_hindrance_factors}.

\subsubsection{Donnan-Manning model}
\label{C:Donnan_Manning}
The Donnan-Manning (D-M) model is an adaptation of Manning's counterion condensation theory for polyelectrolyte solutions to describe ion exchange membranes \cite{Kamcev2015}.
Like the SDE model, it describes equilibrium partitioning and transport hindrance, but introduces a distinction between ions \textit{condensed} near the charged polymer backbone and \textit{uncondensed} mobile ions. \par
Here, partitioning applies to uncondensed ions ($\mathrm{u}$) and is modeled through a Donnan equation (equation \ref{E:Partitioning_SDE}) with the Donnan potential determined from charge neutrality (equations \ref{E:Donnan_exclusion} and \ref{E:Charge_neutrality_SDE}). Steric and dielectric effects are often neglected \cite{Bannon2024}, yielding:
\begin{equation}
    \frac{c_i^\mathrm{u}}{c_i^\mathrm{b}} = \exp{\left(-\frac{z_i e}{k_\mathrm{B} T} \Phi_\mathrm{D}\right)}
    \quad \text{w.r.t.} \quad 
    \sum_i z_i c_i^\mathrm{u} = - q_\mathrm{eff} 
\end{equation}
Unlike in the SDE model, the effective polymer charge $q_\mathrm{eff}$ is not treated as an adjustable parameter, but is estimated from Manning theory, which introduces an upper limit to the effective polyion line charge density motivated by a Poisson-Boltzmann model \cite{Manning1969, Manning1969a, Manning1969b} (see supporting information \ref{C:Manning_condensation_derivation}). This limit is described by the dimensionless Manning parameter $\xi$ and its critical value $\xi_\mathrm{crit}$, which quantify the interaction of counterions with the polyion relative to thermal motion. Here, supercritical values of $\xi$ are compensated by counterion condensation, yielding:
\begin{equation}
q_\mathrm{eff} = 
\begin{cases}
    c_\mathrm{X} z_\mathrm{X}
        & ,\ \forall \xi \leq \xi_\mathrm{crit}\\
    c_\mathrm{X} z_\mathrm{X} \frac{\xi_\mathrm{crit}}{\xi} 
        & ,\ \text{otherwise}
\end{cases}
\label{E:Manning_effective_concentration}
\end{equation}
Here, $c_\mathrm{X}$ is the fixed-site concentration and $z_\mathrm{X}$ the site charge, both of which are usually known from membrane synthesis. The Manning parameter is defined as the ratio of Bjerrum length $\lambda_\mathrm{B}$ over site distance $L$:
\begin{equation}
\begin{split}
    \xi = \frac{\lambda_\mathrm{B}}{L}
    \quad \text{with} \quad
    \lambda_\mathrm{B} = \frac{ e^2 }{4 \pi k_\mathrm{B} T \varepsilon_0 \varepsilon_\mathrm{r}}
\end{split}
\label{E:Manning_Parameter}
\end{equation}
Site distance can be difficult to determine since it depends on line length. Common approaches are to use the mean volumetric site distance $L \approx L^\mathrm{vol} = (c_\mathrm{X} N_\mathrm{A})^{-1/3}$ or to treat the Manning parameter as a fit parameter entirely \cite{Mareev2022}.\par
The critical value of the Manning parameter is defined by fixed site charge $z_\mathrm{X}$ and counterion charge $z_\mathrm{ct}$. It is not designed to account for mixtures with multiple counterions of different charge $z_i$, thus, in such cases the highest magnitude counterion charge is commonly used \cite{Purpura2024}:
\begin{equation}
    \xi_\mathrm{crit} = \frac{1}{-z_\mathrm{X} z_\mathrm{ct}}
    \left( \overset{\text{or}}{=}
    \frac{1}{\left| z_\mathrm{X} \max \left( - z_i\mathrm{sgn}\left(z_\mathrm{X}\right) \right)\right|}
    \right)
    \label{E:Critical_value_of_the_manning_parameter}
\end{equation}
Transport is modeled analogously to the SDE model by a Nernst-Planck equation with modified transport coefficients that are reduced due to electrostatic interaction. This is expressed by a hindrance factor $k^\mathrm{e}$ \cite{Wang2023}, which depends on membrane charge, Manning parameter, charge numbers and ion concentrations:
\begin{equation}
    \frac{D_i^\mathrm{u}}{D_i^\mathrm{b}} = k^\mathrm{e}_i\left(z_i, c_i, z_\mathrm{X}, c_\mathrm{X}, \xi \right)
\end{equation}
Here, $D_i^\mathrm{u}$ denotes the diffusion coefficient of uncondensed ions. The full expression for $k^\mathrm{e}$ is shown in the supporting information (equation \ref{E:Manning_hindrance_factor}). \par
Condensed ions are often treated as immobile \cite{Manning1969a, Kamcev2018}. Alternatively, the diffusion coefficient of condensed ions $D_i^\mathrm{c}$ can be assumed to relate to the diffusion coefficient of uncondensed ions through a proportionality factor $\alpha$ \cite{Kitto2022, Kamcev2018}:
\begin{equation}
    \text{immobile:} \quad D_i^\mathrm{c} = 0 
    \ , \quad \text{mobile:} \quad
    \frac{D_i^\mathrm{c}}{D_i^\mathrm{u}} = \frac{\alpha}{3}
\end{equation}
In the latter assumption, the factor $1/3$ accounts for the random orientation of polymer chains in three spatial directions \cite{Kamcev2018}. \par
\begin{figure}[h]
    \centering
    \def\svgwidth{7.5cm}
    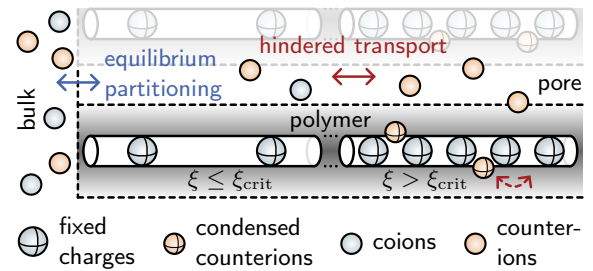
    \vspace{0.2cm}
    \caption{Schematic representation of the Donnan-Manning model}
    \label{F:DM_model_schematic}
\end{figure}

The entirety of this concept with partitioning, condensation and transport is depicted schematically in figure \ref{F:DM_model_schematic}. \par
\par
Note that in multi-ionic environments it can be useful to estimate the condensed concentrations $c_i^\mathrm{c}$ of all counterions. These can be approximated by assuming a proportionality to charge, radius and uncondensed concentration $c_i^\mathrm{u}$ \cite{Purpura2024, Wang2023}, e.g.:
\begin{equation}
c_i^\mathrm{c} = \left(c_\mathrm{X} - \frac{q_\mathrm{eff}}{z_\mathrm{X}}\right) \frac{\lvert z_\mathrm{X} \rvert \, c_i^\mathrm{u} /r_i}{\sum_j \lvert z_j \rvert \, c_j^\mathrm{u} /r_j} \quad \forall\, i,j \text{ with } z_{i/j} z_\mathrm{X} < 0
\label{E:Donnan_Manning_salt_mixture_condensation}
\end{equation}

\subsubsection{low-T* model}
The low-T* model describes membranes in which electrostatic attraction to fixed charges dominates over thermal motion. Thus, all ions are in a condensed state and transport is attributed solely to ion-hopping between adjacent sites. This applies to highly charged, low dielectric membranes, where the distance between the fixed charges is much smaller than the Bjerrum length \cite{Freger2020}:
\begin{equation}
    L \ll \lambda_\mathrm{B} 
    \quad \text{or equally} \quad
    \varepsilon_\mathrm{p} \ll  \frac{e^2}{4\pi \varepsilon_\mathrm{0} k_\mathrm{B} T L}
\end{equation}
The low-T* model assumes that all mobile ions interact with only one fixed charge at a time, which is given if the closest possible approach $b$ is much shorter than the Bjerrum length $\lambda_\mathrm{B}$, i.e. the reduced temperature $T^\ast = b/\lambda_\mathrm{B}$ is low.\par
\begin{figure}[h]
    \centering
    \def\svgwidth{7.5cm}
    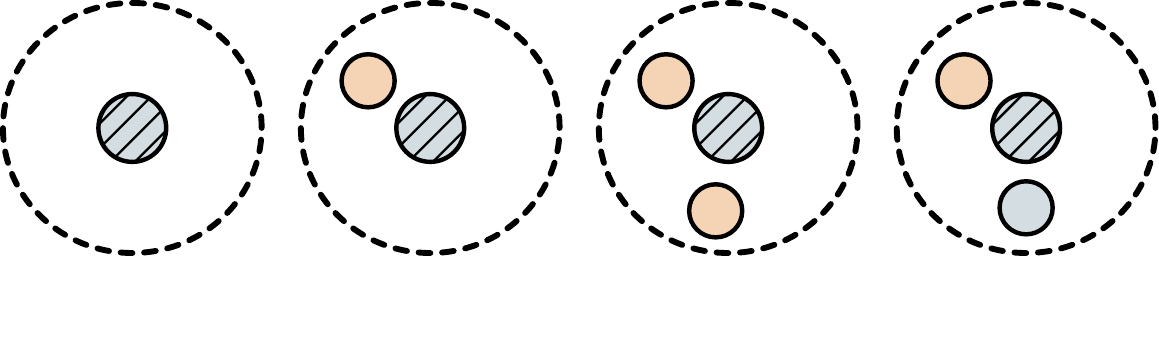
    \caption{Schematic representation of the basic occupational states of fixed charge sites for a monovalent permeating salt \cite{Freger2020}. Colors indicate the sign of ion charge, the hatched pattern denotes fixed sites.}
    \label{F:Freger_X_XM_XM2_XMA}
\end{figure}

Assuming that high local charge imbalances as well as large complexes are unlikely, it is sufficient to only consider complexes with a maximum of two mobile ions around a fixed charge $\mathrm{X}$ with a local charge imbalance of no more than one. For a monovalent $\mathrm{MA}$ salt with counterions $\mathrm{M}$ and coions $\mathrm{A}$, this leaves four possible occupational states for the fixed site (see figure \ref{F:Freger_X_XM_XM2_XMA}) \cite{Freger2020}.
\par
Based on these assumptions, an analytic expression for the partitioning can be derived. For a monovalent salt this yields the following coion and counterion partitioning coefficients $S_\mathrm{A} = c_\mathrm{A}^\mathrm{m} /c_\mathrm{A}^\mathrm{b}$ and $S_\mathrm{M} = c_\mathrm{M}^\mathrm{m} /c_\mathrm{M}^\mathrm{b}$, respectively \cite{Freger2020}:
\begin{equation}
\begin{aligned}
    &S_\mathrm{A} = \frac{c_\mathrm{X}}{c_\mathrm{s}^\mathrm{b}} \frac{K_\mathrm{XMA}\left(S_\mathrm{0} c_\mathrm{s}^\mathrm{b}\right)^2}{\zeta} \, , 
    \quad
    S_\mathrm{M} = \frac{c_\mathrm{X}}{c_\mathrm{s}^\mathrm{b}} + S_\mathrm{A} \\
    &\text{with} \quad
    \zeta =
     K_\mathrm{XM} \left(\frac{1 + K_\mathrm{XMA}\left(S_\mathrm{0} c_\mathrm{s}^\mathrm{b}\right)^2}{K_\mathrm{XM_2}} \right)^{1/2} \\
    & \hspace{1.5cm}+ 2\left(1 + K_\mathrm{XMA}\left(S_\mathrm{0} c_\mathrm{s}^\mathrm{b}\right)^2\right)
\end{aligned}
\end{equation}
Here, $S_0$ is a salt partitioning coefficient given by $c_\mathrm{M}c_\mathrm{A} = (S_0 c_\mathrm{s}^\mathrm{b})^2$, where $c_\mathrm{s}^\mathrm{b}$ is the bulk salt concentration. It can, e.g., account for steric or dielectric exclusion. The association constants are assumed to mainly depend on the number of monomers $n$ in a complex and thus are approximated by a geometric series in rough accordance with Bjerrum theory \cite{Freger2020, Aqua2005}:
\begin{equation}
\begin{split}
    &\frac{K_n}{K_{n-1}} \approx 10 b^3 {T^{*}}^m \exp\left( \Delta \alpha_n / T^*\right)\\
    &\text{with} \quad \Delta \alpha_n \in \left(0,1\right) \quad \text{and} \quad m \in \left[1,2 \right]
    \label{E:low-T_association_constants_geometric_series}
\end{split}
\end{equation}
Diffusion is understood as ions escaping the associated range of one fixed charge site and reattaching at an available neighboring site ("hopping"), since local interactions dominate over random thermal motion. This is characterized by the mean escape times $\langle \ell \rangle$ \cite{Freger2020, Lifson1962, Jackson1963}:
\begin{equation}
\begin{split}
    &\langle \ell_{\mathrm{M,XM}} \rangle \approx \frac{K_\mathrm{XM}}{4\pi D_\mathrm{M}^\mathrm{b} \lambda_\mathrm{B}} 
    \ , \quad
    \langle \ell_{\mathrm{M,XM_2}} \rangle \approx \frac{K_\mathrm{XM_2}}{4\pi D_\mathrm{M}^\mathrm{b} b K_\mathrm{XM}}\\
    & \text{and} \quad \langle \ell_{\mathrm{A,XMA}} \rangle \approx \frac{K_\mathrm{XMA}}{4\pi D_\mathrm{A}^\mathrm{b} b K_\mathrm{XM}}
\end{split}
\end{equation}
Here, $D_\mathrm{A}^\mathrm{b}$ and $D_\mathrm{M}^\mathrm{b}$ denote the bulk diffusion coefficient of the co- and counterion, respectively.
The escape of a counterion $\mathrm{M}$ from an $\mathrm{XMA}$ state is neglected, since this would yield an unlikely same-sign $\mathrm{XA}$ state.\par
The probability $\mathcal{P}$ of a neighboring site being in a state in which it can accommodate for the escaped $\mathrm{M}$ or $\mathrm{A}$ ion is \cite{Freger2020}:
\begin{equation}
    \mathcal{P}_\mathrm{M} = \frac{1 + K_\mathrm{XM} \left[\mathrm{M}\right]}{\zeta} 
    \quad \text{and} \quad
    \mathcal{P}_\mathrm{A} = \frac{K_\mathrm{XM} \left[\mathrm{M}\right]}{\zeta} 
\end{equation}
From that, the diffusion coefficient can be obtained using the Einstein relation with the mean square displacement $\Delta r^2 = L^2$ of a successful hop \cite{Freger2020, Einstein1905, VonSmoluchowski1906}:
\begin{equation}
\begin{split}
    &D_\mathrm{M}^\mathrm{m} = \frac{L^2}{6} \mathcal{P}_\mathrm{M} \left( \frac{1}{\langle \ell_{\mathrm{M,XM}} \rangle} + \frac{1}{\langle \ell_{\mathrm{M,XM_2}} \rangle} \right) \\
    &D_\mathrm{A}^\mathrm{m} = \frac{L^2}{6}  \frac{\mathcal{P}_\mathrm{A}}{\langle \ell_{\mathrm{A,XMA}} \rangle }
\end{split}
\end{equation}
These considerations can be done analogously for divalent $\mathbb{M}\mathrm{A}_2$ salts \cite{Freger2020}. With the resulting diffusion coefficients, pore transport can once again be modeled by the Nernst-Planck equation.

\subsection{Mesoscale hindrance models}
\label{C:Mesoscale_models}
For an accurate macroscopic depiction of membrane behaviour, it is essential to combine appropriate pore and mesoscale models. Here, \textit{mesoscale} refers to the intermediate scale that maps pore-scale transport onto macroscopic effective transport via porosity $\varphi$ and tortuosity $\tau$. In accordance with the terminology above, we denote the hindrance factor arising from mesoscale effects as $k^\mathcal{M}$:
\begin{equation}
    k^\mathcal{M} = \frac{D_i^\mathrm{m}} {D_i^\mathrm{p}} = \frac{\varphi}{\tau^2}
    \label{E:Mesoscale_hindrance_factor}
\end{equation}
Here, $D_i^\mathrm{m}$ denotes the macroscopic membrane diffusion coefficient and $D_i^\mathrm{p}$ the pore diffusion coefficient.
Equation \ref{E:Mesoscale_hindrance_factor} can be simplified e.g. by assuming that tortuosity scales with porosity ($\tau \propto \varphi^{1-\beta}$) with the Bruggeman coefficient $\beta$. It is also not uncommon in membrane modeling to further introduce a percolation threshold $\varphi_\mathrm{pt}$, which yields
\begin{equation}
    k^\mathcal{M} = \left(\varphi -\varphi_\mathrm{pt}\right)^\beta
    \quad \text{or} \quad
    k^\mathcal{M} = \varphi ^\beta
\end{equation}
Another common simplification is the Mackie-Meares model, which assumes random obstruction in a highly swollen polymer network. The resulting estimation depends solely on the volume fraction of the permeating solution $\varphi_\mathrm{w}$ \cite{Mackie1955, Kamcev2017, Kitto2022}:
\begin{equation}
    k^\mathcal{M} = \varphi_\mathrm{w} \left(\frac{\varphi_\mathrm{w}}{2 - \varphi_\mathrm{w}} \right)^2
    \label{E:Mackie_Meares}
\end{equation}
The Mackie-Meares model is a conservative estimate, which is of sufficient accuracy for highly hydrated membranes ($\varphi_w \geq 0.4$). For low hydration, it tends to overestimate the hindrance, since membranes with low water uptake often express distinct water pathways \cite{Marioni2024}.

\subsection{Assessment of modeling approaches}
\label{C:model_assessment}
\begin{figure}[h]
    \centering
    \vspace{-0.1cm}
    \import{2_Background/Figures}{Coulombic_overlay.tex}
    \vspace{-0.1cm}
    \caption{Treatment of the electric potential: Homogeneous in the SDE-model, distinct interactions in the low-T* model}
    \label{F:Approaches_to_the_electric_potential}
\end{figure}
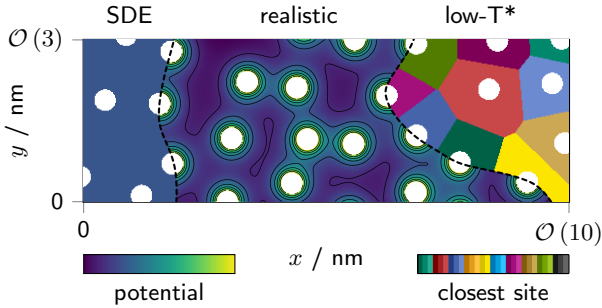

All presented models aim at describing the partitioning and permeability of membranes in the range of nanofiltration based on thermodynamic, kinetic and geometric aspects. The main difference between these models is the treatment of the interactions of fixed charges with mobile ions. The SDE model is a mean-field model that assumes a homogeneous potential within the membrane and neglects distinct interactions of fixed and mobile charges. The low-T* model focuses on these interactions and attributes transport to a hopping mechanism. These approaches are illustrated qualitatively in figure \ref{F:Approaches_to_the_electric_potential}.
\par
The Donnan-Manning model, while also being a mean-field model, introduces a thermodynamic threshold for the inevitable onset of counterion condensation. This threshold arises from limiting laws \cite{Manning1969} and can be interpreted as a non-distinct interaction of fixed sites with mobile ions standing in between the mean-field approach of the SDE model and the distinct interactions in the low-T* model. This approach has proven advantageous, with the Donnan-Manning model being largely consistent for a broad range of membranes and various salts at different concentrations. In contrast, the SDE model fails for such broadly varying conditions due to its sensitive fitting parameters \cite{Santafe-Moros2008}.\par
There are also several extensions to the Donnan-Manning model \cite{Mareev2022}, such as including dielectric exclusion \cite{Bannon2024} or condensation criteria for salt mixtures \cite{Wang2023, Purpura2024}, which have shown to improve accuracy for specific cases. Some of the known drawbacks of Manning condensation stem from the mean-field nature of the approach. For instance, it cannot depict phenomena like charge inversion \cite{Shklovskii1999} or charge oscillation \cite{Deserno2000} observed for sufficient concentrations of multivalent counterions.\par
Since the topic of membrane filtration is broadly studied, there are many more recent models and model extensions available which introduce additional ideas and are able to improve modeling accuracy for various conditions \cite{Wang2024, Yu2019, Purpura2024, Crothers2020}. This wide range of models and model extensions justifies the need for a framework for membrane modeling that allows for consistent combination of suitable assumptions for a broad range of membranes. 
Therefore, in the following we derive a thermodynamically consistent yet descriptive framework for membrane modeling that unifies key assumptions from existing models into a coherent picture. This model avoids imposing restrictions towards multi-ionic environments, making it structurally applicable to membranes in aqueous batteries.

%% file: 2_Background/Figures/DSPM-DE_schematic_concept_new.pdf_tex
\begingroup%
  \makeatletter%
  \providecommand\color[2][]{%
    \errmessage{(Inkscape) Color is used for the text in Inkscape, but the package 'color.sty' is not loaded}%
    \renewcommand\color[2][]{}%
  }%
  \providecommand\transparent[1]{%
    \errmessage{(Inkscape) Transparency is used (non-zero) for the text in Inkscape, but the package 'transparent.sty' is not loaded}%
    \renewcommand\transparent[1]{}%
  }%
  \providecommand\rotatebox[2]{#2}%
  \newcommand*\fsize{\dimexpr\f@size pt\relax}%
  \newcommand*\lineheight[1]{\fontsize{\fsize}{#1\fsize}\selectfont}%
  \ifx\svgwidth\undefined%
    \setlength{\unitlength}{330.27815451bp}%
    \ifx\svgscale\undefined%
      \relax%
    \else%
      \setlength{\unitlength}{\unitlength * \real{\svgscale}}%
    \fi%
  \else%
    \setlength{\unitlength}{\svgwidth}%
  \fi%
  \global\let\svgwidth\undefined%
  \global\let\svgscale\undefined%
  \makeatother%
  \begin{picture}(1,0.47079822)%
    \lineheight{1}%
    \setlength\tabcolsep{0pt}%
    \put(0,0){\includegraphics[width=\unitlength,page=1]{DSPM-DE_schematic_concept_new.pdf}}%
    \put(0.22799708,0.23768975){\color[rgb]{0.2745098,0.39215686,0.66666667}\makebox(0,0)[lt]{\lineheight{1.25}\smash{\begin{tabular}[t]{l}equilibrium\\partitioning\end{tabular}}}}%
    \put(0.71805627,0.28587377){\color[rgb]{0.63529412,0.13333333,0.1372549}\makebox(0,0)[t]{\lineheight{1.25}\smash{\begin{tabular}[t]{c}hindered transport\end{tabular}}}}%
    \put(0,0){\includegraphics[width=\unitlength,page=2]{DSPM-DE_schematic_concept_new.pdf}}%
    \put(0.09543445,0.27844961){\color[rgb]{0,0,0}\rotatebox{90}{\makebox(0,0)[t]{\lineheight{1.25}\smash{\begin{tabular}[t]{c}bulk\end{tabular}}}}}%
    \put(0.99177238,0.27921098){\color[rgb]{0,0,0}\rotatebox{90}{\makebox(0,0)[t]{\lineheight{1.25}\smash{\begin{tabular}[t]{c}pore\end{tabular}}}}}%
    \put(0,0){\includegraphics[width=\unitlength,page=3]{DSPM-DE_schematic_concept_new.pdf}}%
    \put(0.08349704,0.01396167){\color[rgb]{0,0,0}\makebox(0,0)[lt]{\lineheight{1.25}\smash{\begin{tabular}[t]{l}fixed charges\end{tabular}}}}%
    \put(0.49154853,0.01102705){\color[rgb]{0,0,0}\makebox(0,0)[lt]{\lineheight{1.25}\smash{\begin{tabular}[t]{l}counterions\end{tabular}}}}%
    \put(0.90370125,0.01102705){\color[rgb]{0,0,0}\makebox(0,0)[lt]{\lineheight{1.25}\smash{\begin{tabular}[t]{l}coions\end{tabular}}}}%
    \put(0,0){\includegraphics[width=\unitlength,page=4]{DSPM-DE_schematic_concept_new.pdf}}%
  \end{picture}%
\endgroup%

%% file: 2_Background/Figures/Donnan_Manning_schematic.pdf_tex
\begingroup%
  \makeatletter%
  \providecommand\color[2][]{%
    \errmessage{(Inkscape) Color is used for the text in Inkscape, but the package 'color.sty' is not loaded}%
    \renewcommand\color[2][]{}%
  }%
  \providecommand\transparent[1]{%
    \errmessage{(Inkscape) Transparency is used (non-zero) for the text in Inkscape, but the package 'transparent.sty' is not loaded}%
    \renewcommand\transparent[1]{}%
  }%
  \providecommand\rotatebox[2]{#2}%
  \newcommand*\fsize{\dimexpr\f@size pt\relax}%
  \newcommand*\lineheight[1]{\fontsize{\fsize}{#1\fsize}\selectfont}%
  \ifx\svgwidth\undefined%
    \setlength{\unitlength}{384.5105413bp}%
    \ifx\svgscale\undefined%
      \relax%
    \else%
      \setlength{\unitlength}{\unitlength * \real{\svgscale}}%
    \fi%
  \else%
    \setlength{\unitlength}{\svgwidth}%
  \fi%
  \global\let\svgwidth\undefined%
  \global\let\svgscale\undefined%
  \makeatother%
  \begin{picture}(1,0.43932128)%
    \lineheight{1}%
    \setlength\tabcolsep{0pt}%
    \put(0,0){\includegraphics[width=\unitlength,page=1]{Donnan_Manning_schematic.pdf}}%
    \put(0.15525441,0.33662305){\color[rgb]{0.2745098,0.39215686,0.66666667}\makebox(0,0)[lt]{\lineheight{1.25}\smash{\begin{tabular}[t]{l}equilibrium\\partitioning\end{tabular}}}}%
    \put(0,0){\includegraphics[width=\unitlength,page=2]{Donnan_Manning_schematic.pdf}}%
    \put(0.03258426,0.25921775){\color[rgb]{0,0,0}\rotatebox{90}{\makebox(0,0)[t]{\lineheight{1.25}\smash{\begin{tabular}[t]{c}bulk\end{tabular}}}}}%
    \put(0.99690803,0.29479334){\color[rgb]{0,0,0}\makebox(0,0)[rt]{\lineheight{1.25}\smash{\begin{tabular}[t]{r}pore\end{tabular}}}}%
    \put(0.55333905,0.22936218){\color[rgb]{0,0,0}\makebox(0,0)[t]{\lineheight{1.25}\smash{\begin{tabular}[t]{c}polymer\end{tabular}}}}%
    \put(0.59389268,0.35851439){\color[rgb]{0.63529412,0.13333333,0.1372549}\makebox(0,0)[t]{\lineheight{1.25}\smash{\begin{tabular}[t]{c}hindered transport\end{tabular}}}}%
    \put(0,0){\includegraphics[width=\unitlength,page=3]{Donnan_Manning_schematic.pdf}}%
    \put(0.37823953,0.12425845){\color[rgb]{0,0,0}\makebox(0,0)[t]{\lineheight{1.25}\smash{\begin{tabular}[t]{c}$\xi\leq \xi_\mathrm{crit}$\end{tabular}}}}%
    \put(0.72158129,0.12425845){\color[rgb]{0,0,0}\makebox(0,0)[t]{\lineheight{1.25}\smash{\begin{tabular}[t]{c}$\xi > \xi_\mathrm{crit}$\end{tabular}}}}%
    \put(0.07507681,0.04049437){\color[rgb]{0,0,0}\makebox(0,0)[lt]{\lineheight{1.25}\smash{\begin{tabular}[t]{l}fixed\\charges\end{tabular}}}}%
    \put(0.84695663,0.03872289){\color[rgb]{0,0,0}\makebox(0,0)[lt]{\lineheight{1.25}\smash{\begin{tabular}[t]{l}counter-\\ions\end{tabular}}}}%
    \put(0.6283949,0.01318302){\color[rgb]{0,0,0}\makebox(0,0)[lt]{\lineheight{1.25}\smash{\begin{tabular}[t]{l}coions\end{tabular}}}}%
    \put(0,0){\includegraphics[width=\unitlength,page=4]{Donnan_Manning_schematic.pdf}}%
    \put(0.31620714,0.03797371){\color[rgb]{0,0,0}\makebox(0,0)[lt]{\lineheight{1.25}\smash{\begin{tabular}[t]{l}condensed\\counterions\end{tabular}}}}%
  \end{picture}%
\endgroup%

%% file: 2_Background/Figures/Freger_X_XM_XM2_XMA.pdf_tex
\begingroup%
  \makeatletter%
  \providecommand\color[2][]{%
    \errmessage{(Inkscape) Color is used for the text in Inkscape, but the package 'color.sty' is not loaded}%
    \renewcommand\color[2][]{}%
  }%
  \providecommand\transparent[1]{%
    \errmessage{(Inkscape) Transparency is used (non-zero) for the text in Inkscape, but the package 'transparent.sty' is not loaded}%
    \renewcommand\transparent[1]{}%
  }%
  \providecommand\rotatebox[2]{#2}%
  \newcommand*\fsize{\dimexpr\f@size pt\relax}%
  \newcommand*\lineheight[1]{\fontsize{\fsize}{#1\fsize}\selectfont}%
  \ifx\svgwidth\undefined%
    \setlength{\unitlength}{556.05167419bp}%
    \ifx\svgscale\undefined%
      \relax%
    \else%
      \setlength{\unitlength}{\unitlength * \real{\svgscale}}%
    \fi%
  \else%
    \setlength{\unitlength}{\svgwidth}%
  \fi%
  \global\let\svgwidth\undefined%
  \global\let\svgscale\undefined%
  \makeatother%
  \begin{picture}(1,0.30110462)%
    \lineheight{1}%
    \setlength\tabcolsep{0pt}%
    \put(0,0){\includegraphics[width=\unitlength,page=1]{Freger_X_XM_XM2_XMA.pdf}}%
    \put(0.11406062,0.00619603){\color[rgb]{0,0,0}\makebox(0,0)[t]{\lineheight{1.25}\smash{\begin{tabular}[t]{c}\textbf{$[\mathrm{X}]$}\end{tabular}}}}%
    \put(0.3712587,0.00619603){\color[rgb]{0,0,0}\makebox(0,0)[t]{\lineheight{1.25}\smash{\begin{tabular}[t]{c}\textbf{$[\mathrm{XM}]$}\end{tabular}}}}%
    \put(0.62845676,0.00619603){\color[rgb]{0,0,0}\makebox(0,0)[t]{\lineheight{1.25}\smash{\begin{tabular}[t]{c}\textbf{$[\mathrm{XM_2}]$}\end{tabular}}}}%
    \put(0.88565494,0.00619603){\color[rgb]{0,0,0}\makebox(0,0)[t]{\lineheight{1.25}\smash{\begin{tabular}[t]{c}\textbf{$[\mathrm{XMA}]$}\end{tabular}}}}%
  \end{picture}%
\endgroup%

%% file: 2_Background/Figures/Coulombic_overlay.tex
\pgfplotsset{
    colormap={KIT30}{
        rgb255=(0, 97, 70) 
        rgb255=(0, 135, 108) 
        rgb255=(38, 173, 146) 
        rgb255=(124, 0, 0) 
        rgb255=(162, 34, 35) 
        rgb255=(200, 72, 73) 
        rgb255=(32, 62, 132) 
        rgb255=(70, 100, 170) 
        rgb255=(108, 138, 208) 
        rgb255=(185, 117, 0) 
        rgb255=(223, 155, 27) 
        rgb255=(255, 193, 65) 
        rgb255=(214, 191, 0) 
        rgb255=(252, 229, 0) 
        rgb255=(255, 255, 38) 
        rgb255=(0, 120, 184) 
        rgb255=(7, 158, 222) 
        rgb255=(45, 196, 255) 
        rgb255=(125, 0, 86) 
        rgb255=(163, 16, 124) 
        rgb255=(201, 54, 162) 
        rgb255=(129, 92, 8) 
        rgb255=(167, 130, 46) 
        rgb255=(205, 168, 84) 
        rgb255=(81, 124, 0) 
        rgb255=(119, 162, 0) 
        rgb255=(157, 200, 38) 
        rgb255=(26, 26, 26) 
        rgb255=(64, 64, 64) 
        rgb255=(102, 102, 102) 
    },
}

\begin{tikzpicture}
    \begin{axis}[
        name=mainaxis,
        width=8cm,  
        colorbar horizontal,
        colormap/viridis,
        colorbar style={
            at = {(mainaxis.south west)},
            anchor = north west,
            yshift = -0.4cm,
            width = 2cm,
            height = 2.5mm,
            point meta min=0, 
            point meta max=1, 
            ticks = none,
            xlabel=potential,
        }  ,
        axis line style={thick}, 
        axis equal image,
        xtick align=outside,
        ytick align=outside,
        scaled ticks=false,
        xtick pos=bottom,
        ytick pos=left,
        xtick={0, 7.5e-9},  
        ytick={0, 2.5e-9}, 
        xticklabels={0, $\mathcal{O}\left(10\right)$},  
        yticklabels={0, $\mathcal{O}\left(3\right)$}, 
        xlabel= {$x$ / nm}, 
        ylabel= {$y$ / nm},
        xlabel style={yshift=0.2cm},
        ylabel style={yshift=-0.5cm},
        xmin=0, xmax=7.5e-9, 
        ymin=0, ymax=2.5e-9, 
        title = {\hspace{-0.5cm} SDE \hspace{1.2cm} realistic \hspace{1.2cm} low-T*},
        title style={yshift=-0.1cm},
        ]
        \addplot [draw=none, point meta=explicit] coordinates {(0,0) [0] (7.5e-9,0) [1]};
        \addplot graphics [xmin=0,xmax=7.5e-9,ymin=0,ymax=2.5e-9] {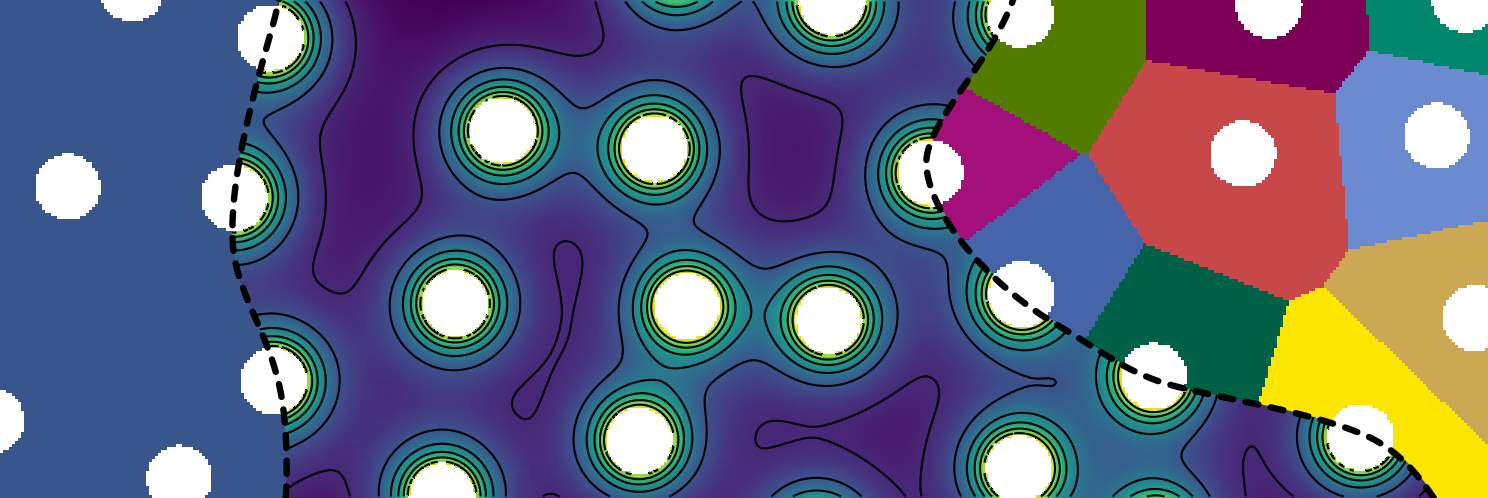};
    \end{axis}
    \begin{axis}[
        at={(mainaxis.south west)}, 
        anchor=south west, 
        width=8cm,
        colorbar horizontal,
        colormap name=KIT30,
        colorbar sampled,
        colormap access=piecewise constant,
        hide axis, 
        colorbar style={
            at = {(mainaxis.south east)},
            samples=30,
            point meta min=0, 
            point meta max=1, 
            yshift = -0.4cm,
            anchor = north east,
            width=2cm,
            height=2.5mm,
            ticks=none,
            xlabel=closest site,
        },
    ]
    \addplot [draw=none, point meta=explicit] coordinates {(0,0) [0] (1,0) [1]};
    \end{axis}
\end{tikzpicture}

%% file: 3_Modeling/0_main.tex
\section{Modeling}
\label{C:Modeling}
As shown in the previous sections, membranes are inhomogeneous on multiple scales. These inhomogeneities can be, e.g., varying monomers along the polymer chain, hydrophobic and hydrophilic domains, or pores (hydrated domains) of various sizes. In this section, we derive a model to describe membranes on length scales at which they can be considered statistically homogeneous. This requires a suitable combination of submodels including a pore model and a porosity-tortuosity model. However, the following derivations focus on the pore model, since the porosity-tortuosity models shown above are commonly considered sufficient. The resulting membrane model highlights connections between the presented existing approaches and remains structurally applicable for multi-ionic environments.

\subsection{Approach}
\label{C:Approach}
In analogy to the Donnan-Manning model, we distinguish two states for ions in membrane. One state is dominated by electrostatic interaction with fixed charges, the other one is governed by random thermal motion. In accordance with the Donnan-Manning model, these states are labeled \textit{condensed} and \textit{uncondensed}. This approach is supported e.g. by studies on the dielectric properties of ionomers, which point out the existence of distinct states of permeating species including bulk-like, loosely bound and strongly bound states \cite{Lu2010, Lu2008}. Given that these states are in dynamic equilibrium \cite{Manning1979, Schurr2002}, it is likely that the resulting overall partitioning and permeability are also rooted in the interplay of both states \cite{Fong2019}.\par
\begin{figure}[h]
    \centering
    \def\svgwidth{7.5cm}
    \hspace{-0.2cm}
    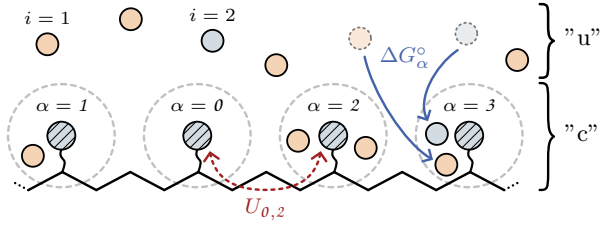
    \caption{Illustration of model variables and indexing. For the depicted system, the occupation fractions $\theta_{\alpha\in\{\mathpzc{0},\mathpzc{1},\mathpzc{2},\mathpzc{3}\}}$ would be $\frac{1}{4}$ each.}
    \label{F:My_model_variables}
\end{figure}

Therefore, we introduce separate concentrations for the two states, denoting the concentration of uncondensed ions as $c_i^\mathrm{u}$ and the concentration of condensed ions as $c_i^\mathrm{c}$ for $i \in \{1,\dots,N\}$. Both concentrations are referenced to the total hydrated domain within the membrane, since distinct volume fractions associated with each state would be difficult to determine. Thus, the total concentration in the membrane is:
\begin{equation}
    c_i^\mathrm{m} = c_i^\mathrm{u} + c_i^\mathrm{c}
    \label{E:Total_concentration}
\end{equation}
For the neutral solvent, we assume positional indifference and simply write $c_0$ \cite{Schammer2021, Kilchert2023}. As before, the fixed site concentration is denoted by $c_\mathrm{X}$. These fixed sites can be in various occupational states $\alpha \in \{ \mathpzc{0},\mathpzc{1}, \dots, N_\mathrm{\theta} \}$ including the empty state $\alpha=\mathpzc{0}$. Since multi-occupation is possible in general, the state indices do not correspond to the species indices $\alpha \nsim i$. Each occupational state accounts for a fraction $\theta_\alpha$ of the total site concentration, which we normalize to fulfill:
\begin{equation}
    \sum_\alpha \theta_\alpha = 1
    \label{E:Theta_normalization}
\end{equation}
Consequently, the occupational states relate to the condensed concentrations by:
\begin{equation}
    c_i^\mathrm{c} = c_\mathrm{X} \sum_\alpha \nu_{i\alpha} \theta_\alpha
    \label{E:Condensed_concentrations}
\end{equation}
Here, $\nu_{i\alpha}$ is the stoichiometric coefficient of ions $i$ in state $\alpha$. This indexing is illustrated in figure \ref{F:My_model_variables}. Moreover, we introduce the molar free energy change associated with the formation of a certain state $\Delta G_\alpha^\circ$ and a dimensionless symmetric nearest neighbor interaction energy $U_{\alpha\gamma}$ (where both $\alpha$ and $\gamma$ represent all occupational states). Both these quantities are introduced with further detail in the following sections, but are also already depicted schematically in figure \ref{F:My_model_variables}.
\par

\subsection{Thermodynamics}
\label{C:Thermodynamics}
For the bulk electrolyte surrounding the membrane, we write the free energy density $g^\mathrm{b}$ as the sum of ideal mixing, electrostatic, and remaining linear \textit{excess} contributions. Assuming sufficient abundance of neutral solvent to eliminate its entropic contribution, we have:
\begin{equation}
\begin{split}
    g^\mathrm{b} = &
    \underbrace{R T \sum_i \left( c_i^\mathrm{b} \ln \frac{c_i^\mathrm{b}}{c^\circ} - c_i^\mathrm{b} \right)}_\text{entropy} 
    + \underbrace{F \sum_i c_i^\mathrm{b} z_i \Phi^\mathrm{b}}_\text{electric potential} \\&
    + \underbrace{\sum_i c_i^\mathrm{b} \mu_i^\mathrm{b,ex}}_\text{excess} 
\end{split}
\label{E:Free_energy_bulk}
\end{equation}
The free energy density of the membrane $g^\mathrm{m}$ is constructed similarly, with the added distinction of condensed and uncondensed states. The entropic contribution of uncondensed ions is described by ideal mixing with sufficiently abundant neutral solvent (see supporting information \ref{C:Uncondensed_entropy}), whereas for condensed states it results from occupation statistics. Additionally, we introduce a mean-field nearest-neighbor interaction term to account for the low electrostatic shielding along the polymer backbone. Moreover, both condensed and uncondensed states contribute individually to electrostatics and the excess. For these contributions, we assume that the offsets between condensed and
uncondensed states are independent of local concentrations and absorb them into effective standard formation free energies of the occupation states, $\Delta G_\alpha^\circ$ (see supporting information \ref{C:Electric_potential_contribution} - \ref{C:Excess_free_energy_contribution} and equation \ref{E:Effective_formation_free_energy}). Thus, the remaining electrostatic and excess contributions refer to the uncondensed state, using $\Phi^\mathrm{u}$ and $\mu^{\mathrm{u,ex}}$, but apply to the total concentration $c_i^\mathrm{m} = c_i^\mathrm{u} + c_i^\mathrm{c}$. Overall, the free energy density in the membrane is:
\begin{equation}
\begin{split}
    g^\mathrm{m} =
    &\underbrace{R T \sum_i \left( c_i^\mathrm{u} \ln \frac{c_i^\mathrm{u}}{c^\circ}  - c_i^\mathrm{u} \right)}_\text{entropy uncondensed} 
    + \underbrace{c_\mathrm{X} R T \sum_\alpha \theta_\alpha \ln \theta_\alpha}_\text{entropy condensed} \\
    &+ \underbrace{c_\mathrm{X} \sum_\alpha \theta_\alpha \Delta G_\alpha^\circ }_\text{state formation (effective)} 
    +\underbrace{\frac{1}{2} R T c_\mathrm{X} n_\mathrm{n} \sum_{\alpha,\gamma} \theta_\alpha \theta_\gamma U_{\alpha\gamma}}_\text{mean-field interaction} \\
    &+ \underbrace{F \sum_i c_i^\mathrm{m} z_i \Phi^\mathrm{u}}_\text{electric potential} 
    + \underbrace{ \sum_i c_i^\mathrm{m} \mu_i^\mathrm{u,ex} }_\text{excess} 
\end{split}
\label{E:Free_energy_membrane}
\end{equation}
Here, the index $i \in \{1,\dots,N\}$ denotes ionic species, while $\alpha,\gamma \in \{ \mathpzc{0},\mathpzc{1}, \dots, N_\mathrm{\theta} \}$ denote occupational states including the empty state (see section \ref{C:Approach}). The number of nearest neighbors is denoted by $n_\mathrm{n}$ and nearest-neighbor interaction is assumed to be symmetric $U_{\alpha\gamma} = U_{\gamma\alpha}$.\par
From this free energy density, we now derive the electrochemical potentials of ions in the membrane. Since the condensed concentrations $c_i^\mathrm{c}$ depend on the normalized occupation fractions $\theta_\alpha$ (see equations \ref{E:Condensed_concentrations} and \ref{E:Theta_normalization}), they are not independent variables. We therefore impose local equilibrium by constrained minimization instead of explicitly constructing $\overline{\mu}_i^\mathrm{c}$ to impose $\overline{\mu}_i^\mathrm{c} = \overline{\mu}_i^\mathrm{u}$. \par
For this minimization approach, we first eliminate $\theta_\mathpzc{0}$ through the normalization constraint (equation \ref{E:Theta_normalization}) and use the concentration relations (equations \ref{E:Total_concentration} and \ref{E:Condensed_concentrations}) to regard the free energy density as:
\begin{equation}
    g^\mathrm{m} = g^\mathrm{m} \left(c_1^\mathrm{m},\dots,c_N^\mathrm{m},\theta_{\mathpzc{1}},\dots,\theta_{N_\theta}\right)
\end{equation}
Assuming fast equilibration of condensed and uncondensed states, the occupation fractions adjust instantaneously to locally minimize $g^\mathrm{m}$ for any fixed composition $c^\mathrm{m}$. Therefore, the equilibrium condition is:
\begin{equation}
    \left.\frac{\partial g^\mathrm{m}}{\partial \theta_\alpha}\right|_{c^\mathrm{m}} 
    = 0  
    \qquad \forall \, \alpha \neq \mathpzc{0}
    \label{E:Stationarity_condition}
\end{equation}
For occupation fractions satisfying this condition, the common electrochemical potential $\overline{\mu}_i^\mathrm{m}$ of an ionic species $i$ in the membrane is obtained by the envelope theorem while using $\left.\frac{\partial c_i^\mathrm{u}}{\partial c_i^\mathrm{m}}\right|_\theta = 1$:
\begin{equation}
    \overline{\mu}_i^\mathrm{m}
    =
    \left. \frac{\partial g^\mathrm{m}}{\partial c_i^\mathrm{m}}\right|_{\theta(c^\mathrm{m})}
    =
    R T \ln \left(\frac{c_i^\mathrm{u}}{c^\circ} \right) + z_i F \Phi^\mathrm{u} + \mu_i^\mathrm{u,ex}
    \label{E:Electrochemical_potential_membrane}
\end{equation}
Note that at local equilibrium, this common membrane potential corresponds to the state-resolved potentials $\overline{\mu}_i^\mathrm{m} \equiv \overline{\mu}_i^\mathrm{c}= \overline{\mu}_i^\mathrm{u}$.\par
The occupation fractions of the condensed phase follow from the stationarity conditions (equation \ref{E:Stationarity_condition}), which yield for $\alpha \neq 0$:
\begin{equation}
\begin{split}
    c_\mathrm{X} R T\bigg[ & 
        \ln \left(\frac{\theta_\alpha}{\theta_\mathpzc{0}} \right) - \sum_i \nu_{i\alpha} \ln \left( \frac{c_i^\mathrm{u}}{c^\circ} \right) \\
        & + \frac{\Delta G_\alpha^\circ}{R T} + n_\mathrm{n}  \sum_{\gamma} \theta_\gamma \left(U_{\alpha\gamma} - U_{\mathpzc{0}\gamma}\right)
    \bigg]
    =
    0
\end{split}
\label{E:Occupation_condition}
\end{equation}
This system of equations is non-trivial and has to be solved simultaneously with the charge neutrality condition:
\begin{equation}
    \sum_i z_i c_i^\mathrm{m} = - z_\mathrm{X} c_\mathrm{X}
    \label{E:IOM_charge_neutrality}
\end{equation}
However, the system of condensation conditions in equation \ref{E:Occupation_condition} can further be reduced to a single implicit equation, as shown in \ref{C:Single_equation_form}.

\subsubsection{Interaction of adjacent sites}
\label{C:nn_Interaction}
We assume that site spacing is independent of the occupational states and approximate all occupational states as monopoles. The total charge $q$ of a site in state $\alpha$ is:
\begin{equation}
    q_\alpha = z_\mathrm{X} + \sum_i \nu_{i\alpha} z_i
\end{equation}
Along these assumptions, all interaction energies $U_{\alpha\gamma}$ in equations \ref{E:Free_energy_membrane} and \ref{E:Occupation_condition} can be written as a multiple of a dimensionless constant $U$, which is a function of site spacing:
\begin{equation}
\begin{gathered}
    U_{\alpha\gamma} = q_\alpha q_\gamma U\\
\end{gathered}
\label{E:Interaction_terms_U}
\end{equation}

\subsubsection{Single-equation form of the condensation conditions}
\label{C:Single_equation_form}
To rewrite the stationarity conditions for condensation (equation \ref{E:Occupation_condition}) in a compact form, we introduce mass-action association constants $K_\alpha = \exp (-\Delta G_\alpha^\circ/RT)$. For occupational states $\alpha\neq \mathpzc{0}$, this yields:
\begin{equation}
    \theta_\alpha = \theta_\mathpzc{0} K_\alpha \prod_i \left(\frac{c_i^\mathrm{u}}{c^\circ}\right)^{\nu_{i\alpha}} \exp \left( - n_\mathrm{n} \sum_{\gamma} \theta_\gamma \left(U_{\alpha\gamma} - U_{\mathpzc{0}\gamma}\right)\right)
    \label{E:Equilibrium_equation_system}
\end{equation}
The definition of the interaction terms in section \ref{C:nn_Interaction} gives:
\begin{equation}
    U_{\alpha\gamma}-U_{\mathpzc{0}\gamma} = (q_\alpha-q_\mathpzc{0}) q_\gamma U = \tilde{q}_\alpha q_\gamma U 
\end{equation}
Here, we introduced $\tilde{q}_\alpha = (q_\alpha-z_\mathrm{X}) = \sum_i \nu_{i\alpha} z_i$ for compactness, using $q_\mathpzc{0} = z_\mathrm{X}$.\par
Next, we introduce a mean effective site valence $Z$:
\begin{equation}
    Z \equiv \sum_\alpha q_\alpha \theta_\alpha
    \label{E:Definition_of_abbreviation_Z}
\end{equation}
This reduces the system of equations defined by equation \ref{E:Equilibrium_equation_system} to:
\begin{equation}
    \theta_\alpha = \theta_\mathpzc{0} K_\alpha \prod_i \left(\frac{c_i^\mathrm{u}}{c^\circ}\right)^{\nu_{i\alpha}}  \exp \left(-n_\mathrm{n} \tilde{q}_\alpha Z U\right)
    \label{E:System_of_Mass_action_equations_simplified} \quad \forall \, \alpha \neq \mathpzc{0}
\end{equation}
To eliminate the dependency on the fraction of empty sites $\theta_\mathpzc{0}$, we use the normalization $\theta_\mathpzc{0} = 1 - \sum_{\alpha \neq \mathpzc{0}} \theta_\alpha$ and define for all $\alpha \neq 0$:
\begin{equation}
\begin{split}
    &\chi_\alpha =
    \frac{\theta_\alpha}{\theta_\mathpzc{0}} =
    K_\alpha \prod_i \left(\frac{c_i^\mathrm{u}}{c^\circ}\right)^{\nu_{i\alpha}} \exp \left(-n_\mathrm{n} \tilde{q}_\alpha Z U\right)\\
    &\text{and} \quad 
    \zeta = 1 + \sum_{\alpha\neq \mathpzc{0}} \chi_\alpha
\end{split}
\label{E:Definition_of_chi_and_zeta}
\end{equation}
After rearranging, this gives explicit expressions for the occupation fractions $\theta_\alpha$:
\begin{equation}
    \theta_\mathpzc{0} = \frac{1}{\zeta}
    \quad \text{and} \quad
    \theta_\alpha = \frac{\chi_\alpha}{\zeta} 
    \quad \forall \alpha\neq \mathpzc{0} .
    \label{E:Occupation_fractions_rearranged}
\end{equation}
This allows all occupation fractions to be expressed in terms of $Z$, since $\chi_\alpha = \chi_\alpha(Z, c_i^\mathrm{u})$ and $\zeta = \zeta(\chi_\alpha)$. Substitution of relations \ref{E:Occupation_fractions_rearranged} into the definition of $Z$ (equation \ref{E:Definition_of_abbreviation_Z}) yields:
\begin{equation}
    Z = \frac{1}{\zeta} \left(z_\mathrm{X} + \sum_{\alpha \neq \mathpzc{0}} q_\alpha \chi_\alpha \right)
\end{equation}
This reduces the system of equations defining condensation (equation \ref{E:Occupation_condition}) to a single implicit equation. Using $\zeta = \zeta(\chi_\alpha)$, this can be rewritten to:
\begin{impeq}
\begin{equation}
    Z - z_\mathrm{X} + \sum_{\alpha \neq \mathpzc{0} } \left(Z - q_\alpha \right)\chi_\alpha = 0
    \label{E:IOM_Residual_charge}
\end{equation}
\end{impeq}

\subsubsection{Surface partitioning}
The membrane is considered to be in equilibrium with the surrounding bulk at its surface. With equations \ref{E:Free_energy_bulk} and \ref{E:Electrochemical_potential_membrane} the equilibrium condition $\overline{\mu}_i^\mathrm{b} = \overline{\mu}_i^\mathrm{m}$ at the membrane surface is:
\begin{equation}
    R T \ln \left(c_i^\mathrm{b}\right) 
    + z_i F \Phi^\mathrm{b}
    + \mu_i^{\mathrm{b},\mathrm{ex}}
    =
    R T \ln \left(c_i^\mathrm{u}\right) 
    + z_i F \Phi^\mathrm{u}
    + \mu_i^{\mathrm{u},\mathrm{ex}}
\end{equation}
Rearranging yields the partition relation:
\begin{equation}
    \frac{c_i^\mathrm{u}}{c_i^\mathrm{b}}
    =
    \exp \left( \frac{\mu_i^{\mathrm{b},\mathrm{ex}} - \mu_i^{\mathrm{u},\mathrm{ex}}} {R T} \right)
    \exp \left( \frac{z_i F} {R T} \left(\Phi^\mathrm{b} - \Phi^\mathrm{u}\right) \right)
\end{equation}
With the Donnan potential $\Phi_\mathrm{D} = \Phi^\mathrm{u} - \Phi^\mathrm{b}$ and the excess exclusion factor $S_i^\mathrm{ex} = \exp \left(\left(\mu_{i}^\mathrm{b,ex} - \mu_{i}^\mathrm{u,ex}\right) / RT \right)$, this relation can be written as:
\begin{impeq}
\begin{equation}
    \frac{c_i^\mathrm{u}}{c_i^\mathrm{b}} = S_i^\mathrm{ex} \exp \left( -\frac{z_i F}{R T} \Phi_\mathrm{D} \right)
    \label{E:IOM_partitioning}
\end{equation}
\end{impeq}
This equation is explicit if $\mu_i^\mathrm{u,ex}$ is independent of the membrane concentrations. Otherwise, it remains implicit through the concentration dependence of $S_i^\mathrm{ex}$. 

\subsection{Kinetics}
\label{C:Kinetics}
For the transport equations to be thermodynamically consistent, entropy production has to be non-negative. To ensure this, we derive transport equations considering condensed and uncondensed states in presence of fixed charges from nonequilibrium thermodynamics (see supporting information \ref{C:Derivation_transport_equations}). This derivation is based on the work of Latz and Zausch \cite{Latz2011}, Schammer et al. \cite{Schammer2021} and Stamm et al. \cite{Stamm2017}. \par 
Since condensed and uncondensed states are assumed to be in local equilibrium, the total species flux is driven by the common electrochemical potential $\overline{\mu}_i^\mathrm{m}$ (see equation \ref{E:Electrochemical_potential_membrane}). The resulting transport equations take the typical Nernst-Planck form:
\begin{impeq}
\begin{equation}
    \mathbf{N}_i^\mathrm{m} =
    - D_i^\mathrm{m}\nabla c_i^\mathrm{u}
    - \frac{F}{RT} D_i^\mathrm{m} z_i c_i^\mathrm{u} \nabla\Phi^\mathrm{u}
\label{E:IOM_kinetic_equations}
\end{equation}
\end{impeq}
Here, $\mathbf{N}_i^\mathrm{m}$ represents the total species flux density. The coefficient $D_i^\mathrm{m}$ is an effective diffusion coefficient, defined from the total mobility with respect to the uncondensed concentration $c_i^\mathrm{u}$. It combines the mobility contributions of condensed and uncondensed ions through:
\begin{equation}
    D_i^\mathrm{m} =
    \frac{1}{c_i^\mathrm{u}} \left(D_{i}^\mathrm{u} c_i^\mathrm{u} + D_{i}^\mathrm{c} c_i^\mathrm{c} \right)
    \label{E:Mobility_coefficients_membrane_total}
\end{equation} 
This highlights the dependence of the effective transport coefficients on the state equilibria, making it hard to state explicit values or dependencies due to the nonlinear ties in the equilibrium conditions (see equation \ref{E:Equilibrium_equation_system}).

\subsection{Model summary}
\label{C:Model_summary}
The model introduced above consists of $2+N$ equilibrium equations (equations \ref{E:IOM_charge_neutrality}, \ref{E:IOM_Residual_charge} and \ref{E:IOM_partitioning}) and $N$ kinetic equations (equation \ref{E:IOM_kinetic_equations}), with $N$ being the number of ionic species.
The parameters required to solve these equations are the concentration of fixed sites $c_\mathrm{X}$, the exclusion factor $S^\mathrm{ex}$, the equilibrium constants $K$, the dimensionless interaction energy $U$ and the diffusion coefficients $D$.\par
Although this parameter set may appear large at first glance, essentially the same set of quantities underlies the other presented membrane models as well. Those models merely adopt closures and fix certain parameters already at the derivation level, which we avoided. Meaningful closures and parameter choices are discussed in section \ref{C:Parameter_estimation}. \par

\subsection{Single-occupation approximation}
The framework as introduced above allows for arbitrary occupation states. However, parameterization of the association constants $K_\alpha$ becomes elusive when fixed sites can coordinate with more than one mobile ion. For the practical analysis below, we therefore restrict the model to single-occupation, leaving a discussion of multi-occupation to future work. This allows us to replace the state indices $\alpha$ by species indices $i$ (except for the empty state $\mathpzc{0}$) and the equations relating the model variables (eqs. \ref{E:Total_concentration}, \ref{E:Definition_of_abbreviation_Z} and \ref{E:Definition_of_chi_and_zeta}) become:
\begin{equation}
\begin{aligned}
    & c_i^\mathrm{m} = && c_i^\mathrm{u} + c_\mathrm{X}\theta_i\\
    & \chi_i = && K_i \frac{c_i^\mathrm{u}}{c^\circ} \exp \left(-n_n z_i Z U\right)\\
    & Z =  && z_\mathrm{X}\theta_\mathpzc{0} + \sum_{i} \left(z_\mathrm{X} + z_i\right) \theta_i
\end{aligned}
\label{E:Model_closure}
\end{equation}
In contrast to the low-T* model, this assumption is not overly restrictive for the presented model, since ions can also occupy uncondensed states.

%% file: 3_Modeling/Figures/My_model_variables.pdf_tex
\begingroup%
  \makeatletter%
  \providecommand\color[2][]{%
    \errmessage{(Inkscape) Color is used for the text in Inkscape, but the package 'color.sty' is not loaded}%
    \renewcommand\color[2][]{}%
  }%
  \providecommand\transparent[1]{%
    \errmessage{(Inkscape) Transparency is used (non-zero) for the text in Inkscape, but the package 'transparent.sty' is not loaded}%
    \renewcommand\transparent[1]{}%
  }%
  \providecommand\rotatebox[2]{#2}%
  \newcommand*\fsize{\dimexpr\f@size pt\relax}%
  \newcommand*\lineheight[1]{\fontsize{\fsize}{#1\fsize}\selectfont}%
  \ifx\svgwidth\undefined%
    \setlength{\unitlength}{659.99998558bp}%
    \ifx\svgscale\undefined%
      \relax%
    \else%
      \setlength{\unitlength}{\unitlength * \real{\svgscale}}%
    \fi%
  \else%
    \setlength{\unitlength}{\svgwidth}%
  \fi%
  \global\let\svgwidth\undefined%
  \global\let\svgscale\undefined%
  \makeatother%
  \begin{picture}(1,0.37422168)%
    \lineheight{1}%
    \setlength\tabcolsep{0pt}%
    \put(0,0){\includegraphics[width=\unitlength,page=1]{My_model_variables.pdf}}%
    \put(0.09621622,0.18427564){\color[rgb]{0,0,0}\makebox(0,0)[t]{\lineheight{1.25}\smash{\begin{tabular}[t]{c}\scriptsize{$\alpha=\mathpzc{1}$}\end{tabular}}}}%
    \put(0.57561172,0.18427564){\color[rgb]{0,0,0}\makebox(0,0)[t]{\lineheight{1.25}\smash{\begin{tabular}[t]{c}\scriptsize{$\alpha=\mathpzc{2}$}\end{tabular}}}}%
    \put(0.33591398,0.18427564){\color[rgb]{0,0,0}\makebox(0,0)[t]{\lineheight{1.25}\smash{\begin{tabular}[t]{c}\scriptsize{$\alpha=\mathpzc{0}$}\end{tabular}}}}%
    \put(0.8153095,0.18427564){\color[rgb]{0,0,0}\makebox(0,0)[t]{\lineheight{1.25}\smash{\begin{tabular}[t]{c}\scriptsize{$\alpha=\mathpzc{3}$}\end{tabular}}}}%
    \put(0.98398123,0.30019244){\color[rgb]{0,0,0}\makebox(0,0)[lt]{\lineheight{1.25}\smash{\begin{tabular}[t]{l}$"\mathrm{u}"$\end{tabular}}}}%
    \put(0.98398123,0.13093997){\color[rgb]{0,0,0}\makebox(0,0)[lt]{\lineheight{1.25}\smash{\begin{tabular}[t]{l}$"\mathrm{c}"$\end{tabular}}}}%
    \put(0.07164532,0.33637586){\color[rgb]{0,0,0}\makebox(0,0)[t]{\lineheight{1.25}\smash{\begin{tabular}[t]{c}\scriptsize{$i=1$}\end{tabular}}}}%
    \put(0.36263427,0.3428928){\color[rgb]{0,0,0}\makebox(0,0)[t]{\lineheight{1.25}\smash{\begin{tabular}[t]{c}\scriptsize{$i=2$}\end{tabular}}}}%
    \put(0,0){\includegraphics[width=\unitlength,page=2]{My_model_variables.pdf}}%
    \put(0.70233545,0.2690535){\color[rgb]{0.2745098,0.39215686,0.66666667}\makebox(0,0)[t]{\lineheight{1.25}\smash{\begin{tabular}[t]{c}\footnotesize{$\Delta G_\alpha^\circ$}\end{tabular}}}}%
    \put(0,0){\includegraphics[width=\unitlength,page=3]{My_model_variables.pdf}}%
    \put(0.45576287,0.00357334){\color[rgb]{0.63529412,0.13333333,0.1372549}\makebox(0,0)[t]{\lineheight{1.25}\smash{\begin{tabular}[t]{c}\footnotesize{$U_{\mathpzc{0},\mathpzc{2}}$}\end{tabular}}}}%
    \put(0,0){\includegraphics[width=\unitlength,page=4]{My_model_variables.pdf}}%
  \end{picture}%
\endgroup%

%% file: 4_Parameters/0_main.tex
\section{Parameterization}
\label{C:Parameter_estimation}
The following sections discuss how the model parameters can be estimated from theory and how robust each estimate is. As an example, parameter values are calculated for a commercial CR61 membrane \cite{Veolia2020}.

\subsection{Fixed charge concentration}
\label{C:Ion_exchange_capacity}
The fixed charge concentration $c_\mathrm{X}$ is defined as fixed sites per solution volume in the membrane. Therefore, it directly depends on water uptake $\omega_\mathrm{w}$:
\begin{equation}
    c_\mathrm{X} 
    =
    \frac{\rho_\mathrm{w}}{\omega_\mathrm{w}} M_\mathrm{X}^\mathrm{dry} 
    \label{E:IEC_depends_on_water_uptake}
\end{equation}
Here, $\omega_\mathrm{w}$ is dimensionless ($\mathrm{kg_{water} / kg_{mem,dry}}$). $M_\mathrm{X}^\mathrm{dry}$ denotes molality of fixed sites per dry polymer mass, which is a constant given from membrane synthesis. Further, $\rho_\mathrm{w}$ represents the density of water. \par
For a CR61 membrane, $M_\mathrm{X}^\mathrm{dry}$ is approximated as \cite{Galizia2017, Oren2024}:
\begin{impeq}
\begin{equation}
    M_\mathrm{X}^\mathrm{dry} \approx 2.5 \, {\textstyle \frac{\mathrm{mol}}{\mathrm{kg}}}
\end{equation}
\end{impeq}

\subsection{Water uptake}
\begin{figure}[h]
    \centering
    \input{4_Parameters/Figures/Water_uptake_IEC}
    \vspace{-0.3cm}
    \caption{Water uptake and the subsequent ion exchange capacity $c_\mathrm{X}$ depending on the bulk salt concentration $c_\mathrm{s}^\mathrm{b}$ for a CR61 membrane. }
    \label{F:Water_uptake_IEC}
\end{figure}
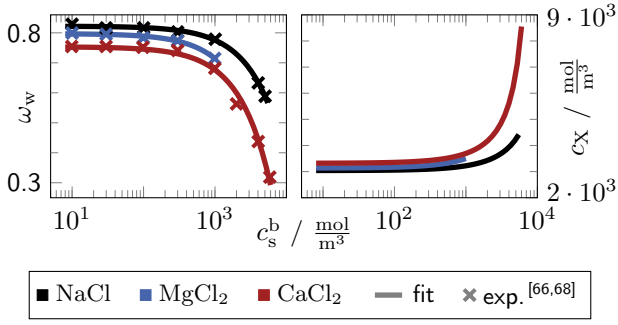

The calculation of the fixed charge concentration $c_\mathrm{X}$ from $M_\mathrm{X}^\mathrm{dry}$ requires information on water uptake $\omega_\mathrm{w}$ (see equation \ref{E:IEC_depends_on_water_uptake}). Since water uptake in turn depends on the bulk salt concentration $c_\mathrm{s}^\mathrm{b}$, this dependence also has to be estimated. Using the data from Galizia et al.\cite{Galizia2017, Galizia2019}, we obtain the following fits
\begin{impeq}
\setlength{\abovedisplayskip}{0pt}
\setlength{\belowdisplayskip}{0pt}
\begin{equation}
\begin{aligned}
    \mathrm{NaCl}: \quad \omega_\mathrm{w} \approx 0.82 - 4.66 \cdot 10^{-5} c_\mathrm{s}^\mathrm{b} \, / \, {\textstyle \frac{\mathrm{mol}}{\mathrm{m^3}}}\\
    \mathrm{CaCl_2}: \quad \omega_\mathrm{w} \approx 0.75 - 7.68 \cdot 10^{-5} c_\mathrm{s}^\mathrm{b} \, / \, {\textstyle \frac{\mathrm{mol}}{\mathrm{m^3}}}\\
    \mathrm{MgCl_2}: \quad \omega_\mathrm{w} \approx 0.80 - 8.32 \cdot 10^{-5} c_\mathrm{s}^\mathrm{b} \, / \, {\textstyle \frac{\mathrm{mol}}{\mathrm{m^3}}}\\
\end{aligned}
\label{E:Water_uptake_fit_polynoms}
\end{equation}
\end{impeq}
Figure \ref{F:Water_uptake_IEC} shows close agreement between the fits and the experimental data. It also shows that the deviations $c_\mathrm{X}$ from its dilute-limit value are minor at salt concentrations lower than $1$ mol/l, which justifies the assumption of constant $c_\mathrm{X}$ often used in other models up to this limit. 
\par

\subsection{Association constants}
\label{C:Association_constants}
The experimental determination of association constants is a complicated matter \cite{Marcus2006, Hefter2006, Peshkova2008}. Since the present model can require multiple association constants (see equation \ref{E:Definition_of_chi_and_zeta}), approximation from theory can be useful. To do so, we assume that the association constants are dominated by the electrostatic contribution (compare supporting information \ref{C:Derivation_free_energy}, equation \ref{E:Effective_formation_free_energy}) and estimate them from ion-association \cite{Bjerrum1926, Fuoss1933, Fuoss1958, Eigen1962, Justice1976, Barthel1979, Ebeling1980, Krienke1998, Krienke1998a, Mulder2000}. For this, the Bjerrum treatment \cite{Bjerrum1926} is the most widely employed \cite{Marcus2006}. In this approach, the association constant is obtained from the excess configurational integral of the Boltzmann factor:
\begin{equation}
    K_{ij} = 4 \pi c^\circ N_\mathrm{A} \int_{b_{ij}}^{R} r^2 \left[ \exp \left(- U(r)\right) -1 \right] \mathrm{d}r
    \label{E:Association_constant_Bjerrum}
\end{equation}
If the dimensionless interaction energy $U(r)$ is approximated from unscreened Coulomb interaction $\lambda_\mathrm{B} z_i z_j / r$, a suitable choice for the cutoff radius $R$ is required, since the integral diverges for unscreened interaction. A common choice is the Bjerrum cutoff $R = - z_i z_j \lambda_\mathrm{B} /2 $, which corresponds to the minimum of the integrand \cite{Bjerrum1926}. This however links association constants to permittivity through $\lambda_\mathrm{B} \propto \varepsilon_\mathrm{r}^{-1}$, which itself can be problematic (as discussed in section \ref{C:Exclusion_factor}). \par
For the association constant of two oppositely charged monovalent ions, $K_1$, a rough estimation with a hydrated radius of $r^\mathrm{h}_1\approx 3\,\text{\AA}$ \cite{Nightingale1959} and a relative permittivity of $\varepsilon_\mathrm{r} \approx 40$ \cite{Galizia2019, Bannon2024} yields $K_1 \approx 1$. Similarly, we obtain the association constant of a monovalent ion with an oppositely charged divalent ion, $K_2$, with the hydrated radius of the divalent ion $r^\mathrm{h}_1\approx 4\,\text{\AA}$ as $K_2 \approx \mathcal{O}(10) K_1$. Like-charged pairs are assumed to be unlikely, $K_- \leq \mathcal{O}(10^{-3})$. 
Overall, we use the following estimates:
\begin{impeq}
\setlength{\abovedisplayskip}{0pt}
\begin{equation}
    K_- \leq \mathcal{O}(10^{-3})
    \ , \quad
    K_1 \approx 1 \,
    \quad \text{and} \quad
    K_2 \approx \mathcal{O}(10) K_1
\end{equation}
\end{impeq}


\subsection{The excess exclusion factor}
\label{C:Exclusion_factor}
The excess exclusion factor $S^\mathrm{ex}$ (equation \ref{E:IOM_partitioning}) collects several poorly separable contributions \cite{Crothers2020, Freger2023}. The Donnan-Manning model commonly neglects it entirely \cite{Bannon2024}, whereas the SDE model includes contributions from steric confinement (st) and dielectric mismatch (de) in it (see sections \ref{C:SDE_model} and \ref{C:Donnan_Manning}). This section shows how the assumptions of the SDE model can be incorporated, as an example how to generally account for various exclusion effects and discusses how $S^\mathrm{ex}$ can be parameterized meaningfully.\par
\begin{figure}[h]
    \centering
    \input{4_Parameters/Figures/Exclusion_factors}
    \caption{Left: The influence of the cavity radius $r_\mathrm{cav}$ on the dielectric exclusion factor $S^\mathrm{de}$ for a divalent ion $\lvert z \rvert = 2$. Right: The influence of the geometry parameter $g$ on the steric exclusion factor $S^\mathrm{st}$.}
    \label{F:Exclusion_influence}
\end{figure}
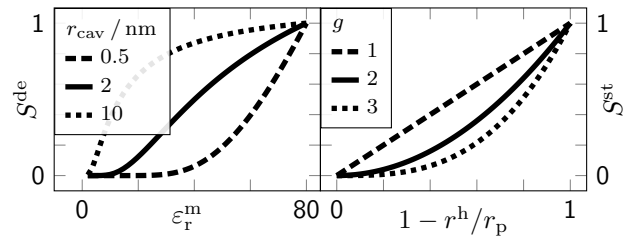

Accounting for steric (st) and dielectric (de) effects, the excess chemical potential of uncondensed ions $\mu^\mathrm{u,ex}$ (equation \ref{E:Free_energy_membrane}) can be expressed as:
\begin{equation}
    \mu^\mathrm{u,ex}
    = 
    \mu_i^\mathrm{u, st} + \mu_i^\mathrm{u, de}
\end{equation}
The contribution from confinement is usually assumed to depend on the ratio of hydrated particle radius $r_i^\mathrm{h}$ over pore radius $r_\mathrm{p}$ and pore geometry \cite{Giddings1968}. In analogy to the known relations shown in the supporting information \ref{C:Steric_exclusion_Giddings} we account for pore geometry by introduction of a geometry factor $g$ representing the surface to volume ratio $g = r_\mathrm{p} A_\mathrm{p}^\mathrm{surf} / V_\mathrm{p}$. Here, a value of $g=1$ corresponds to a slit pore, $g=2$ to a cylindrical pore and $g=3$ to a spherical pore. This gives:
\begin{equation}
    \mu_i^{\mathrm{u,st}} = - R T \ln\left(1 - \frac{r_i^\mathrm{h}}{r_\mathrm{p}}\right)^g
\end{equation}
As in the SDE model (section \ref{C:SDE_model}), we estimate dielectric mismatch using the Born model \cite{Born1920} with the bulk as reference:
\begin{equation}
    \mu_i^\mathrm{u,de} = \frac{N_A e^2 z_i^2 }{8\pi\varepsilon_0 r_{i,\mathrm{cav}}} \left(\frac{1}{\varepsilon_r^\mathrm{b}}-\frac{1}{\varepsilon_r^\mathrm{u}}\right)
\end{equation}
Here, the Born hydration energy is written using a generic cavity radius $r_{i,\mathrm{cav}}$. This is usually approximated with the pore radius $r_p$ instead of the ionic radius $r_i$, following discussions on suitable choices for this radius \cite{Freger2023, Duignan2013, Hennequin2021, Yaroshchuk2000, Bannon2024}. \\ 
The excess exclusion factor $S_i^\mathrm{ex}$ (see equation \ref{E:IOM_partitioning}) therefore becomes:
\begin{equation}
    S_i^\mathrm{ex} \approx \left(1 - \frac{r_i^\mathrm{h}}{r_\mathrm{p}}\right)^g \exp\left( \frac{N_A e^2 z_i^2}{8\pi\varepsilon_0 r_\mathrm{p} R T} \left(\frac{1}{\varepsilon_r^\mathrm{b}}-\frac{1}{\varepsilon_r^\mathrm{u}}\right) \right)
\end{equation}
Even though these approximations are common, it already shows that both contributions depend strongly on poorly defined quantities ($g$ and $r_{i,\mathrm{cav}}$), whose impacts are illustrated in figure \ref{F:Exclusion_influence}.
\par
\begin{figure}[h]
    \centering
    \input{4_Parameters/Figures/Permittivity_analysis}
    \vspace{-0.3cm}
    \caption{Left: The dependence of bulk permittivity on salt concentration $c_\mathrm{s}^\mathrm{b}$ according to Mollerup et al. \cite{Mollerup2015}; Right: The membrane static permittivity depending on the water uptake with either the polymer as the continuous phase ($\mathbb{C} \widehat{=} \mathrm{p}$) or the solute as the continuous phase ($\mathbb{C} \widehat{=} \mathrm{s}$) according to equation \ref{E:Maxwell_Garnett} compared to experimental data for Nafion 117 and XL-pGMA-z \cite{Chang2019, Chang2020, Paddison1998}; the polymer relative permittivity is $\varepsilon_\mathrm{r}^\mathrm{p} \approx 1.8$, the solute relative permittivity is $\varepsilon_\mathrm{r}^\mathrm{w} \approx 80$.}
    \label{F:Permittivity_analysis}
\end{figure}
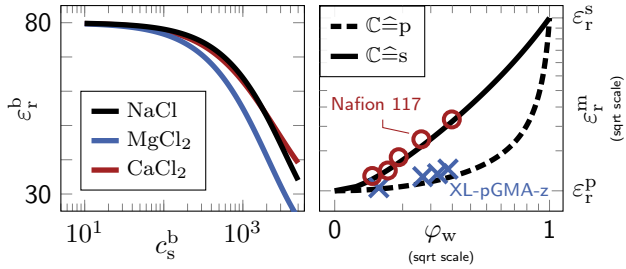

But the model uncertainty does not end here. Generally, neither the bulk permittivity, nor the membrane permittivity are constant. The permittivity of the bulk phase varies with salt concentration \cite{Mollerup2015} (figure \ref{F:Permittivity_analysis}, left). This dependence on concentration is less prominent in confined nanopores \cite{Chogani2024}, but membrane permittivity still varies with hydration \cite{Paddison1998, Chang2019, Chang2020, Chang2021, Chogani2024}, requiring additional models such as the Maxwell–Garnett permittivity model \cite{MaxwellGarnett1904, Chang2021}:
\begin{equation}
    \varepsilon_\mathrm{r}^\mathrm{m} \approx 
    \varepsilon^\mathbb{C} 
    \frac
        {\varepsilon^\mathbb{I} + 2 \varepsilon^\mathbb{C} + 2\varphi^\mathbb{I} \left(\varepsilon^\mathbb{I} - \varepsilon^\mathbb{C} \right)}
        {\varepsilon^\mathbb{I} + 2 \varepsilon^\mathbb{C} - \varphi^\mathbb{I} \left(\varepsilon^\mathbb{I} - \varepsilon^\mathbb{C} \right)}
    \label{E:Maxwell_Garnett}
\end{equation}
Here, $\varphi$ denotes volume fractions, the indices $\mathbb{I}$ and $\mathbb{C}$ the intrusive and the continuous phase, respectively. Whether the polymer or the aqueous phase is the intrusive also depends on the actual membrane and its hydration \cite{Chang2021}, as shown in figure \ref{F:Permittivity_analysis} on the right. \par
\par
This illustrates how detailed considerations of the exclusion factor can introduce many unknown parameters, which are often poorly defined and hard to assess. \par 
To circumvent this, we treat the excess exclusion factor as concentration-independent and use only ballpark estimates motivated by dielectric mismatch, which is commonly considered the dominant contribution \cite{Bannon2024, Freger2023}. The applied estimates for the excess exclusion factors of monovalent and divalent ions, $S^\mathrm{ex}_1$ and $S^\mathrm{ex}_2$ are:
\begin{impeq}
\setlength{\abovedisplayskip}{0pt}
\begin{equation}
    S^\mathrm{ex}_\mathrm{1} \approx 0.75
    \quad \text{and} \quad
    S^\mathrm{ex}_\mathrm{2} \approx 0.25
\end{equation}
\end{impeq}


\subsection{Mesoscale porosity and tortuosity effects}
The CR61 membrane has a water uptake of $\varphi_w\approx [0.4, 0.5]$ at low salt concentrations \cite{Galizia2017}. In this range, the Mackie-Meares model (see equation \ref{E:Mackie_Meares}) is considered sufficiently accurate \cite{Marioni2024,Sireci2023}. However, at water uptakes below $0.4$, the Mackie-Meares model is known to overestimate the dependence of transport hindrance on hydration since it does not account for preferentially percolated pathways \cite{Marioni2024}.\par 
To circumvent this, the dependence on hydration is omitted here and the mesoscale hindrance factor $k^\mathcal{M}$ is subsequently approximated by the Mackie-Meares value in the low-salt limit:
\begin{impeq}
\begin{equation}
    k^\mathcal{M} \approx 0.05
\end{equation}
\end{impeq}


\subsection{Diffusion of uncondensed ions}
Equation \ref{E:IOM_kinetic_equations} shows that transport only depends on the effective coefficient $D_i^\mathrm{m}$, even though transport coefficients for the states can differ (see equation \ref{E:Mobility_coefficients_membrane_total}). Diffusion of uncondensed ions can e.g. be influenced by wall effects, suggesting the introduction of a diffusion hindrance coefficient $k^\mathrm{d} = D_i^\mathrm{u}/D_i^\mathrm{b}$ similarly to the SDE model (equation \ref{E:Nernst_Planck_SDE}).\par
In practice, however, this factor cannot be clearly distinguished from mesoscale hindrance $k^\mathcal{M}$ when differences in $k^\mathrm{d}$ for various species are minor. Given that hydrated ion radii are usually similar ($r^\mathrm{h} \approx [3,5] \,$\AA) \cite{Nightingale1959} and pore radii are about $\mathcal{O}(1\, \mathrm{nm})$, the differences in $k^\mathrm{d}$ are likely smaller than the model uncertainty. Also, hindrance factors usually account for steric exclusion \cite{Dechadilok2006}, which is already accounted for in the excess exclusion factor.\par
Therefore, additional microscopic hindrance is neglected:
\begin{equation}
     D_i^\mathrm{u} \approx D_i^\mathrm{b}
\end{equation}

\subsection{Diffusion of condensed ions}
Transport in the condensed state is attributed to a hopping process, where mobile ions escape a bound state and reattach at an available neighboring site \cite{Freger2020,Lei2020,Kitto2022}. This can be treated as a statistical first-passage process, where an ion reaching the boundary of the volume associated with a certain site represents a successful escape \cite{Freger2020, Redner2008}. For overdamped Brownian motion in a dimensionless conservative potential $U$, the mean first-passage time (MFPT) $\ell$ to such an absorbing boundary fulfills a backward Smoluchowski equation \cite{Gardiner2004}:
\begin{equation}
    \nabla \cdot \left(e^{-U}\nabla \ell \right) 
    =
    -\frac{e^{-U}}{D^\mathrm{u}}
\end{equation}
For a spherical site with an inner hard-shell radius $b$ and absorbing outer boundary at $r=R$, the boundary conditions are:
\begin{equation}
    \left. \ell \right|_{r=R}=0 
    \quad \text{and} \quad
    \left. \frac{\partial \ell}{\partial r} \right|_{r=b}=0 
\end{equation}
With this, the MFPT for an ion at a starting position $r$ becomes \cite{Lifson1962,Jackson1963,Freger2020}:
\begin{equation}
    \ell(r)
    =
    \frac{1}{D^\mathrm{u}}
    \int_r^R \mathrm{d} r' \, r'^{-2} \,e^{U(r')}
    \int_b^{r'} \mathrm{d} r'' \, r''^{2} \, e^{-U(r'')}
    \label{E:Lifson}
\end{equation}
The effective macroscopic diffusion coefficient of condensed ions $D^\mathrm{c}$ is proportional to this MFPT averaged over the starting positions $\langle \ell(r) \rangle$. The proportionality factors are the mean square displacement (MSD) per hop $\Delta r^2$, the dimensionality $d \in \{1,2,3\}$ (for 1D, 2D, 3D) and the probability that a destination site is available $\mathcal{P}_\mathrm{free}$ \cite{Freger2020}:
\begin{equation}
    D^\mathrm{c} = \frac{\Delta r^2}{2 d \langle \ell(r) \rangle} \mathcal{P}_\mathrm{free}
    \label{E:Condensed_Diffusion_Einstein}
\end{equation}
The MFPT averaged over the volume $\langle \ell(r) \rangle$ can be calculated using the probability $\mathcal{P}(r)$ of finding an ion at a given value of $r$:
\begin{equation}
    \langle \ell(r) \rangle = \int_b^R \ell(r) \mathcal{P}\left(r\right) \mathrm{d}r 
\end{equation}
This probability is:
\begin{equation}
    \mathcal{P}\left(r\right) 
    = 
    \frac{4 \pi r^2 \mathrm{e}^{-U(r)}}{\mathcal{K}}
    \quad \text{with} \quad
    \mathcal{K} = 4\pi \int_b^R r^2 \mathrm{e}^{-U(r)}\mathrm{d}r
\label{E:Probability_over_r}
\end{equation}
Comparison to equation \ref{E:Association_constant_Bjerrum} shows that the radial normalization $\mathcal{K}$ occurring here is related to the association constants for the considered site-ion pair $K$:
\begin{equation}
    \mathcal{K} = \frac{K}{N_\mathrm{A}c^\circ} + V_\mathrm{int}
\end{equation}
Here, $V_\mathrm{int} = 4\pi \int_b^R r^2 \mathrm{d}r$ denotes the integration volume. \\
Since the diffusion coefficient should be unimpeded by homogeneous potentials, the MSD is calibrated to yield $D^\mathrm{c}(\nabla \varphi = 0)= D^\mathrm{u} \mathcal{P}_\mathrm{free}$. In 3D, this gives:
\begin{equation}
    \Delta r^2 = \left(R^2 - b^2\right) + 2 b^3 \left(\frac{1}{R} - \frac{1}{b} \right) 
\end{equation}
In the deep, narrow well limit, the integrals in equations \ref{E:Lifson} and \ref{E:Probability_over_r} are dominated by values of $r\approx b$. The assumption of single occupation further gives $\mathcal{P}_\mathrm{free} = \theta_\mathpzc{0}$. Thus, equation \ref{E:Condensed_Diffusion_Einstein} simplifies to:
\begin{equation}
    D^\mathrm{c} \approx D^\mathrm{u} \, \theta_\mathpzc{0} \, \frac{V_\mathrm{int}}{\frac{K}{N_\mathrm{A}c^\circ} + V_\mathrm{int}} 
\label{E:Condensed_diffusion_coefficient_association_constant}
\end{equation}
Using the association constants and the cutoff radius as stated in section \ref{C:Association_constants}, the condensed diffusion coefficient of monovalent and divalent counterions $D^\mathrm{c}_\mathrm{ct,1}$ and $D^\mathrm{c}_\mathrm{ct,2}$ become:
\begin{impeq}
\setlength{\abovedisplayskip}{0pt}
\begin{equation}
    D^\mathrm{c}_\mathrm{ct,1} \approx 0.5 \, \theta_\mathpzc{0} D^\mathrm{u}_\mathrm{ct,1}
    \quad \text{and} \quad
    D^\mathrm{c}_\mathrm{ct,2} \approx 0.3 \, \theta_\mathpzc{0} D^\mathrm{u}_\mathrm{ct,2}
\end{equation}
\end{impeq}
Note that at zero electric current, the influence of condensed diffusion is minor, since neutral salt transport in highly charged membranes is mostly limited by the coions which have low association constants and therefore low condensed concentrations. It mostly becomes relevant at nonzero currents, where counterion transport contributes directly to the ionic current.

\subsection{Nearest neighbor interaction}
Nearest neighbor interaction is given as the product of the number of nearest neighbors $n_\mathrm{n}$ and the dimensionless interaction energy $U$. While $n_\mathrm{n}$ is given from polymer architecture, the magnitude of $U$ depends on the distance of neighboring fixed sites $L$ and an effective dielectric constant along the polymer backbone $\varepsilon_p$:
\begin{equation}
    U = \frac{\lambda_B}{L}
    \qquad \text{with} \qquad 
    \lambda_B = \frac{e^2}{4 \pi \varepsilon_0 \varepsilon_\mathrm{p} k_\mathrm{B} T}
\end{equation}
A rough estimation from Coulombic interaction gives:
\begin{impeq}
\begin{equation}
    1 \lesssim n_\mathrm{n}U \lesssim 20
\end{equation}
\end{impeq}

\subsection{Parameter summary}
The sections above state a variety of parameters and show reasonable estimates obtained for a CR61 membrane. Table \ref{T:Parameter_summary} lists these parameters, how they are obtained and which values are used for the simulations in the following sections.
\begin{table}[h]
\centering
\begin{tabular}{|l|l|l|}
\hline
Parameter & Determination & Used value \\
\hline
$M_X^\mathrm{dry}$ & membrane property & $2.5\,\mathrm{mol\,kg^{-1}}$ \\
$\omega_w$ & experimental fit & see eq. \ref{E:Water_uptake_fit_polynoms} \\
$k^\mathcal{M}$ & derived & $0.05$ \\
$D_\mathrm{Na}^\mathrm{b}$ & literature \cite{Lide2004} & $1.3 \cdot 10^{-9} \mathrm{m}^2\,\mathrm{s}^{-1}$\\
$D_\mathrm{Ca}^\mathrm{b}$ & literature \cite{Lide2004} & $0.8 \cdot 10^{-9} \mathrm{m}^2\,\mathrm{s}^{-1}$\\
$D_\mathrm{Mg}^\mathrm{b}$ & literature \cite{Lide2004} & $0.7 \cdot 10^{-9}\mathrm{m}^2\,\mathrm{s}^{-1}$\\
$D_\mathrm{Cl}^\mathrm{b}$ & literature \cite{Lide2004} & $2.0 \cdot 10^{-9} \mathrm{m}^2\,\mathrm{s}^{-1}$\\
$D_i^\mathrm{u}$ & derived & $D_i^\mathrm{b}$\\
$D_1^\mathrm{c}$ & derived & $0.5\, \theta_\mathpzc{0} D_1^\mathrm{u}$ \\
$D_2^\mathrm{c}$ & derived & $0.3\, \theta_\mathpzc{0} D_2^\mathrm{u}$ \\
$S_1^\mathrm{ex}$ & estimated & $0.75$ \\
$S_2^\mathrm{ex}$ & estimated & $0.25$ \\
$K_1$ & estimated & $1$ \\
$K_2$ & estimated & $5$ \\
$n_\mathrm{n}U$ & estimated & $3$ \\
\hline
\end{tabular}
\caption{Summary of parameter estimates for a CR61 membrane.}
\label{T:Parameter_summary}
\end{table}


%% file: 4_Parameters/Figures/Water_uptake_IEC.tex
\begin{tikzpicture}

\begin{groupplot}[
    group style={
        group name=my plots,
        group size=2 by 2,
        horizontal sep=0.2cm,
    },
    tick align=inside,
    height = 4cm,
    width = 4.7cm,
]
\nextgroupplot[
    xlabel={$c_\mathrm{s}^\mathrm{b} \ / \ \frac{\mathrm{mol}}{\mathrm{m^3}}$},
    xlabel style = {yshift = 0.4cm, xshift = 1.8cm},
    xmin=5, xmax=10000,
    xmode = log,
    xtick={1,10,100,1000,10000},
    xticklabels={,$10^{1}$,,$10^{3}$,},
    ylabel={$\omega_\mathrm{w}$},
    ylabel style = {yshift = -0.6cm, xshift = 0cm},
    ymin= 0.25, ymax=0.86,
    ytick={0.3,0.8},
    yticklabels={0.3,0.8},
    extra y ticks={0.4,0.5,0.6,0.7},
    extra y tick style={major tick length = \pgfkeysvalueof{/pgfplots/minor tick length}},
    extra y tick labels={},
    legend style={
        nodes={anchor=west},
        anchor = north, 
        at={(1.1,-0.4)}, 
        legend columns = 5,
        /tikz/every even column/.append style={column sep=9pt},
        /tikz/every odd column/.append style={column sep=1pt},
        font=\small,
    },
]
\addlegendimage{only marks, mark=square*, color=KITblack}
\addlegendentry{$\mathrm{NaCl}$}
\addlegendimage{only marks, mark=square*, color=KITblue}
\addlegendentry{$\mathrm{MgCl_2}$}
\addlegendimage{only marks, mark=square*, color=KITred}
\addlegendentry{$\mathrm{CaCl_2}$}


\addlegendimage{KITgrey, line legend, line width=2pt, legend image post style={scale=0.6}}
\addlegendentry{fit}
\addlegendimage{KITgrey, only marks, mark=x, mark size=3, mark options={line width=1.5pt}}
\addlegendentry{exp. \cite{Galizia2017,Galizia2019}}


\addplot [line width = 2pt, KITblack, domain=8:5500, samples = 50]{-4.65624182e-05*x + 8.22055712e-01};
\addplot [line width = 2pt, KITblue, domain=8:1000, samples = 50]{-8.31718964e-05*x + 7.97154172e-01};
\addplot [line width = 2pt, KITred, domain=8:6000, samples = 50]{-7.68385052e-05*x + 7.53235507e-01};


\addplot [KITblack, only marks, mark = x, mark size=3, mark options={line width=1.5pt}]
table {%
10 0.8286677673578873
29.891531455721648 0.8165261159937516
100.3890252891329 0.8165261159937516
300.07817072393556 0.803035412841034
1007.7956398502166 0.7787521101127628
4030.8471562816794 0.6330522937431355
5029.414488130898 0.5885329054079717
};
\addplot [KITblue, only marks, mark = x, mark size=3, mark options={line width=1.5pt}]
table {%
10 0.7993174088094424
29.935772947204903 0.7952218460427642
99.09040990033273 0.7836177723682859
299.357729472049 0.7740614488304823
990.9040990033277 0.7146757887136502
};
\addplot [KITred, only marks, mark = x, mark size=3, mark options={line width=1.5pt}]
table {%
9.884194603948472 0.7544688073844914
29.891531455721648 0.7544688073844914
98.07726558638025 0.7517706420516952
300.07817072393556 0.7409781042317738
996.1248225290318 0.6816188991996779
2003.8031111346086 0.5629005508911185
4030.8471562816794 0.4374368201283838
5851.737837231454 0.318718410064192
};


\nextgroupplot[
    xmode = log,
    xmin=5, xmax=10000,
    ymin = 2000, ymax = 9000,
    xtick={1,10,100,1000,10000},
    xticklabels={,,$10^{2}$,,$10^{4}$},
    yticklabel pos=right,
    ytick={2000,9000},
    yticklabels={\raisebox{0.2cm}{$2\cdot10^{3}$},$9\cdot10^{3}$},
    extra y ticks={3000,4000,5000,6000,7000,8000},
    extra y tick style={major tick length = \pgfkeysvalueof{/pgfplots/minor tick length}},
    extra y tick labels={},
    ylabel = {$c_\mathrm{X} \ / \ \frac{\mathrm{mol}}{\mathrm{m}^3}$},
    ylabel style = {yshift = 1cm},
    xlabel = {},]
\addplot [line width = 2pt, KITblack, domain=8:5500, samples = 50]{1000/(-4.65624182e-05*x + 8.22055712e-01)*2.5};
\addplot [line width = 2pt, KITblue, domain=8:1000, samples = 50]{1000/(-8.31718964e-05*x + 7.97154172e-01)*2.5};
\addplot [line width = 2pt, KITred, domain=8:6000, samples = 50]{1000/(-7.68385052e-05*x + 7.53235507e-01)*2.5};

\end{groupplot}

\end{tikzpicture}

%% file: 4_Parameters/Figures/Exclusion_factors.tex
\begin{tikzpicture}

\begin{groupplot}[
    group style={
        group name=my plots,
        group size=2 by 1,
        horizontal sep=0pt,
    },
    width=4cm,
    height=4cm,
    tick align=inside,
    ytick pos=both,
    height = 4cm,
    width = 5.1cm,
    legend style={
        fill opacity=0.9,
        text opacity=1,
        draw opacity=1,
        nodes={anchor=west},
        anchor = north west, 
        at={(0.0,1.0)}, 
        font=\footnotesize,
        },
    legend image post style={scale=0.6},
]
\nextgroupplot[
    xmin=-10, xmax=85,
    xtick={0,20,40,60,80},
    xticklabels={0,,,,80},
    xlabel={$\varepsilon_\mathrm{r}^\mathrm{m}$},
    xlabel style = {yshift = 0.4cm},
    ymin= -0.1, ymax=1.1,
    ylabel={$S^\mathrm{de}$},
    ylabel style = {yshift = -0.3cm},
    ytick={0,0.2,0.4,0.6,0.8,1},
    yticklabels={0,,,,,1},
    domain = 2:80,
    samples= 50, 
]
\addlegendimage{empty legend}
\addlegendentry{\hspace{-.55cm} $r_\mathrm{cav} \, / \, \mathrm{nm} $}
\addplot [line width = 2pt, KITblack, dashed]{exp(-222.80126238006955*(1/x - 1/80))};
\addlegendentry{$0.5$}
\addplot [line width = 2pt, KITblack]{exp(-55.700315595017386*(1/x - 1/80))};
\addlegendentry{$2$}
\addplot [line width = 2pt, KITblack, dotted]{exp(-11.140063119003479*(1/x - 1/80))};
\addlegendentry{$10$}
\nextgroupplot[
    xmin = -0.07, xmax = 1.07,
    xtick={0,0.2,0.4,0.6,0.8,1},
    xticklabels={0,,,,,1},
    yticklabel pos=right,
    ylabel = {$S^\mathrm{st}$},
    ylabel style = {yshift = 0.35cm},
    ytick={0,0.2,0.4,0.6,0.8,1},
    yticklabels={0,,,,,1},
    xlabel = {$1 - r^\mathrm{h} / r_\mathrm{p}$},
    xlabel style = {yshift = 0.4cm},
    domain = 0:1,
    samples= 50, 
    ]
\addlegendimage{empty legend}
\addlegendentry{\hspace{-.65cm}$g$}
\addplot [line width = 2pt, KITblack, dashed]{x};
\addlegendentry{$1$}
\addplot [line width = 2pt, KITblack]{x^2};
\addlegendentry{$2$}
\addplot [line width = 2pt, KITblack, dotted]{x^3};
\addlegendentry{$3$}

\end{groupplot}

\end{tikzpicture}

%% file: 4_Parameters/Figures/Permittivity_analysis.tex
\begin{tikzpicture}

\begin{groupplot}[
    group style={
        group name=my plots,
        group size=2 by 1,
        horizontal sep=0.2cm,
    },
    width=4cm,
    height=4cm,
    tick align=inside,
    ytick pos=both,
    height = 4.3cm,
    width = 4.8cm,
    legend style={
        font=\footnotesize,
        nodes={anchor=west},
        fill opacity=0.9,
        text opacity=1,
        draw opacity=1,
        },
    legend image post style={scale=0.6},
]
\nextgroupplot[
    xmin=5, xmax=6000,
    xmode = log,
    xtick={1,10,100,1000,10000},
    xticklabels={ ,$10^1$, , $10^3$, },
    xlabel={$c_\mathrm{s}^\mathrm{b}$},
    xlabel style = {yshift = 0.4cm, xshift = -0.2cm},
    ymin= 25, ymax=85,
    ylabel={$\varepsilon_\mathrm{r}^\mathrm{b}$},
    ylabel style = {yshift = -0.4cm},
    ytick={0,10,20,30,40,50,60,70,80},
    yticklabels={,,,30,,,,,80},
    reverse legend ,
    legend style = {
        anchor = south west, 
        at={(0.08,0.1)}, }
]

\addplot [line width = 2pt, KITred]
table {%
10.0 79.76233927813375
10.64785977823349 79.74700652848144
11.337691785692261 79.7306889975029
12.072215234288088 79.7133240228043
12.854325502735373 79.69484503283523
13.687105549689702 79.67518131341373
14.57383806629774 79.65425776170063
15.518018416061985 79.63199462714056
16.523368413027306 79.6083072388962
17.59385099260072 79.58310571931843
18.733685832834666 79.55629468301663
19.94736598775028 79.52777292112324
21.2396755982669 79.49743307038443
22.61570875055135 79.46516126675706
24.0808895561239 79.43083678325141
25.640993532873463 79.39433165183075
27.302170371262854 79.35551026926686
29.070968175464785 79.31422898695196
30.954359274983734 79.27033568479081
32.95976770850881 79.22366932943797
35.09509848833501 79.17405951731118
37.36876876070854 79.12132600300284
39.78974098492568 79.06527821393132
42.36755826197188 79.00571475232479
45.11238195196144 78.94242288591616
48.03503172865967 78.87517802905074
51.1470282289765 78.80374321627338
54.460638465549216 78.72786857086984
57.98892418142371 78.64729077129542
61.745793337441306 78.56173251893
65.74605493528587 78.47090201116262
70.00547739229599 78.37449242442676
74.54085069814627 78.27218141248647
79.37005259841 78.16363062601337
84.51211906588864 78.04848526029629
89.98731933749553 77.9263736387891
95.81723581247715 77.7969068411266
102.02484912691901 77.65967838522033
108.634628739886 77.51426397407998
115.67262938827614 77.36022131908537
123.16659379059347 77.19709005254758
131.1460620044883 77.02439174353262
139.64248786913063 76.84163003205973
148.68936299141743 76.64829089790776
158.32234876474746 76.44384308134013
168.57941694076075 76.22773867406168
179.50099930815816 75.99941389961104
191.1301470686055 75.7582901031287
203.51270053796563 75.5037749709743
216.69746984178835 75.23526400094283
230.7364273173341 74.95214224378911
245.68491238055393 74.65378633634705
261.60184966557193 74.33956684565759
278.5499812965528 74.00885094212506
296.5961142075259 73.66100541773487
315.8113834850658 73.29540006271684
336.2715327718908 72.91141141066508
358.0572128366741 72.50842685797221
381.25429948700105 72.08584915846288
405.9542320786224 71.64310128829857
432.2543739553627 71.17963166957388
460.258396240481 70.69491973356921
490.0766864923265 70.18848179643317
521.8267838351593 69.6598772112518
555.6338422803334 69.10871475117492
591.6311240642094 68.53465916871771
629.960524947437 67.93743786678944
670.7731335462663 67.31684760772568
714.2298269006938 66.67276117795438
760.5019046270565 66.0051339182965
809.7717641548397 65.31401002368247
862.2336197093495 64.59952851167364
918.0942678743853 63.86192875699945
977.5739027526402 63.10155548972099
1040.9069839370575 62.318863157895876
1108.3431607145728 61.51441956195508
1180.1482561452879 60.68890867749076
1256.6053148961805 59.84313259573693
1338.015718959747 58.97801252648661
1424.700375655567 58.09458882613757
1517.0009825977056 57.19402003343626
1615.2813746142795 56.27758091656959
1719.928957928509 55.34665955668121
1831.3562372546012 54.40275351369834
1950.0024418280298 53.44746513953601
2076.3352567797792 52.48249612029127
2210.8526666793514 51.50964134201864
2354.084918513532 50.53078218329502
2506.5966118386305 49.54787934145635
2668.98892434529 48.56296529779197
2841.901981608688 47.57813652010411
3026.0173803653233 46.59554548919373
3222.06087526274 45.61739261967006
3430.8052396729927 44.64591812596155
3653.073311846677 43.68339386276203
3889.741238415044 42.73211514680116
4141.741928025567 41.79439254532342
4410.068728724671 40.87254359754267
4695.779343583273 39.96888442006316
5000.0 39.08572113707848
};
\addlegendentry{$\mathrm{CaCl_2}$}

\addplot [line width = 2pt, KITblue]
table {%
10.0 79.65621664757761
10.64785977823349 79.63403327652586
11.337691785692261 79.61042458580106
12.072215234288088 79.58529980982141
12.854325502735373 79.55856251126437
13.687105549689702 79.5301102410877
14.57383806629774 79.49983418011067
15.518018416061985 79.46761876142222
16.523368413027306 79.43334127289388
17.59385099260072 79.39687143909299
18.733685832834666 79.35807098191982
19.94736598775028 79.31679315932928
21.2396755982669 79.27288228154845
22.61570875055135 79.22617320426447
24.0808895561239 79.17649079833792
25.640993532873463 79.1236493956939
27.302170371262854 79.06745221116293
29.070968175464785 79.0076907401855
30.954359274983734 78.94414413246344
32.95976770850881 78.8765785418401
35.09509848833501 78.8047464529239
37.36876876070854 78.72838598523994
39.78974098492568 78.64722017600656
42.36755826197188 78.56095624299161
45.11238195196144 78.46928482931366
48.03503172865967 78.37187923251948
51.1470282289765 78.26839462079722
54.460638465549216 78.15846723978217
57.98892418142371 78.04171361407991
61.745793337441306 77.91772974838294
65.74605493528587 77.78609033389036
70.00547739229599 77.64634796666657
74.54085069814627 77.49803238559737
79.37005259841 77.34064973872562
84.51211906588864 77.17368188797779
89.98731933749553 76.99658576363016
95.81723581247715 76.80879278131096
102.02484912691901 76.60970833589113
108.634628739886 76.39871138827993
115.67262938827614 76.17515416290519
123.16659379059347 75.93836197551445
131.1460620044883 75.68763321286694
139.64248786913063 75.42223948787985
148.68936299141743 75.14142599582252
158.32234876474746 74.84441209918644
168.57941694076075 74.53039217086014
179.50099930815816 74.19853672716013
191.1301470686055 73.84799388405584
203.51270053796563 73.4778911715126
216.69746984178835 73.08733774218736
230.7364273173341 72.67542701166056
245.68491238055393 72.24123976787646
261.60184966557193 71.78384778738868
278.5499812965528 71.30231799524815
296.5961142075259 70.7957172038056
315.8113834850658 70.26311746319882
336.2715327718908 69.70360205272226
358.0572128366741 69.11627213750566
381.25429948700105 68.50025410882834
405.9542320786224 67.85470761885807
432.2543739553627 67.17883431153099
460.258396240481 66.47188724061354
490.0766864923265 65.73318095367675
521.8267838351593 64.9621022067773
555.6338422803334 64.15812125914289
591.6311240642094 63.32080368022416
629.960524947437 62.44982258330073
670.7731335462663 61.54497118067992
714.2298269006938 60.606175535757004
760.5019046270565 59.63350736724743
809.7717641548397 58.6271967412599
862.2336197093495 57.587644468146344
918.0942678743853 56.51543400387731
977.5739027526402 55.4113426407467
1040.9069839370575 54.276351760222795
1108.3431607145728 53.111655912452065
1180.1482561452879 51.918670482979024
1256.6053148961805 50.69903770830186
1338.015718959747 49.45463080848051
1424.700375655567 48.18755601756926
1517.0009825977056 46.900152311425074
1615.2813746142795 45.59498865752941
1719.928957928509 44.27485864274275
1831.3562372546012 42.942772372054264
1950.0024418280298 41.60194557385022
2076.3352567797792 40.25578589423372
2210.8526666793514 38.90787641351962
2354.084918513532 37.56195647104814
2506.5966118386305 36.22189993860678
2668.98892434529 34.891691136602404
2841.901981608688 33.57539863919775
3026.0173803653233 32.277147263408246
3222.06087526274 31.001088581170944
3430.8052396729927 29.75137033125563
3653.073311846677 28.532105138343073
3889.741238415044 27.347338968585156
4141.741928025567 26.20101976367593
4410.068728724671 25.09696669835554
4695.779343583273 24.038840499116063
5000.0 23.030115244770755
};
\addlegendentry{$\mathrm{MgCl_2}$}

\addplot [line width = 2pt, KITblack]
table {%
10.0 79.80661322186866
10.64785977823349 79.7941114407911
11.337691785692261 79.78080331995199
12.072215234288088 79.76663709991396
12.854325502735373 79.75155773185405
13.687105549689702 79.73550667293947
14.57383806629774 79.71842166956807
15.518018416061985 79.70023652783391
16.523368413027306 79.68088087055665
17.59385099260072 79.66027988019252
18.733685832834666 79.63835402692385
19.94736598775028 79.61501878120684
21.2396755982669 79.59018431004085
22.61570875055135 79.56375515620987
24.0808895561239 79.53562989973686
25.640993532873463 79.50570080078728
27.302170371262854 79.47385342325684
29.070968175464785 79.43996623828575
30.954359274983734 79.40391020695381
32.95976770850881 79.36554834143334
35.09509848833501 79.32473524390761
37.36876876070854 79.28131662260613
39.78974098492568 79.23512878436401
42.36755826197188 79.18599810318445
45.11238195196144 79.13374046437232
48.03503172865967 79.07816068391638
51.1470282289765 79.0190519029292
54.460638465549216 78.95619495711217
57.98892418142371 78.88935772140022
61.745793337441306 78.81829443016125
65.74605493528587 78.7427449735826
70.00547739229599 78.6624341711767
74.54085069814627 78.57707102368296
79.37005259841 78.4863479450412
84.51211906588864 78.38993997656652
89.98731933749553 78.28750398597339
95.81723581247715 78.1786778544856
102.02484912691901 78.06307965593216
108.634628739886 77.94030683247756
115.67262938827614 77.80993537247271
123.16659379059347 77.67151899684818
131.1460620044883 77.52458836151216
139.64248786913063 77.3686502843666
148.68936299141743 77.2031870068247
158.32234876474746 77.02765550110571
168.57941694076075 76.8414868361034
179.50099930815816 76.64408561627624
191.1301470686055 76.43482950979109
203.51270053796563 76.21306888406541
216.69746984178835 75.97812656889418
230.7364273173341 75.72929776950392
245.68491238055393 75.46585015413883
261.60184966557193 75.18702414313071
278.5499812965528 74.89203342881082
296.5961142075259 74.5800657580563
315.8113834850658 74.25028401168058
336.2715327718908 73.90182761722664
358.0572128366741 73.533814333941
381.25429948700105 73.14534245071336
405.9542320786224 72.73549343948042
432.2543739553627 72.30333510789933
460.258396240481 71.84792529588341
490.0766864923265 71.36831616072345
521.8267838351593 70.86355909484075
555.6338422803334 70.33271031857475
591.6311240642094 69.77483718761731
629.960524947437 69.1890252505903
670.7731335462663 68.57438608663033
714.2298269006938 67.93006594550876
760.5019046270565 67.25525520360029
809.7717641548397 66.54919863775332
862.2336197093495 65.81120650568373
918.0942678743853 65.04066640581051
977.5739027526402 64.23705587144485
1040.9069839370575 63.39995563395861
1108.3431607145728 62.52906346711137
1180.1482561452879 61.6242085003213
1256.6053148961805 60.685365862651174
1338.015718959747 59.71267149210593
1424.700375655567 58.70643691709095
1517.0009825977056 57.667163789295515
1615.2813746142795 56.59555792071204
1719.928957928509 55.49254255297968
1831.3562372546012 54.35927056585689
1950.0024418280298 53.19713531457347
2076.3352567797792 52.007779774316475
2210.8526666793514 50.793103665390085
2354.084918513532 49.555268235810736
2506.5966118386305 48.29669839027524
2668.98892434529 47.02008187638718
2841.901981608688 45.728365271282655
3026.0173803653233 44.42474655455275
3222.06087526274 43.11266410640782
3430.8052396729927 41.79578203270685
3653.073311846677 40.47797178965666
3889.741238415044 39.16329015906632
4141.741928025567 37.85595370797796
4410.068728724671 36.560309951853725
4695.779343583273 35.28080552554529
5000.0 34.02195174808034
};
\addlegendentry{$\mathrm{NaCl}$}

\nextgroupplot[
    xmin = 0, xmax = 1.13,
    enlarge x limits={abs=0.005,lower},
    x coord trafo/.code=\pgfmathparse{sqrt(#1)},
    x coord inv trafo/.code=\pgfmathparse{(#1)^2},
    y coord trafo/.code=\pgfmathparse{sqrt(#1)},
    y coord inv trafo/.code=\pgfmathparse{(#1)^2},
    ymin = 0.2, ymax = 90,
    xtick={0,1},
    xticklabels={0,1},
    extra x ticks={0.1,0.2,0.3,0.4,0.5,0.6,0.7,0.8,0.9},
    extra x tick style={major tick length = \pgfkeysvalueof{/pgfplots/minor tick length}},
    extra x tick labels={},
    yticklabel pos=right,
    ylabel= \shortstack{$\varepsilon_\mathrm{r}^\mathrm{m}$ \\ \tiny{(sqrt scale)}},
    ylabel style = {yshift = 0.5cm},
    ytick={1.8,80},
    extra y ticks={9.62,17.44,25.26,33.08,40.9,48.72,56.54,64.36, 72.18},
    extra y tick style={major tick length = \pgfkeysvalueof{/pgfplots/minor tick length}},
    extra y tick labels={},
    yticklabels={$\varepsilon_\mathrm{r}^\mathrm{p}$, $\varepsilon_\mathrm{r}^\mathrm{s}$ },
    xlabel = \shortstack{$\varphi_\mathrm{w}$ \\ \tiny{(sqrt scale)}},
    xlabel style = {yshift = 0.4cm},
    legend style = {
        anchor = north west, 
        at={(0,1)}, }
    ]
\addplot [line width = 2pt, KITblack, dashed]
table {%
0.0 1.8
0.010101010101010102 1.8515088678002487
0.020202020202020204 1.9040098522167486
0.030303030303030304 1.957531895844214
0.04040404040404041 2.0121050781053795
0.05050505050505051 2.06776067162097
0.06060606060606061 2.1245312019674145
0.07070707070707072 2.182450511062233
0.08080808080808081 2.241553824436661
0.09090909090909091 2.3018778226764915
0.10101010101010102 2.363460717335611
0.11111111111111112 2.426342331652329
0.12121212121212122 2.4905641864268193
0.13131313131313133 2.5561695914488003
0.14141414141414144 2.6232037428985184
0.15151515151515152 2.6917138271813497
0.16161616161616163 2.7617491316973184
0.17171717171717174 2.8333611630919826
0.18181818181818182 2.906603773584906
0.19191919191919193 2.9815332960268606
0.20202020202020204 3.058208688397593
0.21212121212121213 3.136691688523108
0.22222222222222224 3.217046979865772
0.23232323232323235 3.2993423693229182
0.24242424242424243 3.383648978061129
0.25252525252525254 3.4700414465150136
0.26262626262626265 3.558598154792415
0.27272727272727276 3.649401459854015
0.2828282828282829 3.7425379509758834
0.29292929292929293 3.838098725160603
0.30303030303030304 3.9361796843383257
0.31313131313131315 4.036881856395886
0.32323232323232326 4.140311742293038
0.33333333333333337 4.246581691772886
0.3434343434343435 4.355810310452863
0.3535353535353536 4.4681229013972645
0.36363636363636365 4.5836519446275545
0.37373737373737376 4.702537618428386
0.38383838383838387 4.824928366762179
0.393939393939394 4.9509815176214
0.4040404040404041 5.080863957734443
0.4141414141414142 5.214752869709284
0.42424242424242425 5.352836538461538
0.43434343434343436 5.495315234645286
0.4444444444444445 5.642402183803459
0.4545454545454546 5.794324631101023
0.4646464646464647 5.951325012822706
0.4747474747474748 6.113662247337539
0.48484848484848486 6.281613159989388
0.494949494949495 6.455474058407957
0.5050505050505051 6.6355624770978405
0.5151515151515152 6.822219112914514
0.5252525252525253 7.015809976247033
0.5353535353535354 7.2167287864853105
0.5454545454545455 7.425399644760217
0.5555555555555556 7.6422800221361396
0.5656565656565657 7.867864107564409
0.5757575757575758 8.10268656716418
0.5858585858585859 8.347326775021388
0.595959595959596 8.60241358597718
0.6060606060606061 8.868630733177103
0.6161616161616162 9.14672294792083
0.6262626262626263 9.437502917152862
0.6363636363636365 9.741859215475555
0.6464646464646465 10.060765374740193
0.6565656565656566 10.395290286215321
0.6666666666666667 10.74661016949153
0.6767676767676768 11.11602239051696
0.686868686868687 11.504961470866574
0.696969696969697 11.915017704644871
0.7070707070707072 12.347958892377969
0.7171717171717172 12.805755818221868
0.7272727272727273 13.290612244897964
0.7373737373737375 13.805000389438437
0.7474747474747475 14.351703084832911
0.7575757575757577 14.933864145309785
0.7676767676767677 15.55504885993486
0.7777777777777778 16.219317073170735
0.787878787878788 16.93131201764059
0.797979797979798 17.696369007910043
0.8080808080808082 18.52064937636114
0.8181818181818182 19.411306765523648
0.8282828282828284 20.37669527896998
0.8383838383838385 21.426632321648583
0.8484848484848485 22.572733661279006
0.8585858585858587 23.828844973609932
0.8686868686868687 25.21160391954617
0.8787878787878789 26.74118126272916
0.888888888888889 28.442271293375427
0.8989898989898991 30.34543521190949
0.9090909090909092 32.48895348837212
0.9191919191919192 34.92142734011379
0.9292929292929294 37.70550831792982
0.9393939393939394 40.92337118948003
0.9494949494949496 44.684961106309544
0.9595959595959597 49.14080717488802
0.9696969696969697 54.50265210608426
0.9797979797979799 61.078089725036314
0.98989898989899 69.33172323759804
1.0 80.00000000000013
};
\addlegendentry{$\mathbb{C} \widehat{=} \mathrm{p}$}
\addplot [line width = 2pt, KITblack]
table {%
0.0 1.8000000000000018
0.010101010101010102 2.334282022481399
0.020202020202020204 2.8721042552830998
0.030303030303030304 3.413502002091362
0.04040404040404041 3.958511037565256
0.05050505050505051 4.507167615216741
0.06060606060606061 5.059508475449532
0.07070707070707072 5.615570853760464
0.08080808080808081 6.175392489107134
0.09090909090909091 6.739011632445938
0.10101010101010102 7.306467055444338
0.11111111111111112 7.877798059371704
0.12121212121212122 8.453044484172894
0.13131313131313133 9.032246717729103
0.14141414141414144 9.615445705310345
0.15151515151515152 10.20268295922433
0.16161616161616163 10.794000568666485
0.17171717171717174 11.389441209775935
0.18181818181818182 11.989048155902731
0.19191919191919193 12.592865288091158
0.20202020202020204 13.20093710578483
0.21212121212121213 13.813308737758733
0.22222222222222224 14.430025953284092
0.23232323232323235 15.05113517353175
0.24242424242424243 15.676683483220007
0.25252525252525254 16.306718642513193
0.26262626262626265 16.94128909917706
0.27272727272727276 17.58044400099776
0.2828282828282829 18.224233208470874
0.29292929292929293 18.87270730776747
0.30303030303030304 19.52591762398431
0.31313131313131315 20.18391623468537
0.32323232323232326 20.84675598374229
0.33333333333333337 21.514490495481464
0.3434343434343435 22.18717418914564
0.3535353535353536 22.864862293678357
0.36363636363636365 23.547610862839466
0.37373737373737376 24.235476790660673
0.38383838383838387 24.928517827249816
0.393939393939394 25.626792594953216
0.4040404040404041 26.330360604885616
0.4141414141414142 27.03928227383744
0.42424242424242425 27.753618941569478
0.43434343434343436 28.4734328885054
0.4444444444444445 29.198787353832824
0.4545454545454546 29.92974655402401
0.4646464646464647 30.666375701787462
0.4747474747474748 31.4087410254623
0.48484848484848486 32.156909788867566
0.494949494949495 32.91095031161872
0.5050505050505051 33.67093198992444
0.5151515151515152 34.436925317876984
0.5252525252525253 35.209001909249714
0.5353535353535354 35.98723451981607
0.5454545454545455 36.771697070204546
0.5555555555555556 37.562464669304696
0.5656565656565657 38.359613638239914
0.5757575757575758 39.16322153492276
0.5858585858585859 39.97336717920976
0.595959595959596 40.79013067867254
0.6060606060606061 41.61359345500315
0.6161616161616162 42.44383827107185
0.6262626262626263 43.280949258656065
0.6363636363636365 44.12501194686037
0.6464646464646465 44.97611329124723
0.6565656565656566 45.83434170369974
0.6666666666666667 46.69978708303763
0.6767676767676768 47.57254084640886
0.686868686868687 48.452695961480075
0.696969696969697 49.34034697944915
0.7070707070707072 50.23559006890519
0.7171717171717172 51.13852305056076
0.7272727272727273 52.04924543288325
0.7373737373737375 52.967858448652194
0.7474747474747475 53.89446509247102
0.7575757575757577 54.829170159262375
0.7676767676767677 55.77208028377711
0.7777777777777778 56.72330398114846
0.787878787878788 57.68295168852348
0.797979797979798 58.65113580780562
0.8080808080808082 59.627970749542975
0.8181818181818182 60.61357297799814
0.8282828282828284 61.60806105743735
0.8383838383838385 62.61155569967689
0.8484848484848485 63.62417981292755
0.8585858585858587 64.64605855197803
0.8686868686868687 65.67731936976072
0.8787878787878789 66.7180920703443
0.888888888888889 67.76850886339938
0.8989898989898991 68.82870442018549
0.9090909090909092 69.89881593110873
0.9191919191919192 70.97898316490225
0.9292929292929294 72.0693485294828
0.9393939393939394 73.1700571345391
0.9494949494949496 74.28125685591012
0.9595959595959597 75.40309840181253
0.9696969696969697 76.5357353809805
0.9797979797979799 77.67932437278202
0.98989898989899 78.83402499937876
1.0 80.0
};
\addlegendentry{$\mathbb{C} \widehat{=}  \mathrm{s}$}

\addplot [KITblue, only marks, mark = x, mark size=5, mark options={line width=1.5pt}]
table {%
0.042105275092420055 2.1891064186684908
0.16800826843010258 3.854060023736336
0.2291021801000311 4.3782128373369815
0.2782249632485243 5.272353082714847
}; 
\addplot [KITred, only marks, mark = o, mark size=3, mark options={line width=1.5pt}]
table {%
0.030546955834964265 3.977390296781005
0.06109391166992853 4.994860815203539
0.0887512933594284 7.954780593562012
0.16264190672755516 13.011306545287885
0.2972135998260328 20.07194350611525
}; 
\node[anchor=south east] (lbl) at (axis cs: 0.2, 20) {\textcolor{KITred}{\scriptsize{Nafion 117}}};
\draw[KITred]  (axis cs: 0.05, 20) -- (axis cs: 0.05, 17) -- (axis cs: 0.12, 15);
\node[anchor=north west] (lbl) at (axis cs: 0.24, 4.3) {\textcolor{KITblue}{\scriptsize{XL-pGMA-z}}};
\end{groupplot}

\end{tikzpicture}

%% file: 5_Results/0_main.tex
\section{Results}
\label{C:Results}
From here on, we refer to the model derived above as the \textit{interaction-occupation model} (IOM). Section \ref{C:Theoretical_model_analysis} analyzes the IOM on a theoretical basis. Section \ref{C:Comparison_to_experiments} then compares experimentally determined static and dynamic properties of a commercial CR61 membrane in single-salt environments to estimates from all presented models. 

\subsection{Model analysis}
\label{C:Theoretical_model_analysis}
For the theoretical analysis, we discuss the low-salt limiting behavior of the IOM, as well as the influence of the interaction parameter and the association constants.

\subsubsection{Condensation in the low-salt limit}
\label{C:Analytical_low_salt}
Manning's limiting law predicts a saturation of the effective polymer charge $q_\mathrm{eff}$ for large bare polymer charge $\sim \text{(}c_\mathrm{X} \to \infty\text{)}$ through counterion condensation (see equation \ref{E:Manning_effective_concentration}). Since this prediction has proven remarkably accurate in the low-salt limit, it is a useful first benchmark for the IOM.\par
For a single-salt, single-occupation case, the limiting value of the residual charge $Z$ is obtained as (see supporting information \ref{C:Appendix_limiting_Z}):
\begin{equation}
    \lim_{c_\mathrm{X} \to \infty} Z = 0
\end{equation}
Using this, the limiting occupation fractions of co- and counterions $\theta_\mathrm{co}$ and $\theta_\mathrm{ct}$ become (see supporting information \ref{C:Appendix_limiting_theta}):
\begin{equation}
    \lim_{c_\mathrm{X}\to\infty}\theta_\mathrm{ct} \approx
    -\frac{z_\mathrm{X}}{z_\mathrm{ct}}
    \quad \text{and} \quad
    \lim_{c_\mathrm{X}\to\infty}\theta_\mathrm{co} \approx
    0
\end{equation}
For counterions with larger charge magnitude than the fixed sites $\lvert z_\mathrm{ct} \rvert > \lvert z_\mathrm{X} \rvert$, the low-salt limiting value of the effective charge follows directly. For equal charge magnitudes $\lvert z_\mathrm{ct} \rvert = \lvert z_\mathrm{X} \rvert$, the strict single-occupation closure becomes singular. We therefore relax this assumption by allowing for higher occupation states approximated by penalized individual association to regularize the limit for this case (see supporting information \ref{C:Appendix_limiting_q_eff}). The resulting limits are: 
\begin{equation}
    \lim_{c_\mathrm{X}\to\infty} q_\mathrm{eff}
    \approx
    \begin{cases}
        \dfrac{z_\mathrm{X}c^\circ}
        {K_\mathrm{ct}} \ ,
        &
        \lvert z_\mathrm{X}\rvert = \lvert z_\mathrm{ct}\rvert \\[1.2em]
        \dfrac{z_\mathrm{X}z_\mathrm{ct}c^\circ}
        {K_\mathrm{ct}\left(z_\mathrm{X}+z_\mathrm{ct}\right)} \ ,
        &
        \lvert z_\mathrm{X}\rvert < \lvert z_\mathrm{ct}\rvert
    \end{cases}
    \label{E:Low_salt_qeff_limit}
\end{equation}
Thus, the IOM reproduces limiting laws similar to those obtained by Manning for the saturation of the effective membrane charge (equation \ref{E:Manning_effective_concentration}). \par
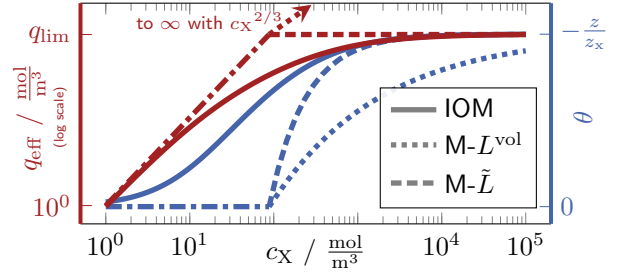
\begin{figure}[h]
    \centering
    \input{5_Results/Figures/salt_free_theta}
    \caption{The occupation fraction $\theta$ in the low-salt limit, computed for $z =2$ and $z_\mathrm{X} = -1$ with $K$ according to eq. \ref{E:K_equal_manning}. The labels M-$L^\mathrm{vol}$ and M-$\Tilde{L}$ refer to the Manning model with $L$ from an average volumetric distance and a corrected $\Tilde{L}$ according to eq. \ref{E:Manning_L_corrected}.}
    \label{F:Comparison_to_Manning}
\end{figure}

In a three-dimensional implementation of Manning condensation, an assumption on line length has to be made to link the fixed-site concentration $c_\mathrm{X}$ to line charge density (see section \ref{C:Donnan_Manning} and supporting information \ref{C:Manning_condensation_derivation}) \cite{Mareev2022}. If the line length is not specifically limited, e.g., when simple mean volumetric site-to-site distance $L^\mathrm{vol} = (c_\mathrm{X} N_\mathrm{A})^{-1/3}$ is applied, the effective charge $q_\mathrm{eff}$ diverges in the Manning model for $c_\mathrm{X} \to \infty$. A simple, non-diverging approach is to relate the site distance $L$ to the critical volumetric site-to-site distance:
\begin{equation}
    \Tilde{L} = \frac{{\xi_\mathrm{crit}}^3}{c_\mathrm{X} {\lambda_B}^2 N_\mathrm{A}} 
    \label{E:Manning_L_corrected}
\end{equation}
With this, the effective membrane charge in the Manning model (M-$\Tilde{L}$) converges to:
\begin{equation}
    q_\mathrm{eff}^\mathrm{M-\Tilde{L}} =
    \mathrm{sgn}(z_\mathrm{X}) \frac{\xi_\mathrm{crit} }{ \lambda_B^3 N_\mathrm{A}} 
\end{equation}
This allows to state a condition on the association constant $K_\mathrm{ct}$ for the equality of the IOM and the D-M model in the high $c_\mathrm{X}$ limit:
\begin{equation}
    K_\mathrm{ct}^\text{D-M} = 
    \begin{cases}
        \dfrac{c^\circ \lambda_B^3 N_\mathrm{A}}
        {\xi_\mathrm{crit}} \lvert z_\mathrm{X}\rvert \ ,
        &
        \lvert z_\mathrm{X}\rvert = \lvert z_\mathrm{ct}\rvert \\[1.2em]
        \dfrac{c^\circ \lambda_B^3 N_\mathrm{A}}
        {\xi_\mathrm{crit}}
        \dfrac{\lvert z_\mathrm{X}z_\mathrm{ct}\rvert}
        {\lvert z_\mathrm{X}+z_\mathrm{ct}\rvert} \ ,
        &
        \lvert z_\mathrm{X}\rvert < \lvert z_\mathrm{ct}\rvert
    \end{cases}
    \label{E:K_equal_manning}
\end{equation}
Figure \ref{F:Comparison_to_Manning} compares both theories along this premise. It shows that the Manning condensation in the D-M model can be regarded as a discontinuous piecewise approximation of the IOM.

\subsubsection{The role of interaction}
\label{C:Interaction_role}
To isolate the influence of nearest-neighbor interactions, we vary the dimensionless interaction parameter $n_\mathrm{n}U$ for the CR61 membrane at otherwise fixed parameters. Its most direct effect is on the effective membrane charge $q_\mathrm{eff}$, through which it propagates to other model quantities.\par
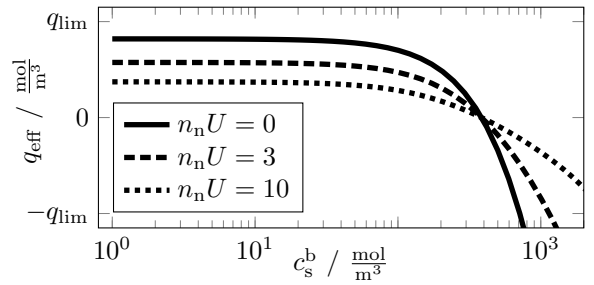
\begin{figure}[h]
    \centering
    \input{5_Results/Figures/Interaction_deviation}
    \caption{The influence of the interaction term $n_\mathrm{n}U$ on the effective charge $q_\mathrm{eff}$ for divalent counterions $z_\mathrm{ct}=2$ at constant water uptake.}
    \label{F:Interaction_deviation}
\end{figure}

Therefore, figure \ref{F:Interaction_deviation} depicts the influence of $n_\mathrm{n}U$ on $q_\mathrm{eff}$ for the CR61 membrane in an environment with divalent counterions. It shows that $n_\mathrm{n}U$ controls both the steepness of the charge reversal at elevated bulk salt concentrations $c_\mathrm{s}^\mathrm{b}$ and the deviation from the high-$c_\mathrm{X}$ limiting value of $q_\mathrm{eff}$ (equation \ref{E:Low_salt_qeff_limit}) in the low-salt limit. \par
Here, high values of $n_\mathrm{n}U$ reduce the steepness of the charge reversal, since interaction penalizes like-charged neighboring sites. The salt concentration at which $q_\mathrm{eff}$ crosses zero is independent of $n_\mathrm{n}U$. This is due to interaction entering the exponential of the condensation equation only through the product $n_\mathrm{n} z Z U$ (equations \ref{E:System_of_Mass_action_equations_simplified} and \ref{E:Model_closure}), which vanishes for $Z=0$.\par
The deviation from $q_\mathrm{lim}$ in the low-salt limit also follows from this condensation equation. Since $z_\mathrm{ct}Z < 0$ and $n_\mathrm{n}U \geq 0$, the exponential fulfills $\exp(-n_\mathrm{n}U z_\mathrm{ct}Z) > 1$. Therefore, stronger interactions favour counterion occupation and shift the self-consistent solution towards smaller $|Z|$, thereby reducing the magnitude of $q_\mathrm{eff}=c_\mathrm{X}Z$.
\subsubsection{Sensitivity to association constants}
\label{C:Association_role}
\begin{figure}[h]
    \centering
    \vspace{-0.3cm}
    \input{5_Results/Figures/Association_deviation}
    \caption{The influence of the association constant on the co- and counterion concentrations for divalent counterions $z_\mathrm{ct}=2$ with an interaction of $n_\mathrm{n}U = 3$.}
    \label{F:Association_deviation}
\end{figure}
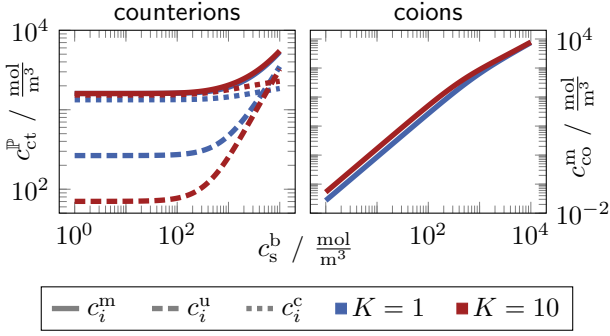

To assess the sensitivity of the IOM to association constants, we vary the counterion association constant $K_\mathrm{ct}$ within a physically reasonable range while keeping all other parameters fixed. As seen from equation \ref{E:Definition_of_chi_and_zeta}, $K_\mathrm{ct}$ enters the occupation factor $\chi_\mathrm{ct}$ and therefore controls the equilibrium between condensed and uncondensed ions. Here, larger association constants favour occupation and shift counterions towards the condensed state, as shown in figure \ref{F:Association_deviation} (left).\par
Since stronger condensation reduces the effective charge $q_\mathrm{eff}$ (cf. equation \ref{E:Low_salt_qeff_limit}), larger values of $K_\mathrm{ct}$ also weaken partitioning (equation \ref{E:IOM_partitioning}), resulting in higher coion uptake. This effect, however, is rather weak. Figure \ref{F:Association_deviation} (right) shows that even order-of-magnitude variations in $K_\mathrm{ct}$ only lead to moderate changes of the coion concentration in the membrane. This indicates that the influence of condensation on coion uptake is indirect and, to some extent, buffered by charge neutrality. \par

\subsection{Comparison to experiments}
\label{C:Comparison_to_experiments}
The previous sections examined the IOM from a theoretical perspective, focusing on properties that are hardly accessible experimentally. The following sections compare model predictions to literature data for experimentally measurable membrane properties and assess how well the IOM, the SDE model, the Donnan-Manning model, and the low-T* model reproduce the experimental data. For this purpose, we consider partitioning \cite{Galizia2019}, permeability \cite{Kamcev2017}, and the Donnan potential \cite{Gokturk2022}.\par
The IOM parameters are applied as stated in section \ref{C:Parameter_estimation}.\par

\subsubsection{Partitioning}
\label{C:Partitioning}
Partitioning is usually analyzed through the concentrations of co- and counterions in the membrane at varying ambient bulk salt concentrations $c_\mathrm{s}^\mathrm{b}$ \cite{Galizia2017, Galizia2019}. As stated above, the IOM is parameterized according to section \ref{C:Parameter_estimation}. For the SDE model, the exclusion factor is treated as a fitting parameter. The Donnan-Manning (D-M) model uses the condensation threshold from theory and neglects excess exclusion. For the low-T* model, the pair and triplet formation constants are fitted to best reproduce the experimental data.\par
\begin{figure}[h]
    \centering
    \input{5_Results/Figures/PartitioningNaCl}
    \vspace{-0.2cm}
    \caption{Co- and counterion concentrations in the CR61 membrane depending on ambient bulk salt concentration (partitioning) for aqueous solutions of $\mathrm{NaCl}$.}
    \label{F:PartitioningNaCl}
\end{figure}
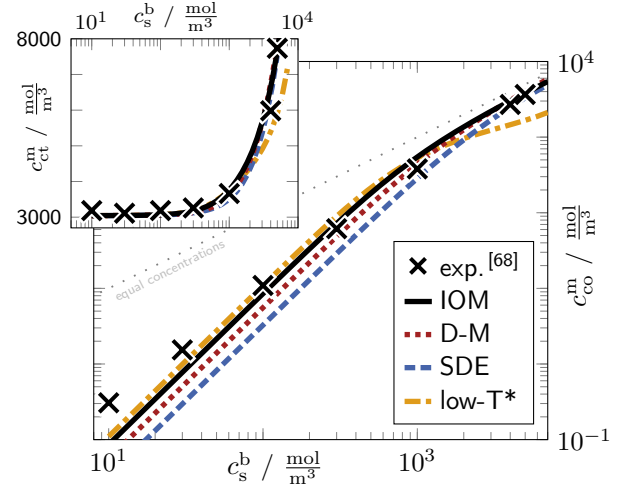

Figure \ref{F:PartitioningNaCl} compares the concentrations of co- and counterions in a CR61 membrane equilibrated with aqueous $\mathrm{NaCl}$ solutions, as predicted by the four models, to experimental data \cite{Galizia2019}. It shows, that the experimental data is reproduced with appropriate accuracy by all four models. Only the low-T* model shows noticeable deviations from the experimental data at high salt concentrations, because the active sites saturate in the triplet state and larger multiplets are neglected. In this saturated state, the remaining increase in counterion concentration is entirely caused by decreasing water uptake, which in turn increases $c_\mathrm{X}$.
\par
\begin{figure}[h]
    \centering
    \input{5_Results/Figures/PartitioningCaCl2}
    \vspace{-0.2cm}
    \caption{Co- and counterion concentrations in the CR61 membrane depending on ambient bulk salt concentration (partitioning) for aqueous solutions of $\mathrm{CaCl_2}$.}
    \label{F:PartitioningCaCl2}
\end{figure}
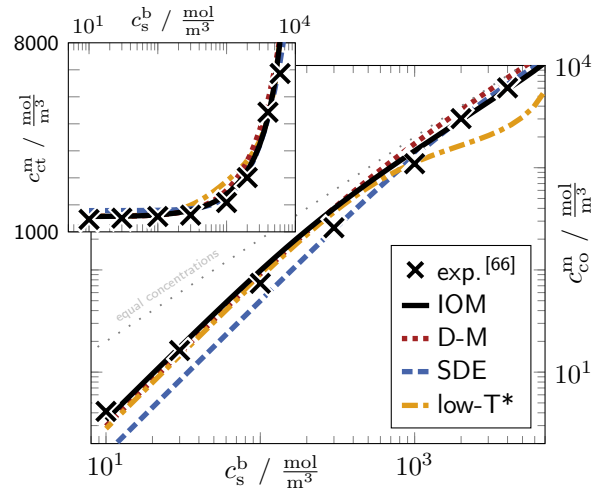

For divalent counterions, a similar level of agreement is found. All models adequately reproduce the experimentally determined co- and counterion concentrations in a membrane equilibrated with aqueous $\mathrm{CaCl_2}$ solutions, as shown in Figure \ref{F:PartitioningCaCl2}.\par
However, both the SDE and the low-T* model require refitted parameters to reproduce this case with comparable accuracy. The Donnan-Manning model again uses a Manning parameter $\xi$ estimated from theory (see equation \ref{E:Manning_Parameter}), and the IOM retains the parameterization described in section \ref{C:Parameter_estimation}.
\subsubsection{Donnan potential}
\label{C:Donnan_Potential}
The partitioning analyzed above is directly linked to the Donnan potential through equation \ref{E:IOM_partitioning}. Recent studies have shown that the Donnan potential can also be accessed experimentally by tender ambient-pressure X-ray photoelectron spectroscopy \cite{Gokturk2022}.\par
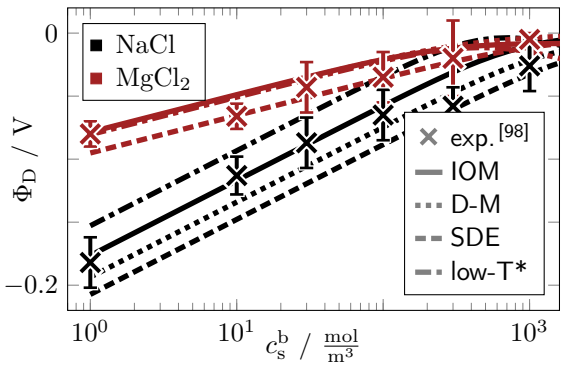
\begin{figure}[h]
    \centering
    \input{5_Results/Figures/Donnan_potentials}
    \caption{Donnan potentials as predicted by the presented models compared to experimental data from Gokturk et al. \cite{Gokturk2022}}
    \label{F:Donnan_potentials}
\end{figure}

Figure \ref{F:Donnan_potentials} compares the model predictions to experimental data for the CR61 membrane \cite{Gokturk2022}. It shows that the IOM and the Donnan-Manning model reproduce the experimental data most accurately, supporting the distinction between condensed and uncondensed ion states underlying both approaches.
\par

\subsubsection{Permeability}
\label{C:Permeability}
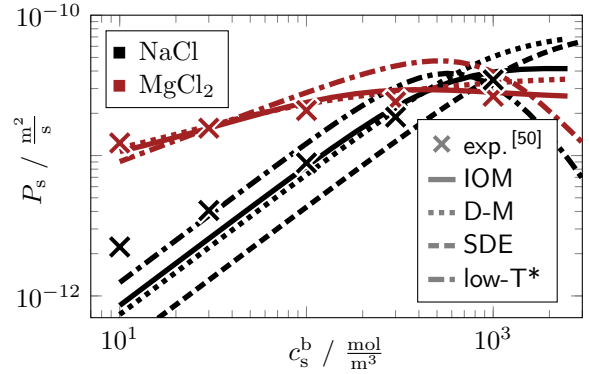
\begin{figure}[h]
    \centering
    \input{5_Results/Figures/Permeability}
    \caption{CR61 membrane permeability of $\mathrm{NaCl}$ and $\mathrm{MgCl_2}$ as predicted by the presented models compared to experimental data from Kamcev et al. \cite{Kamcev2017}}
    \label{F:Permeability}
\end{figure}

While partitioning is a static property, permeability as a dynamic property is usually the more relevant quantity for technical applications. A common experimental setup for the determination of permeability is a diffusion cell with two reservoirs separated by a membrane, where the upstream compartment has a defined solute concentration and the downstream compartment is initially filled with deionized water. By measuring the increase in conductivity of the downstream compartment over time, the flux across the membrane and subsequently the apparent permeability can be estimated. Since concentrations vary along the membrane, a 1D simulation is preferable over 0D property estimates. Figure \ref{F:Permeability} shows the resulting permeability estimates for $\mathrm{NaCl}$ and $\mathrm{MgCl_2}$ of the four discussed models compared to experimental data. Note that $\mathrm{MgCl_2}$ is used here over $\mathrm{CaCl_2}$ for reasons of data availability. However, since both salts partition and permeate similarly, these differences are not expected to affect the qualitative analysis presented here. \par
Again, all models reproduce the experimental data \cite{Kamcev2017} with reasonable accuracy. As before, the SDE model shows the highest deviation at low concentrations, and the low-T* model is the least accurate at higher concentrations due to saturation. The Donnan-Manning model and the IOM exhibit only minor differences, with the IOM showing only slightly better agreement at high salt concentrations. \par
Also, parameterization raises issues similar to those observed for partitioning in section \ref{C:Partitioning}. The SDE model requires different values for the diffusive hindrance coefficient of the coion $\mathrm{Cl^-}$ to fit both the $\mathrm{NaCl}$ and $\mathrm{MgCl_2}$ dataset. For the D-M model, the Mackie-Meares factor is $0.06$ for $\mathrm{NaCl}$ and $0.03$ for $\mathrm{MgCl_2}$, which is a larger difference than suggested by the difference in water uptake. For the low-T* model, the association constants had to be varied by factors of $\mathcal{O}(10)$ relative to the partitioning studies to reproduce the data accurately. The IOM is still parameterized with the same set of parameters as before for both salts.
\par

\subsection{Model comparison}
Sections \ref{C:Theoretical_model_analysis} and \ref{C:Comparison_to_experiments} analyze the IOM and compare its predictions with those of the SDE, D-M, and low-T* model. This comparison shows that all presented models reproduce the considered experimental data with reasonable accuracy. However, for the SDE model, the parameters do not retain their physical meaning and have to be treated as fit parameters. This is problematic when the ambient salt composition or its concentrations vary, since the predictive value is low and refitting is required for different conditions. From a theoretical standpoint, the low-T* model is consistent for varying conditions and salt mixtures thanks to its mass-action perspective. Yet, for membranes that do not strictly fulfill the low-T* rule, such as the CR61 membrane considered above, a consistent set of parameters could not be found. Also, the limitation to triplets introduces an upper limit for ion uptake, which is justified for some membranes, but not generally valid. The Donnan-Manning model is largely consistent and matches the experimental data equally well as the IOM does. Yet, its reliance on limiting laws becomes questionable at higher concentrations and some of its extensions are of a rather empirical nature. \par
The IOM consistently links static and dynamic properties through its mass-action mean-field approach. It circumvents the limitations of the low-T* model at high concentrations by accounting for both condensed and uncondensed ions. This makes it conceptually similar to the D-M model, to which its parameters also closely relate. The interaction parameter $n_\mathrm{n}U$ is related to the Manning parameter $\xi$ and the association constants $K$ correspond to the critical values of the Manning parameter $\xi_\mathrm{crit}$ (see equations \ref{E:Critical_value_of_the_manning_parameter} and \ref{E:K_equal_manning}). Thus, we argue that the IOM does not necessarily add further adjustable parameters while providing a more coherent approach to salt mixtures at elevated concentrations.\par
The most fundamental difference in the model equations is that the IOM employs an implicit condensation equation which is not limited in terms of salt concentration, whereas the D–M model relies on an explicit low-salt-limit condition. Also, mass-action handles salt mixtures more naturally than a purely mean-field based approach. The following section analyses the relevance of these differences for membranes in multi-ionic environments as commonly encountered in, e.g., aqueous batteries. \par

\section{Discussion: Salt mixtures}
\label{C:Multi-ionic}
For real membrane applications, salt mixtures are oftentimes more relevant than single salts. As an illustrative example, consider a mixture of $\mathrm{NaCl}$ and $\mathrm{MgCl_2}$, which share the common coion $\mathrm{Cl^-}$ and provide mono- and divalent counterions. To compare the IOM and the D-M model for this case, we examine their predictions for effective charge, partitioning, and permeability. Since comparable experimental data is scarce, this section is intended as a theoretical demonstration and plausibility test for the proposed framework.\par
\begin{figure}[h]
    \centering
    \input{5_Results/Figures/Mixture_qeff}
    \vspace{-0.5cm}
    \caption{The effective membrane charge $q_\mathrm{eff}$ of the CR61 membrane as predicted by the IOM (left) and the D-M model (right) depending on the concentrations of $\mathrm{NaCl}$ and $\mathrm{MgCl_2}$ in the bulk salt mixture.}
    \label{F:Salt_mixture_q_eff}
\end{figure}
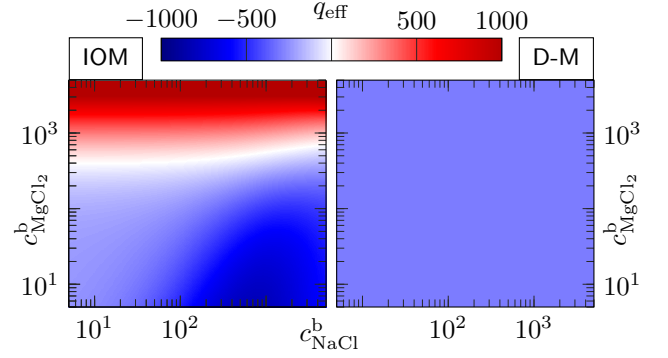

For the effective charge $q_\mathrm{eff}$, the IOM predicts an increase in magnitude at larger concentrations of monovalent $\mathrm{Na^+}$ counterions and charge reversal at larger concentrations of divalent $\mathrm{Mg^{2+}}$ counterions. The D-M model on the other hand predicts a constant $q_\mathrm{eff}$. This is depicted in figure \ref{F:Salt_mixture_q_eff}.\par
For the IOM, the charge reversal along increasing $\mathrm{MgCl_2}$ concentrations aligns with the discussion in section \ref{C:Interaction_role}; the increasing magnitude at higher relative $\mathrm{NaCl}$ concentrations is explained by lower relative uptake of $\mathrm{Mg^{2+}}$, shifting condensation towards $\mathrm{Na^+}$.
The constant value predicted by the D-M model is determined by equation \ref{E:Critical_value_of_the_manning_parameter}, which suggests that the critical value of the Manning parameter $\xi_\mathrm{crit}$ is set by the highest magnitude counterion valence. This can be problematic at trace amounts of such higher-valence counterions in an otherwise lower-valence electrolyte. \par
\par
\begin{figure}[h]
    \centering
    \input{5_Results/Figures/Mixture_partitioning}
    \vspace{-0.5cm}
    \caption{Left: Membrane coion ($Cl^-$) concentrations as predicted by the IOM for a CR61 membrane depending on the concentrations of $\mathrm{NaCl}$ and $\mathrm{MgCl_2}$ in the bulk salt mixture. Right: The relative differences $\eta_\mathrm{\,D\text{-M}}^\mathrm{\,IOM} = c_\mathrm{co}^\mathrm{m,IOM} / c_\mathrm{co}^\mathrm{m,D\text{-}M}$ of the predictions by the IOM and the D-M model.}
    \label{F:Salt_mixture_partitioning}
\end{figure}
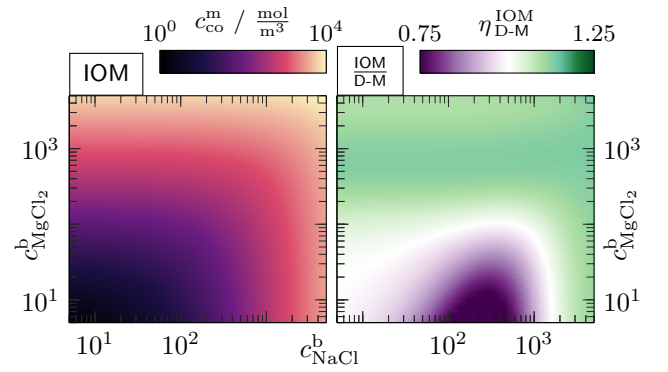

For the coion uptake (partitioning) as seen in figure \ref{F:Salt_mixture_partitioning}, the predictions of both models are similar in accordance with section \ref{C:Partitioning}. The IOM deviates from the D-M model at high salt concentrations due to charge reversal, as well as for low $\mathrm{MgCl_2}$ concentrations at intermediate $\mathrm{NaCl}$ concentrations. This is coherent with the differences in the predicted effective charge $q_\mathrm{eff}$.
\par
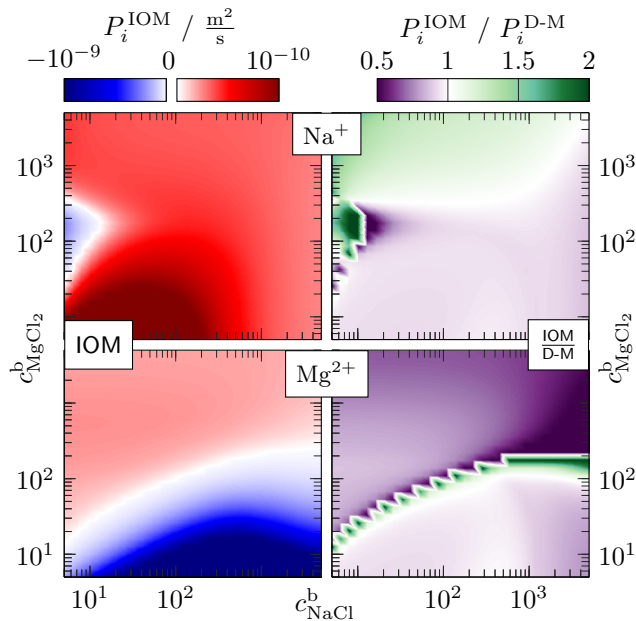
\begin{figure}[h]
    \centering
    \input{5_Results/Figures/Mixture_permeability}
    \vspace{-0.5cm}
    \caption{The permeabilities of $\mathrm{Na^+}$ and $\mathrm{Mg^{2+}}$ in a CR61 membrane depending on the concentrations of $\mathrm{NaCl}$ and $\mathrm{MgCl_2}$ in the bulk salt mixture.}
    \label{F:Salt_mixture_permeability}
\end{figure}

For the permeability of mono- and divalent counterions, the IOM and the D-M model are also in qualitative agreement with minor quantitative differences, as seen in figure \ref{F:Salt_mixture_permeability}. \par
Both models correctly reproduce the uphill transport of the divalent counterion at low divalent salt fractions known from literature (and likewise to a lesser extent for the monovalent ion at low monovalent salt fractions). The predicted onsets differ minorly, showing as neighboring opposing values in the relative predictions in the right panel.
This uphill transport is attributed to the electric potential, which is established primarily by the excess counterion and drives the small fraction of minority counterions against their concentration gradient \cite{Vermaas2014,Avci2016,Moya2020,Rijnaarts2017,Higa1988}. 
\par
In summary, both models are highly similar. The key advantage of the IOM towards salt mixtures is its internal consistency. While the Donnan–Manning model requires empirical extensions \cite{Bannon2024, Kamcev2017, Purpura2024, Wang2023} to achieve comparable predictions (such as equations \ref{E:Donnan_Manning_salt_mixture_condensation} and \ref{E:Critical_value_of_the_manning_parameter}), the IOM is naturally applicable.
For instance, the extension of the D-M model for the fractions of condensed ions (equation \ref{E:Donnan_Manning_salt_mixture_condensation}) is conceptually similar to the relations for the occupation fractions in the IOM model (equation \ref{E:Occupation_fractions_rearranged}), it just directly relates charge and size instead of association constants which are functions of these parameters. 

%% file: 5_Results/Figures/salt_free_theta.tex
\begin{tikzpicture}

\begin{axis}[
    tick align=inside,
    y axis line style={line width=1.5pt},
    axis y line*=right,
    y axis line style={KITblue},
    ytick style={KITblue},
    xlabel={$c_\mathrm{X} \ / \ \frac{\mathrm{mol}}{\mathrm{m^3}}$},
    xmin=5e-1, xmax=2e5,
    xmode=log,
    xtick={1e0,1e1,1e2,1e3,1e4,1e5},
    xticklabels={$10^{0}$,$10^{1}$,,,$10^{4}$,$10^{5}$},
    xlabel style = {yshift = 0.5cm},
    ylabel={$\textcolor{KITblue}\theta$},
    ymin=-0.05, ymax=0.59,
    ytick={0,0.5},
    yticklabels={$\textcolor{KITblue} 0$, $\textcolor{KITblue}{ -\frac{z}{z_\mathrm{x}}}$},
    ylabel style = {xshift = -0.05cm, yshift = 0.7cm},
    height = 4.5cm,
    width = 7.8cm,
]

\addplot [line width = 2pt, KITblue]
table {%
1 0.0134169099697296
1.02334021219164 0.013737664646504
1.04722518988843 0.0140658619732338
1.07166764803286 0.0144016604216562
1.09668059833687 0.0147452210469153
1.12227735620851 0.0150967074871059
1.14847154784029 0.0154562859601433
1.17527711746295 0.0158241252578152
1.20270833476851 0.016200396736869
1.23077980250667 0.0165852743069827
1.25950646425836 0.0169789344154671
1.28890361239089 0.0173815560285432
1.31898689619867 0.0177933206090345
1.34977233023394 0.0182144120903158
1.38127630283201 0.0186450168463539
1.4135155848354 0.0190853236576773
1.44650733852165 0.0195355236731079
1.48026912673951 0.0199958103670904
1.51481892225835 0.020466379492452
1.55017511733577 0.0209474290284258
1.58635653350859 0.0214391591237731
1.62338243161228 0.0219417720348407
1.66127252203429 0.0224554720583885
1.70004697520672 0.0229804654590311
1.7397264323438 0.0235169603911332
1.78033201643011 0.0240651668150073
1.82188534346517 0.0246252964072669
1.86440853397049 0.0251975624651888
1.90792422476527 0.0257821798049516
1.95245558101686 0.0263793646536185
1.99802630857255 0.0269893345347431
2.04466066657912 0.0276123081474866
2.09238348039698 0.0282485052391424
2.14122015481573 0.0288981464709763
2.19119668757815 0.0295614532773028
2.24233968321985 0.0302386477177284
2.29467636723194 0.0309299523225106
2.34823460055428 0.0316355899309925
2.40304289440697 0.0323557835230893
2.45913042546804 0.0330907560438236
2.51652705140539 0.0338407302209196
2.5752633267712 0.0346059283754881
2.63537051926739 0.0353865722258537
2.69688062639069 0.036182882684596
2.75982639246618 0.0369950796489004
2.82424132607844 0.0378233817843339
2.89015971790951 0.0386680063021874
2.95761665899325 0.0395291687305508
3.02664805939569 0.0404070826793087
3.09729066733141 0.0413019595992771
3.16958208872612 0.0422140085357182
3.24356080723581 0.0431434358765089
3.31926620473319 0.0440904450952549
3.39673858227221 0.0450552364896771
3.47601918154198 0.0460380069156206
3.55715020682139 0.047038949517067
3.64017484744614 0.0480582534525546
3.72513730080021 0.0490961036184424
3.81208279584389 0.0501526803694766
3.90105761719099 0.0512281592371458
3.99210912974805 0.0523227106463387
4.08528580392856 0.0534364996308383
4.18063724145576 0.0545696855482115
4.27821420176762 0.0557224217946787
4.37806862903817 0.0568948555205604
4.48025367982949 0.0580871273469271
4.5848237513891 0.0592993710840853
4.6918345106078 0.0605317134525571
4.80134292365345 0.0617842738072182
4.91340728629636 0.0630571638652719
5.02808725494248 0.0643504874387447
5.14544387839093 0.0656643401721989
5.26553963033276 0.0669988092863524
5.38843844260822 0.0683539733283086
5.5142057392403 0.0697299019290854
5.64290847126254 0.0711266555691373
5.77461515235982 0.0725442853525472
5.90939589534097 0.0739828327905645
6.04732244946265 0.0754423295951413
6.18846823862439 0.0769227974831087
6.33290840045512 0.0784242479916132
6.48071982631197 0.0799466823054084
6.63198120221267 0.0814900910965774
6.7867730507233 0.0830544543772197
6.9451777738237 0.0846397413656267
7.10727969677342 0.0862459103664068
7.27316511300146 0.087872908665018
7.44292233004376 0.0895206724370923
7.61664171655289 0.0911891266729295
7.79441575040495 0.0928781851174635
7.97633906792928 0.0945877502259866
8.16250851428723 0.0963177131358518
8.35302319502678 0.0980679536543362
8.54798452884041 0.0998383402627946
8.74749630155442 0.101628730137192
8.9516647213783 0.103438969185038
9.16059847544371 0.10526889209872
9.374408787663 0.107118322425151
9.59320947793824 0.108987072651633
9.81711702275219 0.110874944307755
10.0462506171734 0.11278172808313
10.2807322383087 0.11470720396069
10.5206867102362 0.116651141365245
10.7662417704549 0.118613299326958
11.0175281378839 0.120593426659326
11.2746795824495 0.122591262151257
11.5378329962966 0.12460653477274
11.8071284666619 0.126638963893634
12.0827093504478 0.128688259515003
12.3647223505371 0.130754122512439
12.6533175938894 0.132836244890772
12.948648711459 0.134934310049527
13.2508729199795 0.137047993058497
13.5601511056563 0.139176960942741
13.8766479098131 0.141320872976337
14.2005318165368 0.14347938098418
14.531975242369 0.14565212965112
14.8711546280895 0.14783875683772
15.2182505326439 0.1500388939019
15.5734477292613 0.152252166025763
15.9369353038177 0.154478192546856
16.3089067554933 0.156716587293161
16.6895600997802 0.158966958921091
17.0790979738943 0.161228911255792
17.4777277446468 0.16350204363305
17.8856616188346 0.165785951242125
18.3031167562061 0.168080225468844
18.7303153850644 0.170384454238307
19.1674849205681 0.172698222356569
19.6148580857943 0.175021111850695
20.0726730356257 0.1773527023066
20.5411734835306 0.179692571204106
21.0206088313016 0.182040294248684
21.5112343018217 0.184395445699365
22.0133110749303 0.186757598692324
22.5271064264598 0.189126325559694
23.0528938705171 0.191501198143169
23.5909533050864 0.193881788101981
24.1415711610302 0.196267667214902
24.7050405545683 0.198658407675897
25.281661443315 0.201053582383116
25.8717407859592 0.203452765220948
26.4755927056707 0.205855531334851
27.0935386573205 0.20826145739874
27.7259075986048 0.210670121874711
28.3730361651621 0.213081105264928
29.0352688497781 0.215493990355512
29.7129581857733 0.217908362452298
30.4064649346707 0.220323809608359
31.1161582782436 0.222739922843201
31.8424160150465 0.225156296353583
32.5856247615322 0.227572527715912
33.3461801578636 0.229988218080192
34.1244870785289 0.232402972355537
34.9209598478727 0.234816399387259
35.7360224606579 0.237228112125569
36.5701088077749 0.239637727785949
37.4236629072198 0.242044868001263
38.2971391404628 0.244449158965678
39.191002494334 0.246850231570522
40.105728808555 0.249247721532149
41.0418050290471 0.251641269511971
41.999729467153 0.254030521228765
42.980012064908 0.256415127563399
43.9831746665022 0.258794744656142
45.0097512960805 0.2611690339967
46.060288442024 0.263537662507146
47.1353453478691 0.265900302617935
48.2354943100147 0.268256632337144
49.3613209823792 0.270606335313155
50.5134246881676 0.272949100890944
51.692418738916 0.275284624162174
52.8989307609815 0.277612606009272
54.1336030296538 0.279932753143699
55.397092811064 0.282244778138591
56.6900727120743 0.284548399455973
58.0132310383338 0.286843341468742
59.3672721606912 0.289129334477604
60.7529168901607 0.291406114723169
62.1709028616383 0.293673424393387
63.6219849265749 0.295931011626525
65.1069355548146 0.298178630509857
66.6265452458115 0.300416041074275
68.1816229494448 0.302643009284976
69.7729964966553 0.304859307028443
71.401513040134 0.307064712095851
73.068039505295 0.309259008163108
74.7734630517759 0.311441984767692
76.5186915457082 0.313613437282432
78.3046540430119 0.31577316688642
80.1323012839689 0.317920980533198
82.0026061993413 0.320056690916373
83.9165644283016 0.322180116432819
85.8751948484517 0.324291081143599
87.8795401182132 0.326389414732755
89.9306672318762 0.328474952464102
92.0296680876042 0.330547535136156
94.1776600686952 0.33260700903532
96.3757866384109 0.334653225887456
98.6252179486878 0.336686042807959
100.927151463057 0.338705322250439
103.282812594103 0.340710931954139
105.693455355799 0.34270274489017
108.160363031071 0.344680639206677
110.684848854941 0.346644498173034
113.268256713615 0.348594210123148
115.911961859889 0.350529668397971
118.617371645248 0.352450771287292
121.385926269063 0.3543574219709
124.219099545262 0.356249528459185
127.118399686903 0.358127003533245
130.085370109057 0.359989764684582
133.121590250431 0.361837734054431
136.228676414165 0.363670838372792
139.408282628258 0.365489008897232
142.662101526074 0.367292181351479
145.991865247398 0.3690802958639
149.399346360526 0.370853296905869
152.886358805873 0.372611133230099
156.45475886161 0.374353757808958
160.106446131832 0.376081127772824
163.843364557798 0.377793204348495
167.66750345277 0.379489952797701
171.580898561 0.381171342355747
175.585633141447 0.382837346170299
179.683839076772 0.384487941240356
183.877698008233 0.38612310835542
188.169442497056 0.38774283203489
192.56135721292 0.389347100467689
197.05578015018 0.39093590545215
201.655103872475 0.392509242336175
206.361776786386 0.39406710995767
211.178304444824 0.395609510585278
216.107250880838 0.39713644985941
221.151239972549 0.398647936733593
226.312956839953 0.400143983416128
231.595149274315 0.401624605312078
237.000629200933 0.403089820965574
242.532274176035 0.404539652002456
248.193028918626 0.405974123073243
253.985906878072 0.407393261796438
259.913991838293 0.408797098702159
265.980439559376 0.410185667176105
272.188479457518 0.411559003403856
278.541416324177 0.412917146315495
285.042632085343 0.414260137530555
291.69558760188 0.415588021303297
298.503824511873 0.416900844468294
305.470967115997 0.418198656386343
312.60072430687 0.419481508890682
319.896891543454 0.420749456233508
327.363352871524 0.422002555032812
335.004082991313 0.4232408642195
342.823149373397 0.424464444984812
350.824714423979 0.425673360728026
359.013037700707 0.426867677004458
367.392478180207 0.428047461473726
375.967496578547 0.429212783848303
384.74265772585 0.430363715842337
393.722632996348 0.431500331120739
402.912202795135 0.43262270524854
412.316259102975 0.43373091564051
421.939808080502 0.434825041511027
431.787972733202 0.435905163824221
441.865995638594 0.436971365244357
452.17924173707 0.438023730086485
462.733201187869 0.439062344267334
473.533492291712 0.440087295256463
484.58586448165 0.441098672027669
495.896201383721 0.442096565010641
507.470523949047 0.443081066042875
519.314993659021 0.444052268321845
531.435915805324 0.445010266357433
543.83974284648 0.445955155924616
556.533077842765 0.446887034016429
569.522677971283 0.447805998797189
582.815458123085 0.448712149555994
596.418494584246 0.449605586660505
610.339028802862 0.450486411511013
624.584471243962 0.451354726494789
639.1624053344 0.452210634940744
654.080591499825 0.453054241074386
669.346971295866 0.453885649973087
684.969671635745 0.454704967521685
700.957009116562 0.455512300368398
717.317494446561 0.456307755881088
734.059836975721 0.45709144210387
751.192949332097 0.457863467714083
768.725952166373 0.458623941979616
786.668179007158 0.459372974716633
805.029181229598 0.460110676247669
823.81873313996 0.460837157360137
843.046837178897 0.461552529265244
862.723729246145 0.462256903557332
882.859884149515 0.462950392173651
903.466021181052 0.463633107354581
924.553109823357 0.464305161604314
946.132375589077 0.464966667652004
968.215305996709 0.465617738413402
990.813656685867 0.466258486952984
1013.93945767529 0.466889026446586
1037.60501976691 0.467509470144562
1061.82294109938 0.468119931335463
1086.6061138546 0.468720523310266
1111.9677311207 0.469311359327156
1137.92129391532 0.469892552576864
1164.48061837269 0.470464216148593
1191.65984309856 0.47102646299652
1219.47343669674 0.4715794059069
1247.93620547131 0.472123157465774
1277.06330130864 0.472657830027296
1306.87022974335 0.473183535682684
1337.3728582125 0.473700386229804
1368.58742450252 0.474208493143402
1400.53054539322 0.474707967545979
1433.21922550357 0.475198920179337
1466.67086634397 0.475681461376777
1500.90327557974 0.476155701035978
1535.9346765109 0.476621748592554
1571.78371777316 0.477079712994287
1608.46948326536 0.477529702676059
1646.01150230855 0.477971825535473
1684.42976004232 0.478406188909163
1723.74470806362 0.478832899549819
1763.97727531404 0.479252063603901
1805.14887922111 0.479663786590061
1847.28143709963 0.480068173378279
1890.39737781922 0.480465328169692
1934.51965374404 0.48085535447714
1979.67175295133 0.481238355106415
2025.87771173502 0.481614432138213
2073.16212740123 0.481983686910795
2121.55017136245 0.482346220003345
2171.06760253726 0.482702131220027
2221.74078106288 0.483051519574741
2273.59668232772 0.483394483276569
2326.66291133146 0.483731119715905
2380.96771738036 0.484061525451268
2436.54000912547 0.484385796196799
2493.40936995188 0.484704026810417
2551.60607372718 0.485016311282651
2611.16110091746 0.485322742726116
2672.10615507943 0.485623413365652
2734.47367973758 0.485918414529095
2798.29687565512 0.486207836638689
2863.60971850812 0.486491769203115
2930.44697697214 0.486770300810144
2998.84423123103 0.487043519119892
3068.83789191764 0.487311510858668
3140.46521949675 0.487574361813418
3213.76434410028 0.487832156826737
3288.77428582551 0.488084979792449
3365.53497550709 0.48833291365174
3444.08727597382 0.488576040389838
3524.47300380158 0.488814441033217
3606.73495157403 0.489048195647334
3690.91691066278 0.489277383334855
3777.06369453936 0.489502082234398
3865.22116263126 0.489722369519744
3955.4362447347 0.489938321399528
4047.75696599732 0.490150013117387
4142.23247248389 0.490357518952552
4238.91305733878 0.490560912220882
4337.85018755899 0.490760265276312
4439.09653139217 0.49095564951271
4542.70598637404 0.491147135366142
4648.73370802026 0.491334792317502
4757.23613918789 0.491518688895532
4868.27104012228 0.491698892680186
4981.89751920516 0.49187547030635
5098.17606442043 0.492048487467891
5217.16857555435 0.492218008922027
5338.93839714735 0.492384098494008
5463.55035221488 0.492546819082094
5591.07077675529 0.492706232662813
5721.56755506325 0.492862400296498
5855.11015586724 0.49301538213308
5991.76966931062 0.493165237418136
6131.61884479577 0.493312024499162
6274.73212971157 0.493455800832092
6421.18570906476 0.493596622988011
6571.05754603627 0.493734546660087
6724.42742348425 0.49386962667069
6881.37698641566 0.494001916978695
7041.98978544929 0.494131470686955
7206.35132129305 0.494258340049937
7374.54909025955 0.494382576481514
7546.67263084389 0.49450423056289
7722.81357138864 0.494623352050666
7903.06567886135 0.494739989885024
8087.52490877045 0.494854192198029
8276.28945624634 0.494966006322026
8469.45980831458 0.495075478798151
8667.13879738923 0.495182655384909
8869.4316560147 0.495287581066848
9076.44607288536 0.495390300063296
9288.2922501725 0.495490855837161
9505.0829621895 0.495589291103794
9726.93361542617 0.495685647839893
9953.96230998423 0.495779967292455
10186.2899024469 0.495872289987756
10424.0400702156 0.495962655740365
10667.3393773486 0.496051103662177
10916.3173419361 0.496137672171461
11171.1065050482 0.49622239900192
11431.8425012915 0.496305321211753
11698.6641310131 0.496386475192712
11971.7134341897 0.496465896679158
12251.1357660412 0.496543620757101
12537.0798744092 0.496619681873224
12829.6979789415 0.496694113843882
13129.1458521247 0.496766949864079
13435.5829022083 0.496838222516411
13749.1722580642 0.496907963779975
14070.0808560268 0.496976205039239
14398.4795287601 0.497042977092873
14734.5430961983 0.497108310162522
15078.4504586105 0.497172233901547
15430.3846918356 0.497234777403696
15790.5331447418 0.497295969211731
16159.0875389592 0.49735583732599
16536.2440709418 0.49741440921289
16922.2035164104 0.497471711813364
17317.1713372335 0.497527771551233
17721.3577908036 0.497582614341509
18134.978041965 0.497636265598623
18558.2522775552 0.497688750244589
18991.4058236194 0.497740092717082
19434.6692653602 0.497790316977448
19888.2785698881 0.497839446518631
20352.4752118358 0.497887504373022
20827.5063019051 0.497934513120228
21313.6247184144 0.497980494894753
21811.0892419152 0.498025471393606
22320.1646929523 0.498069463883813
22841.1220730383 0.498112493209852
23374.2387089181 0.498154579800996
23919.7984002024 0.498195743678574
24478.0915704444 0.498236004463137
25049.415421745 0.498275381381549
25634.0740929652 0.49831389327397
26232.3788216312 0.498351558600772
26844.6481096197 0.498388395449345
27471.2078927081 0.498424421540827
28112.3917140846 0.498459654236743
28768.5409019059 0.498494110545545
29440.0047510003 0.498527807129071
30127.1407088116 0.498560760308913
30830.3145656828 0.498592986072692
31549.9006495809 0.498624500080247
32286.2820253673 0.498655317669728
33039.8506987185 0.498685453863611
33811.0078248068 0.498714923374615
34600.1639218511 0.498743740611531
35407.7390896527 0.49877191968497
36234.1632332315 0.498799474413015
37079.8762916817 0.498826418326793
37945.3284723694 0.498852764675955
38830.980490596 0.498878526434071
39737.303814856 0.498903716303944
40664.7809178185 0.498928346722835
41613.9055331671 0.498952429867603
42585.1829184341 0.498975977659763
43579.1301239703 0.498999001770465
44596.276268191 0.499021513625383
45637.1628192475 0.499043524409533
46702.3438832733 0.499065045071997
47792.3864993559 0.499086086330581
48907.8709413959 0.499106658676386
50049.3910270095 0.499126772378303
51217.5544336424 0.499146437487427
52412.9830220606 0.499165663841407
53636.3131673923 0.499184461068702
54888.1960978967 0.499202838592778
56169.298241638 0.499220805636222
57480.3015812535 0.499238371224788
58821.9040169996 0.499255544191368
60194.8197382727 0.499272333179894
61599.7796038017 0.499288746649169
63037.5315307128 0.499304792876627
64508.8408926769 0.499320479962026
66014.4909273488 0.499335815831078
67555.2831533164 0.499350808239001
69132.0377967813 0.49936546477402
70745.5942281988 0.49937979286079
72396.8114091088 0.499393799763765
74086.5683493957 0.499407492590498
75815.764575221 0.49942087829488
77585.3206078783 0.499433963680323
79396.1784538228 0.499446755402877
81249.3021061405 0.499459259974289
83145.6780577206 0.499471483765007
85086.3158264058 0.499483433007122
87072.2484923992 0.499495113797256
89104.533248215 0.499506532099394
91184.2519614656 0.499517693747664
93312.5117507824 0.499528604449056
95490.4455751808 0.499539269786096
97719.21283718 0.499549695219463
100000 0.499559886090556
};
\addplot [line width = 2pt, KITblue, dashdotted]
table {%
1 0
88 0
};
\addplot [line width = 2pt, KITblue, dashed]
table {%
88 0
112.16189511413755 0.10770990936605547
142.95784904085005 0.19221696958082346
182.20935533934693 0.2585195342025422
232.23802957257348 0.31053921237197646
296.00292630037455 0.35135282089964825
377.2755587861412 0.3833743692764857
480.8629733374863 0.4084978415064633
612.8920724971756 0.4282092199027378
781.1720039967576 0.4436743767379269
995.6560497543754 0.4558080322910159
1269.030334344395 0.4653278579643006
1617.4641733794876 0.4727969245166849
2061.566442789378 0.4786570061062566
2627.604535533728 0.4832547099820474
3349.058003590527 0.4868619773223314
4268.5987787486765 0.4896921677860531
5440.6150966630075 0.49191267913310255
6934.428407139887 0.49365484832827805
8838.393541796817 0.4950217197512281
11265.124652414615 0.4960941399800162
14358.15601945242 0.49693553963751413
18300.43169870733 0.4975956851333126
23325.126144702834 0.4981136222060693
29729.4358200703 0.4985199853684981
37892.157525604714 0.49883880985213713
48296.09315949152 0.49908895322330327
61556.60608383538 0.49928521075479576
78458.01812678891 0.4994391905244293
100000.0 0.49956
};
\addplot [line width = 2pt, KITblue, dotted]
table {%
88.0 0.0
109.85411419875584 0.03390343724174475
138.94954943731375 0.06901486151814847
175.7510624854793 0.10148131431606287
222.29964825261956 0.1315020434377681
281.1768697974231 0.15926128719034438
355.64803062231283 0.18492940506286484
449.8432668969444 0.20866392322878063
568.9866029018299 0.23061050128977942
719.6856730011522 0.25090382619406126
910.2981779915218 0.26966843881504166
1151.3953993264481 0.28701949826322326
1456.3484775012444 0.3030634886218474
1842.0699693267163 0.3178988724435889
2329.951810515372 0.33161669501882396
2947.0517025518097 0.34430114312389093
3727.593720314942 0.3560300616783999
4714.866363457394 0.36687543148233726
5963.623316594642 0.3769038109648548
7543.120063354622 0.38617674465577134
9540.954763499944 0.3947511408865916
12067.92640639329 0.40267962103900795
15264.179671752334 0.4100108424842347
19306.977288832495 0.4167897971950656
24420.5309454865 0.42305808786324844
30888.43596477485 0.4288541832167181
39069.39937054621 0.43421365410357954
49417.13361323838 0.43916939179169784
62505.51925273976 0.44375180982360507
79060.43210907701 0.44798903066551793
100000.0 0.45190705829593664
};
\end{axis}

\begin{axis}[
    height = 4.5cm,
    width = 7.8cm,
    y axis line style={line width=1.5pt},
    xlabel={},
    ylabel= \textcolor{KITred}{\shortstack{$q_\mathrm{eff} \ / \  \frac{\mathrm{mol}}{\mathrm{m^3}}$ \\ \tiny{(log scale)}}},
    axis y line*=left,
    y axis line style={KITred},
    ytick style={KITred},
    axis x line=none, 
    ymin=0.62, ymax=200,
    yticklabel pos=right,
    ytick = {1e0,88},
    yticklabels = {$\textcolor{KITred} {10^0}$,$\textcolor{KITred}{q_\mathrm{lim}}$},
    ylabel style = {xshift = -0.05cm, yshift = -0.75 cm},
    xmin=5e-1, xmax=2e5,
    xmode = log,
    ymode = log,
    legend entries={  
        IOM,
        M-$L^\mathrm{vol}$,
        M-$\Tilde{L}$
    },
    legend style={
        nodes={anchor=west},
        anchor = south east,
        at={(0.97,0.07)},
        fill opacity=0.9,
        text opacity=1,
        draw opacity=1,
        }
]
\addlegendimage{no markers,line width = 2pt, KITgrey}
\addlegendimage{no markers,line width = 2pt, KITgrey, dotted}
\addlegendimage{no markers,line width = 2pt, KITgrey, dashed}

\addplot [line width = 2pt, KITred]
table {%
1 0.973166180060541
1.02334021219164 0.9952236028829
1.04722518988843 1.01776493993671
1.07166764803286 1.04080006092918
1.09668059833687 1.06433900265619
1.12227735620851 1.08839197027634
1.14847154784029 1.11296933851927
1.17527711746295 1.13808165282419
1.20270833476851 1.16373963040453
1.23077980250667 1.18995416123453
1.25950646425836 1.21673630895336
1.28890361239089 1.24409731168256
1.31898689619867 1.27204858275231
1.34977233023394 1.30060171133197
1.38127630283201 1.32976846296046
1.4135155848354 1.35956077997189
1.44650733852165 1.38999078181162
1.48026912673951 1.42107076523843
1.51481892225835 1.45281320440777
1.55017511733577 1.48523075083172
1.58635653350859 1.51833623321073
1.62338243161228 1.55214265713267
1.66127252203429 1.58666320463447
1.70004697520672 1.62191123362178
1.7397264323438 1.65790027714213
1.78033201643011 1.69464404250714
1.82188534346517 1.7321564102594
1.86440853397049 1.77045143297978
1.90792422476527 1.80954333393102
1.95245558101686 1.84944650553359
1.99802630857255 1.89017550766999
2.04466066657912 1.93174506581386
2.09238348039698 1.9741700689804
2.14122015481573 2.01746556749479
2.19119668757815 2.06164677057571
2.24233968321985 2.10672904373111
2.29467636723194 2.15272790596378
2.34823460055428 2.19965902678447
2.40304289440697 2.24753822303071
2.45913042546804 2.29638145548983
2.51652705140539 2.34620482532488
2.5752633267712 2.39702457030267
2.63537051926739 2.44885706082351
2.69688062639069 2.50171879575258
2.75982639246618 2.55562639805333
2.82424132607844 2.61059661022372
2.89015971790951 2.6666462895366
2.95761665899325 2.72379240308599
3.02664805939569 2.78205202264135
3.09729066733141 2.84144231931273
3.16958208872612 2.90198055802983
3.24356080723581 2.96368409183874
3.31926620473319 3.02657035602054
3.39673858227221 3.09065686203644
3.47601918154198 3.15596119130466
3.55715020682139 3.2225009888148
3.64017484744614 3.29029395658578
3.72513730080021 3.35935784697419
3.81208279584389 3.42971045584001
3.90105761719099 3.50136961557751
3.99210912974805 3.57435318801922
4.08528580392856 3.64867905722157
4.18063724145576 3.72436512214099
4.27821420176762 3.80142928920986
4.37806862903817 3.87988946482172
4.48025367982949 3.9597635477359
4.5848237513891 4.0410694214116
4.6918345106078 4.12382494628194
4.80134292365345 4.20804795197874
4.91340728629636 4.29375622951874
5.02808725494248 4.3809675234623
5.14544387839093 4.46969952405569
5.26553963033276 4.55996985936797
5.38843844260822 4.65179608743367
5.5142057392403 4.74519568841225
5.64290847126254 4.84018605677522
5.77461515235982 4.93678449353195
5.90939589534097 5.03500819850445
6.04732244946265 5.13487426266173
6.18846823862439 5.23639966052368
6.33290840045512 5.33960124264419
6.48071982631197 5.44449572818292
6.63198120221267 5.55109969757447
6.7867730507233 5.65942958530361
6.9451777738237 5.76950167279422
7.10727969677342 5.88133208141961
7.27316511300146 5.99493676564071
7.44292233004376 6.11033150627863
7.61664171655289 6.22753190392677
7.79441575040495 6.34655337250778
7.97633906792928 6.46741113297913
8.16250851428723 6.5901202071911
8.35302319502678 6.71469541189981
8.54798452884041 6.84115135293746
8.74749630155442 6.9695024195409
8.9516647213783 7.09976277883941
9.16059847544371 7.23194637050134
9.374408787663 7.36606690153888
9.59320947793824 7.50213784126947
9.81711702275219 7.64017241643147
10.0462506171734 7.78018360645135
10.2807322383087 7.92218413885884
10.5206867102362 8.06618648484578
10.7662417704549 8.21220285496419
11.0175281378839 8.36024519495794
11.2746795824495 8.5103251817225
11.5378329962966 8.66245421938641
11.8071284666619 8.81664343550772
12.0827093504478 8.97290367737822
12.3647223505371 9.13124550842829
12.6533175938894 9.29167920472422
12.948648711459 9.45421475155019
13.2508729199795 9.61886184006676
13.5601511056563 9.78562986403713
13.8766479098131 9.954527916613
14.2005318165368 10.1255647871711
14.531975242369 10.2987489581922
14.8711546280895 10.474088602173
15.2182505326439 10.6515915785641
15.5734477292613 10.8312654307233
15.9369353038177 11.0131173828778
16.3089067554933 11.1971543370867
16.6895600997802 11.3833828701945
17.0790979738943 11.5718092307703
17.4777277446468 11.7624393360232
17.8856616188346 11.9552787686881
18.3031167562061 12.1503327738747
18.7303153850644 12.3476062558733
19.1674849205681 12.5471037749112
19.6148580857943 12.7488295438557
20.0726730356257 12.9527874248556
20.5411734835306 13.1589809259202
21.0206088313016 13.3674131974283
21.5112343018217 13.5780870285659
22.0133110749303 13.7910048436881
22.5271064264598 14.0061686986028
23.0528938705171 14.2235802767744
23.5909533050864 14.4432408854454
24.1415711610302 14.6651514516743
24.7050405545683 14.8893125182903
25.281661443315 15.1157242397638
25.8717407859592 15.3443863779933
26.4755927056707 15.5752982980088
27.0935386573205 15.8084589635952
27.7259075986048 16.0438669328347
28.3730361651621 16.2815203535731
29.0352688497781 16.5214169588106
29.7129581857733 16.7635540620224
30.4064649346707 17.0079285524114
31.1161582782436 17.254536890098
31.8424160150465 17.5033751012507
32.5856247615322 17.754438773164
33.3461801578636 18.0077230492874
34.1244870785289 18.2632226242124
34.9209598478727 18.5209317386237
35.7360224606579 18.7808441742203
36.5701088077749 19.0429532486147
37.4236629072198 19.3072518102163
38.2971391404628 19.5737322331074
39.191002494334 19.8423864119195
40.105728808555 20.1132057567178
41.0418050290471 20.3861811879027
41.999729467153 20.6613031311371
42.980012064908 20.9385615123084
43.9831746665022 21.2179457525343
45.0097512960805 21.4994447632224
46.060288442024 21.7830469411921
47.1353453478691 22.0687401638705
48.2354943100147 22.3565117845707
49.3613209823792 22.6463486278632
50.5134246881676 22.9382369850521
51.692418738916 23.2321626097635
52.8989307609815 23.5281107136612
54.1336030296538 23.8260659622956
55.397092811064 24.1260124711006
56.6900727120743 24.4279338015474
58.0132310383338 24.7318129574662
59.3672721606912 25.0376323815484
60.7529168901607 25.3453739520381
62.1709028616383 25.6550189796264
63.6219849265749 25.9665482045572
65.1069355548146 26.2799417939583
66.6265452458115 26.5951793394061
68.1816229494448 26.9122398547376
69.7729964966553 27.2311017741186
71.401513040134 27.5517429503802
73.068039505295 27.8741406536343
74.7734630517759 28.1982715701783
76.5186915457082 28.5241118017007
78.3046540430119 28.8516368647971
80.1323012839689 29.1808216908068
82.0026061993413 29.511640625982
83.9165644283016 29.8440674319968
85.8751948484517 30.1780752868085
87.8795401182132 30.5136367858787
89.9306672318762 30.8507239437652
92.0296680876042 31.189308196092
94.1776600686952 31.5293604019076
96.3757866384109 31.8708508464401
98.6252179486878 32.2137492442555
100.927151463057 32.5580247428299
103.282812594103 32.9036459265399
105.693455355799 33.2505808210811
108.160363031071 33.5987968983195
110.684848854941 33.9482610815829
113.268256713615 34.2989397513979
115.911961859889 34.6507987516786
118.617371645248 35.0038033963699
121.385926269063 35.357918476553
124.219099545262 35.7131082680137
127.118399686903 36.069336539279
130.085370109057 36.4265665601245
133.121590250431 36.7847611105542
136.228676414165 37.1438824902544
139.408282628258 37.5038925285236
142.662101526074 37.8647525946781
145.991865247398 38.2264236089334
149.399346360526 38.5888660537601
152.886358805873 38.9520399857133
156.45475886161 39.3159050477336
160.106446131832 39.6804204819153
163.843364557798 40.0455451427399
167.66750345277 40.4112375107698
171.580898561 40.7774557067967
175.585633141447 41.1441575064405
179.683839076772 41.5113003551891
183.877698008233 41.8788413838765
188.169442497056 42.246737424586
192.56135721292 42.6149450269734
197.05578015018 42.9834204749991
201.655103872475 43.3521198040591
206.361776786386 43.720998818504
211.178304444824 44.0900131095327
216.107250880838 44.4591180734521
221.151239972549 44.8282689302844
226.312956839953 45.1974207427107
231.595149274315 45.5665284353378
237.000629200933 45.935546814268
242.532274176035 46.3044305869602
248.193028918626 46.6731343823633
253.985906878072 47.0416127713037
259.913991838293 47.4098202871116
265.980439559376 47.7777114464635
272.188479457518 48.145240770424
278.541416324177 48.5123628056663
285.042632085343 48.8790321458519
291.69558760188 49.2452034531445
298.503824511873 49.6108314798423
305.470967115997 49.9758710901033
312.60072430687 50.3402772817384
319.896891543454 50.7040052080584
327.363352871524 51.0670101997414
335.004082991313 51.4292477867039
342.823149373397 51.7906737199487
350.824714423979 52.1512439933685
359.013037700707 52.5109148654775
367.392478180207 52.8696428810492
375.967496578547 53.2273848926359
384.74265772585 53.5840980819438
393.722632996348 53.9397399810415
402.912202795135 54.2942684933753
412.316259102975 54.647641914569
421.939808080502 54.9998189529841
431.787972733202 55.350758750013
441.865995638594 55.7004209000868
452.17924173707 56.04876547037
462.733201187869 56.3957530201217
473.533492291712 56.7413446196982
484.58586448165 57.0855018691783
495.896201383721 57.428186916585
507.470523949047 57.7693624756871
519.314993659021 58.1089918433553
531.435915805324 58.4470389164573
543.83974284648 58.7834682082701
556.533077842765 59.1182448643895
569.522677971283 59.4513346781228
582.815458123085 59.7827041053437
596.418494584246 60.1123202787957
610.339028802862 60.4401510218264
624.584471243962 60.7661648615403
639.1624053344 61.090331041355
654.080591499825 61.4126195329482
669.346971295866 61.7330010475829
684.969671635745 62.0514470468035
700.957009116562 62.3679297524878
717.317494446561 62.6824221562516
734.059836975721 62.9948980281923
751.192949332097 63.3053319249704
768.725952166373 63.6136991972209
786.668179007158 63.9199759962875
805.029181229598 64.2241392802825
823.81873313996 64.5261668194633
843.046837178897 64.8260372009286
862.723729246145 65.1237298326311
882.859884149515 65.4192249467114
903.466021181052 65.7125036021504
924.553109823357 66.0035476867471
946.132375589077 66.2923399184216
968.215305996709 66.5788638458532
990.813656685867 66.8631038484558
1013.93945767529 67.1450451356981
1037.60501976691 67.4246737457779
1061.82294109938 67.7019765436579
1086.6061138546 67.9769412184736
1111.9677311207 68.2495562803215
1137.92129391532 68.5198110564438
1164.48061837269 68.7876956868164
1191.65984309856 69.0532011191544
1219.47343669674 69.3163191033515
1247.93620547131 69.5770421853664
1277.06330130864 69.8353637005657
1306.87022974335 70.0912777665494
1337.3728582125 70.3447792754625
1368.58742450252 70.5958638858218
1400.53054539322 70.844528013866
1433.21922550357 71.0907688244455
1466.67086634397 71.3345842214831
1500.90327557974 71.5759728380047
1535.9346765109 71.814934025773
1571.78371777316 72.0514678445365
1608.46948326536 72.2855750509162
1646.01150230855 72.5172570869434
1684.42976004232 72.7465160682748
1723.74470806362 72.9733547721004
1763.97727531404 73.197776624761
1805.14887922111 73.4197856891046
1847.28143709963 73.6393866515862
1890.39737781922 73.8565848091454
1934.51965374404 74.0713860558697
1979.67175295133 74.2837968694675
2025.87771173502 74.4938242975713
2073.16212740123 74.7014759438854
2121.55017136245 74.9067599541985
2171.06760253726 75.1096850022819
2221.74078106288 75.3102602756863
2273.59668232772 75.5084954614581
2326.66291133146 75.704400731793
2380.96771738036 75.8979867296403
2436.54000912547 76.0892645542787
2493.40936995188 76.2782457468759
2551.60607372718 76.4649422760496
2611.16110091746 76.6493665234461
2672.10615507943 76.8315312693476
2734.47367973758 77.0114496783284
2798.29687565512 77.1891352849637
2863.60971850812 77.3646019796214
2930.44697697214 77.5378639943267
2998.84423123103 77.7089358887362
3068.83789191764 77.8778325362122
3140.46521949675 78.0445691100222
3213.76434410028 78.2091610696701
3288.77428582551 78.3716241473681
3365.53497550709 78.5319743346575
3444.08727597382 78.6902278691972
3524.47300380158 78.8464012217115
3606.73495157403 79.0005110831246
3690.91691066278 79.1525743518773
3777.06369453936 79.3026081214306
3865.22116263126 79.4506296679786
3955.4362447347 79.5966564383549
4047.75696599732 79.74070603816
4142.23247248389 79.8827962200985
4238.91305733878 80.0229448725385
4337.85018755899 80.1611700082949
4439.09653139217 80.2974897536449
4542.70598637404 80.4319223375724
4648.73370802026 80.5644860812466
4757.23613918789 80.6951993877439
4868.27104012228 80.8240807319987
4981.89751920516 80.9511486510096
5098.17606442043 81.0764217342784
5217.16857555435 81.1999186144936
5338.93839714735 81.3216579584643
5463.55035221488 81.441658458294
5591.07077675529 81.559938822799
5721.56755506325 81.676517769175
5855.11015586724 81.7914140149043
5991.76966931062 81.9046462699083
6131.61884479577 82.016233228938
6274.73212971157 82.1261935642102
6421.18570906476 82.2345459182809
6571.05754603627 82.3413088971516
6724.42742348425 82.4465010636163
6881.37698641566 82.5501409308368
7041.98978544929 82.6522469561498
7206.35132129305 82.7528375351036
7374.54909025955 82.8519309957109
7546.67263084389 82.9495455929377
7722.81357138864 83.0456995033938
7903.06567886135 83.1404108202539
8087.52490877045 83.2336975483777
8276.28945624634 83.325577599647
8469.45980831458 83.4160687885053
8667.13879738923 83.5051888276963
8869.4316560147 83.5929553241988
9076.44607288536 83.6793857753621
9288.2922501725 83.7644975652251
9505.0829621895 83.8483079610188
9726.93361542617 83.9308341098562
9953.96230998423 84.0120930355975
10186.2899024469 84.0921016358905
10424.0400702156 84.1708766793738
10667.3393773486 84.2484348030555
10916.3173419361 84.324792509846
11171.1065050482 84.3999661662477
11431.8425012915 84.4739720002103
11698.6641310131 84.5468260991035
11971.7134341897 84.6185444078883
12251.1357660412 84.6891427273629
12537.0798744092 84.7586367126045
12829.6979789415 84.8270418715094
13129.1458521247 84.8943735634753
13435.5829022083 84.9606469981993
13749.1722580642 85.0258772346076
14070.0808560268 85.090079179898
14398.4795287601 85.1532675886954
14734.5430961983 85.2154570623104
15078.4504586105 85.2766620481315
15430.3846918356 85.3368968390636
15790.5331447418 85.3961755731391
16159.0875389592 85.4545122331562
16536.2440709418 85.5119206464631
16922.2035164104 85.5684144847744
17317.1713372335 85.6240072641466
17721.3577908036 85.6787123449499
18134.978041965 85.7325429320018
18558.2522775552 85.7855120746967
18991.4058236194 85.8376326672795
19434.6692653602 85.888917449141
19888.2785698881 85.9393790052035
20352.4752118358 85.9890297663472
20827.5063019051 86.0378820099275
21313.6247184144 86.0859478603205
21811.0892419152 86.1332392895538
22320.1646929523 86.1797681179624
22841.1220730383 86.2255460149048
23374.2387089181 86.270584499535
23919.7984002024 86.3148949416173
24478.0915704444 86.35848856236
25049.415421745 86.4013764353336
25634.0740929652 86.4435694873845
26232.3788216312 86.4850784996254
26844.6481096197 86.5259141084147
27471.2078927081 86.5660868064039
28112.3917140846 86.6056069436079
28768.5409019059 86.6444847284947
29440.0047510003 86.6827302291021
30127.1407088116 86.7203533741828
30830.3145656828 86.7573639543923
31549.9006495809 86.7937716234552
32286.2820253673 86.8295858993844
33039.8506987185 86.8648161657166
33811.0078248068 86.8994716727421
34600.1639218511 86.9335615387809
35407.7390896527 86.9670947514548
36234.1632332315 87.0000801689608
37079.8762916817 87.0325265213944
37945.3284723694 87.0644424120219
38830.980490596 87.0958363186429
39737.303814856 87.1267165948808
40664.7809178185 87.157091471533
41613.9055331671 87.1869690578982
42585.1829184341 87.2163573431318
43579.1301239703 87.245264197576
44596.276268191 87.2736973741215
45637.1628192475 87.3016645095536
46702.3438832733 87.3291731258931
47792.3864993559 87.356230631788
48907.8709413959 87.3828443238113
50049.3910270095 87.4090213878422
51217.5544336424 87.43476890041
52412.9830220606 87.4600938300252
53636.3131673923 87.4850030385689
54888.1960978967 87.5095032825433
56169.298241638 87.533601214505
57480.3015812535 87.5573033843144
58821.9040169996 87.5806162404953
60194.8197382727 87.6035461315587
61599.7796038017 87.6260993072825
63037.5315307128 87.6482819200465
64508.8408926769 87.6701000260869
66014.4909273488 87.6915595868461
67555.2831533164 87.7126664701723
69132.0377967813 87.7334264516699
70745.5942281988 87.7538452158969
72396.8114091088 87.7739283576654
74086.5683493957 87.7936813832453
75815.764575221 87.813109711621
77585.3206078783 87.83221867573
79396.1784538228 87.8510135236213
81249.3021061405 87.8694994197099
83145.6780577206 87.8876814459585
85086.3158264058 87.9055646030281
87072.2484923992 87.9231538115007
89104.533248215 87.940453912959
91184.2519614656 87.957469671248
93312.5117507824 87.9742057734775
95490.4455751808 87.990666831277
97719.21283718 88.0068573817857
100000 88.0227818888635
};
\addplot [line width = 2pt, KITred, -stealth, dotted]
table {%
88 88
109.85411419875584 102.40525006578588
138.94954943731375 119.77038161246308
175.7510624854793 140.08016485853744
222.29964825261956 163.83393225118647
281.1768697974231 191.61568937324415
};
\node[anchor=north east] (lbl) at (axis cs: 200, 210) {\textcolor{KITred}{\scriptsize{to $\infty$ with ${c_\mathrm{X}}^{2/3}$ }}};
\addplot [line width = 2pt, KITred, dashdotted]
table {%
1 1
88 88
};
\addplot [line width = 2pt, KITred, dashed]
table {%
88 88
100000 88
};
\end{axis}

\end{tikzpicture}

%% file: 5_Results/Figures/Interaction_deviation.tex
\begin{tikzpicture}

\begin{axis}[
tick align=inside,
xlabel={$c_\mathrm{s}^\mathrm{b} \ / \ \frac{\mathrm{mol}}{\mathrm{m^3}}$},
xmin=0.8, xmax=2000,
xmode=log,
xtick={1,10,100,1000,10000},
xticklabels={$10^{0}$,$10^{1}$,,$10^{3}$,$10^{4}$},
xlabel style = {yshift = 0.4cm},
ylabel={$q_\mathrm{eff} \ / \ \frac{\mathrm{mol}}{\mathrm{m^3}}$},
ylabel style = {yshift = -0.5cm, xshift = 0.1cm},
ymin= -460, ymax=460,
ytick={-400,0,400},
yticklabels={$-q_\mathrm{lim}$,$0$, $q_\mathrm{lim}$},
height = 4.5cm,
width = 8cm,
legend style={nodes={anchor=west}, anchor = south west, at={(0.03,0.07)}}
]

\addplot [line width = 2pt, KITblack]
table {%
1.0 327.99909128441374
1.1942771623577337 327.9852879977271
1.4262979405292406 327.9671530122779
1.7033950570919412 327.9433413890114
2.034325815157953 327.9120933690758
2.4295488618379237 327.87110699032553
2.901554720525257 327.81737171610905
3.465260538054591 327.74695118544884
4.138481522218071 327.6546995830751
4.942493968824512 327.53389143965177
5.90270767205795 327.3757385782988
7.049468968812594 327.16876002786137
8.419019796202404 326.8979605086523
10.054643072042195 326.5437599335753
12.008030596598394 326.0805993611668
14.340916706410374 325.47512748195385
17.1270293097404 324.68384378797145
20.4544199636545 323.6500401702706
24.428246631866678 322.2998411748428
29.174097068880588 320.53708730105836
34.84195786177179 318.2367483294276
41.61095456614455 315.2364792669742
49.6950127422517 311.3258520125941
59.34961880114778 306.23273990882916
70.87989432884798 299.6062415510748
84.65023906727255 290.9955693181366
101.09584730616609 279.82437727429215
120.73646164695873 265.3602692813265
144.19279880884315 246.6797247367514
172.2061665938449 222.62954581662245
205.66189198020015 191.78725953327915
245.61730075923634 152.4247297042824
293.3351329767067 102.48142030026976
350.3234502312496 39.55589333160709
418.38329604954777 -39.07451754875572
499.6656155839293 -136.39636835928434
596.7392335073057 -255.55895545273216
712.6720384606338 -399.73040119014223
851.127939784467 -571.9240225347351
1016.4826607291778 -774.8268530532042
1213.962027641481 -1010.6918718257841
1449.8071255817094 -1281.4024122869182
1731.4715399057457 -1588.8873938427148
2067.8569173818087 -1936.163759902974
2469.5942914525585 -2329.4226561426835
2949.380062570818 -2781.849416730514
3522.377251841554 -3320.6378594527337
4206.694709082762 -4001.124855352584
5023.959420068655 -4940.336004236244
6000.0 -6415.020651802925
};
\addlegendentry{$n_\mathrm{n}U = 0$}

\addplot [line width = 2pt, KITblack, dashed]
table {%
1.0 229.7831895367278
1.1942771623577337 229.771089243885
1.4262979405292406 229.75511741792178
1.7033950570919412 229.73405833679354
2.034325815157953 229.7063189640993
2.4295488618379237 229.66981281060725
2.901554720525257 229.6218083467884
3.465260538054591 229.5587311234874
4.138481522218071 229.47590599294685
4.942493968824512 229.36722123863407
5.90270767205795 229.22469140175767
7.049468968812594 229.03788953418467
8.419019796202404 228.79321068062364
10.054643072042195 228.47291894640085
12.008030596598394 228.0539179900871
14.340916706410374 227.5061712026583
17.1270293097404 226.79068219657208
20.4544199636545 225.85692965076288
24.428246631866678 224.6396396341772
29.174097068880588 223.05477117383344
34.84195786177179 220.9946046369998
41.61095456614455 218.32185107654223
49.6950127422517 214.86283486519068
59.34961880114778 210.39988188568677
70.87989432884798 204.663427481644
84.65023906727255 197.32466396099895
101.09584730616609 187.99010737187322
120.73646164695873 176.19998783167932
144.19279880884315 161.43277030267384
172.2061665938449 143.11803231640783
205.66189198020015 120.65900820380784
245.61730075923634 93.46408598690289
293.3351329767067 60.98357297518262
350.3234502312496 22.744902867593243
418.38329604954777 -21.622629580819034
499.6656155839293 -72.38245901202185
596.7392335073057 -129.71121825888414
712.6720384606338 -193.7481928674498
851.127939784467 -264.67424194299605
1016.4826607291778 -342.82074223701755
1213.962027641481 -428.81337029838204
1449.8071255817094 -523.7655312911686
1731.4715399057457 -629.5552158578029
2067.8569173818087 -749.2552292287461
2469.5942914525585 -887.8610389112476
2949.380062570818 -1053.6282352944818
3522.377251841554 -1260.751268967667
4206.694709082762 -1535.2988470784096
5023.959420068655 -1930.2110776902916
6000.0 -2570.9632161469417
};
\addlegendentry{$n_\mathrm{n}U = 3$}

\addplot [line width = 2pt, KITblack, dotted]
table {%
1.0 148.95002167151847
1.1942771623577337 148.93815862901536
1.4262979405292406 148.92251568969556
1.7033950570919412 148.90190972125615
2.034325815157953 148.8747917425633
2.4295488618379237 148.83913501251948
2.901554720525257 148.79228938603256
3.465260538054591 148.73079203837503
4.138481522218071 148.6501219441174
4.942493968824512 148.54438222195355
5.90270767205795 148.40589028230195
7.049468968812594 148.2246512847587
8.419019796202404 147.98768533875622
10.054643072042195 147.67817351453778
12.008030596598394 147.27438412036386
14.340916706410374 146.74833914982867
17.1270293097404 146.0641854536514
20.4544199636545 145.17625030155614
24.428246631866678 144.02679402195045
29.174097068880588 142.54352904193948
34.84195786177179 140.63707193712298
41.61095456614455 138.198652469641
49.6950127422517 135.09852302830637
59.34961880114778 131.18579802871898
70.87989432884798 126.29050355063748
84.65023906727255 120.2286447562055
101.09584730616609 112.81076204474205
120.73646164695873 103.85375403869159
144.19279880884315 93.19473427977158
172.2061665938449 80.7046519988532
205.66189198020015 66.29880203866439
245.61730075923634 49.94155892474134
293.3351329767067 31.643738399103178
350.3234502312496 11.452513872347605
418.38329604954777 -10.564840530583941
499.6656155839293 -34.341459640602764
596.7392335073057 -59.83394434700987
712.6720384606338 -87.0472042063815
851.127939784467 -116.06176669738139
1016.4826607291778 -147.06608689116038
1213.962027641481 -180.3988770094859
1449.8071255817094 -216.61032965129894
1731.4715399057457 -256.5580081215863
2067.8569173818087 -301.5668718805705
2469.5942914525585 -353.7123200467621
2949.380062570818 -416.3534735014477
3522.377251841554 -495.21805050058583
4206.694709082762 -600.8395366232766
5023.959420068655 -754.8219621653077
6000.0 -1009.4177686659577
};

\addlegendentry{$n_\mathrm{n}U = 10$}

\end{axis}

\end{tikzpicture}

%% file: 5_Results/Figures/Association_deviation.tex
\begin{tikzpicture}

\begin{groupplot}[
    group style={
        group name=my plots,
        group size=2 by 2,
        horizontal sep=0.2cm,
    },
    tick align=inside,
    height = 4cm,
    width = 4.7cm,
]
\nextgroupplot[
    title = counterions, 
    title style={yshift=-0.2cm},
    xlabel={$c_\mathrm{s}^\mathrm{b} \ / \ \frac{\mathrm{mol}}{\mathrm{m^3}}$},
    xlabel style = {yshift = 0.4cm, xshift = 1.7cm},
    xmin=0.5, xmax=20000,
    xmode=log,
    xtick={1,10,100,1000,10000},
    xticklabels={$10^{0}$,,$10^{2}$,,},
    ylabel={$c_\mathrm{ct}^\mathbb{P}\ / \ \frac{\mathrm{mol}}{\mathrm{m}^3}$},
    ylabel style = {yshift = -0.6cm, xshift = 0.15cm},
    ymin= 50, ymax=10000,
    ymode=log,
    ytick={10,100,1000,10000},
    yticklabels={,$10^{2}$,,$10^{4}$},
    legend style={
        nodes={anchor=west},
        anchor = north, 
        at={(1.05,-0.4)}, 
        legend columns = 5,
        /tikz/every even column/.append style={column sep=10pt}
    },
]
\addlegendimage{line legend, color=KITgrey, line width = 2pt, legend image post style={scale=0.6}}
\addlegendentry{$c_i^\mathrm{m}$}
\addlegendimage{line legend, color=KITgrey, line width = 2pt, dashed, legend image post style={scale=0.6}}
\addlegendentry{$c_i^\mathrm{u}$}
\addlegendimage{line legend, color=KITgrey, line width = 2pt, dotted, legend image post style={scale=0.6}}
\addlegendentry{$c_i^\mathrm{c}$}
\addlegendimage{only marks, mark=square*, color=KITblue}
\addlegendentry{$K = 1$}
\addlegendimage{only marks, mark=square*, color=KITred}
\addlegendentry{$K = 10$}

\addplot [line width = 2pt, KITblue]
table {%
1.0 1600.013605507235
1.0974987654930561 1600.0156429958695
1.2045035402587823 1600.0179855985427
1.321941148466029 1600.020679007744
1.4508287784959397 1600.0237757544994
1.5922827933410924 1600.0273362352123
1.7475284000076838 1600.0314298890676
1.9179102616724888 1600.0361365516083
2.1049041445120205 1600.0415480108122
2.3101297000831598 1600.0477697959248
2.535364493970112 1600.0549232338235
2.782559402207125 1600.0631478128607
3.0538555088334154 1600.0726039000651
3.3516026509388426 1600.0834758644423
3.6783797718286344 1600.0959756668917
4.0370172585965545 1600.1103469862903
4.430621457583881 1600.1268699615478
4.862601580065355 1600.1458666412746
5.3366992312063095 1600.1677072462003
5.857020818056667 1600.1928173649862
6.428073117284322 1600.221686221783
7.054802310718643 1600.2548761741366
7.742636826811272 1600.2930336230243
8.497534359086444 1600.3369015431879
9.326033468832199 1600.3873338720855
10.235310218990262 1600.4453120300739
11.233240329780276 1600.511963883427
12.328467394420665 1600.5885855060633
13.530477745798075 1600.6766661459744
14.84968262254465 1600.777916858971
16.297508346206442 1600.8943033361645
17.886495290574352 1601.028083523226
19.630406500402714 1601.18185070956
21.544346900318846 1601.3585828546863
23.644894126454084 1601.5616990177766
25.950242113997373 1601.7951238647217
28.48035868435802 1602.0633613452853
31.25715849688237 1602.3715787603383
34.30469286314919 1602.725702574867
37.649358067924695 1603.1325274744586
41.32012400115339 1603.5998403083754
45.34878508128585 1604.1365607067287
49.770235643321115 1604.7529002963863
54.62277217684343 1605.4605425616878
59.94842503189412 1606.2728454903963
65.79332246575683 1607.2050691980405
72.20809018385468 1608.2746307164941
79.24828983539177 1609.5013880426668
86.97490026177834 1610.9079553438994
95.45484566618342 1612.520050877348
104.76157527896652 1614.3668786682179
114.97569953977369 1616.4815442726454
126.1856883066021 1618.901503995453
138.48863713938732 1621.6690457205766
151.99110829529346 1624.831798040207
166.81005372000593 1628.4432626638143
183.07382802953697 1632.5633632184276
200.9233002565048 1637.259001641181
220.51307399030455 1642.6046116044054
242.01282647943833 1648.6826962169455
265.6087782946687 1655.5843427033153
291.5053062825179 1663.4096893733456
319.92671377973846 1672.268352802066
351.11917342151344 1682.2797934061714
385.3528593710531 1693.5736335758934
422.9242874389499 1706.2899267578523
464.1588833612782 1720.5794028222183
509.413801481638 1736.6037190403858
559.0810182512229 1754.5357580822604
613.5907273413176 1774.5600245948467
673.4150657750828 1796.8731985452114
739.0722033525783 1821.6849061956598
811.1308307896873 1849.2187675061448
890.2150854450392 1879.7137720324606
977.0099572992257 1913.4260246620809
1072.2672220103243 1950.630889396736
1176.811952434999 1991.6255456444337
1291.549665014884 2036.7319590399404
1417.4741629268062 2086.300259280075
1555.6761439304723 2140.71251190591
1707.352647470692 2200.3868698097554
1873.817422860385 2265.782093299964
2056.5123083486537 2337.402434162508
2257.0197196339213 2415.802888385827
2477.0763559917114 2501.5948330633946
2718.588242732943 2595.4520745422105
2983.64724028334 2698.1173464196736
3274.5491628777318 2810.4093070143185
3593.8136638046294 2933.2300961879623
3944.20605943766 3067.5735208176293
4328.7612810830615 3214.533946897439
4750.810162102798 3375.3159843888598
5214.00828799969 3551.245058785602
5722.3676593502205 3743.778971194077
6280.29144183426 3954.520556826982
6892.612104349702 4185.231560424334
7564.633275546291 4437.847856487897
8302.175681319752 4714.496152549503
9111.627561154895 5017.512325177533
10000.0 5349.461551226788
};

\addplot [line width = 2pt, KITblue, dashed]
table {%
1.0 265.0016072123885
1.0974987654930561 265.00277117124114
1.2045035402587823 265.00410943550975
1.321941148466029 265.00564810940466
1.4508287784959397 265.0074172041205
1.5922827933410924 265.0094512247399
1.7475284000076838 265.0117898430652
1.9179102616724888 265.0144786711563
2.1049041445120205 265.01757015064777
2.3101297000831598 265.0211245751534
2.535364493970112 265.0252112656657
2.782559402207125 265.02990992182237
3.0538555088334154 265.0353121753231
3.3516026509388426 265.04152337571134
3.6783797718286344 265.0486646432192
4.0370172585965545 265.0568752285649
4.430621457583881 265.0663152255077
4.862601580065355 265.07716868878754
5.3366992312063095 265.08964721788936
5.857020818056667 265.103994076034
6.428073117284322 265.12048892408035
7.054802310718643 265.13945326080193
7.742636826811272 265.1612566745057
8.497534359086444 265.1863240264053
9.326033468832199 265.2151437038618
10.235310218990262 265.24827710181756
11.233240329780276 265.2863695138615
12.328467394420665 265.33016264074166
13.530477745798075 265.3805089542079
14.84968262254465 265.4383881883208
16.297508346206442 265.5049262692914
17.886495290574352 265.58141703912014
19.630406500402714 265.66934717836574
21.544346900318846 265.77042478994986
23.644894126454084 265.88661216968393
25.950242113997373 266.0201633608421
28.48035868435802 266.1736671703027
31.25715849688237 266.3500964131492
34.30469286314919 266.55286425170857
37.649358067924695 266.78588860417983
41.32012400115339 267.0536657174349
45.34878508128585 267.36135412806203
49.770235643321115 267.71487037462015
54.62277217684343 268.12099797112785
59.94842503189412 268.5875113049478
65.79332246575683 269.1233162783879
72.20809018385468 269.7386096682481
79.24828983539177 270.4450593254833
86.97490026177834 271.25600747086474
95.45484566618342 272.18669945304254
104.76157527896652 273.25454041225214
114.97569953977369 274.4793823244021
126.1856883066021 275.88384387425856
138.48863713938732 277.4936655116955
151.99110829529346 279.3381018729704
166.81005372000593 281.45035349680074
183.07382802953697 283.8680394386519
200.9233002565048 286.6337120053009
220.51307399030455 289.79541442987215
242.01282647943833 293.40728147825615
265.6087782946687 297.53018599906994
291.5053062825179 302.23242463176524
319.92671377973846 307.5904527015679
351.11917342151344 313.68966068218845
385.3528593710531 320.62520187284247
422.9242874389499 328.5028697359066
464.1588833612782 337.4400346183888
509.413801481638 347.56664840855854
559.0810182512229 359.0263291016615
613.5907273413176 371.97754033097533
673.4150657750828 386.5948832805812
739.0722033525783 403.0705201855306
811.1308307896873 421.61574958459056
890.2150854450392 442.46275367101134
977.0099572992257 465.8665375261969
1072.2672220103243 492.10707910778166
1176.811952434999 521.491707948148
1291.549665014884 554.3577300125212
1417.4741629268062 591.0753164178922
1555.6761439304723 632.0506749452252
1707.352647470692 677.7295255748871
1873.817422860385 728.6009045823696
2056.5123083486537 785.2013258934227
2257.0197196339213 848.1193332027736
2477.0763559917114 918.0004815910372
2718.588242732943 995.5527928444038
2983.64724028334 1081.552734263629
3274.5491628777318 1176.851776382282
3593.8136638046294 1282.383590704798
3944.20605943766 1399.1719543834547
4328.7612810830615 1528.3394347811693
4750.810162102798 1671.116933240944
5214.00828799969 1828.8541742413722
5722.3676593502205 2003.0312336016357
6280.29144183426 2195.2712076442726
6892.612104349702 2407.3541343573193
7564.633275546291 2641.2322877385322
8302.175681319752 2899.0469777653857
9111.627561154895 3183.147000924374
10000.0 3496.1089000602924
};

\addplot [line width = 2pt, KITblue, dotted]
table {%
1.0 1335.0119982948465
1.0974987654930561 1335.0128718246283
1.2045035402587823 1335.013876163033
1.321941148466029 1335.0150308983393
1.4508287784959397 1335.016358550379
1.5922827933410924 1335.0178850104724
1.7475284000076838 1335.0196400460025
1.9179102616724888 1335.021657880452
2.1049041445120205 1335.0239778601645
2.3101297000831598 1335.0266452207713
2.535364493970112 1335.029711968158
2.782559402207125 1335.0332378910384
3.0538555088334154 1335.037291724742
3.3516026509388426 1335.0419524887309
3.6783797718286344 1335.0473110236726
4.0370172585965545 1335.0534717577254
4.430621457583881 1335.06055473604
4.862601580065355 1335.068697952487
5.3366992312063095 1335.0780600283108
5.857020818056667 1335.0888232889522
6.428073117284322 1335.1011972977028
7.054802310718643 1335.1154229133347
7.742636826811272 1335.1317769485186
8.497534359086444 1335.1505775167825
9.326033468832199 1335.1721901682236
10.235310218990262 1335.1970349282562
11.233240329780276 1335.2255943695654
12.328467394420665 1335.2584228653216
13.530477745798075 1335.2961571917665
14.84968262254465 1335.33952867065
16.297508346206442 1335.389377066873
17.886495290574352 1335.446666484106
19.630406500402714 1335.5125035311942
21.544346900318846 1335.5881580647365
23.644894126454084 1335.6750868480926
25.950242113997373 1335.7749605038796
28.48035868435802 1335.8896941749827
31.25715849688237 1336.021482347189
34.30469286314919 1336.1728383231584
37.649358067924695 1336.3466388702786
41.32012400115339 1336.5461745909406
45.34878508128585 1336.7752065786667
49.770235643321115 1337.038029921766
54.62277217684343 1337.3395445905599
59.94842503189412 1337.6853341854485
65.79332246575683 1338.0817529196527
72.20809018385468 1338.536021048246
79.24828983539177 1339.0563287171835
86.97490026177834 1339.6519478730347
95.45484566618342 1340.3333514243054
104.76157527896652 1341.1123382559658
114.97569953977369 1342.0021619482434
126.1856883066021 1343.0176601211942
138.48863713938732 1344.175380208881
151.99110829529346 1345.4936961672367
166.81005372000593 1346.9929091670135
183.07382802953697 1348.6953237797757
200.9233002565048 1350.62528963588
220.51307399030455 1352.8091971745332
242.01282647943833 1355.2754147386893
265.6087782946687 1358.0541567042453
291.5053062825179 1361.1772647415803
319.92671377973846 1364.677900100498
351.11917342151344 1368.5901327239828
385.3528593710531 1372.948431703051
422.9242874389499 1377.7870570219457
464.1588833612782 1383.1393682038295
509.413801481638 1389.0370706318274
559.0810182512229 1395.5094289805988
613.5907273413176 1402.5824842638713
673.4150657750828 1410.2783152646302
739.0722033525783 1418.6143860101292
811.1308307896873 1427.6030179215543
890.2150854450392 1437.2510183614493
977.0099572992257 1447.559487135884
1072.2672220103243 1458.5238102889543
1176.811952434999 1470.1338376962858
1291.549665014884 1482.3742290274192
1417.4741629268062 1495.2249428621828
1555.6761439304723 1508.661836960685
1707.352647470692 1522.6573442348683
1873.817422860385 1537.1811887175945
2056.5123083486537 1552.2011082690856
2257.0197196339213 1567.6835551830534
2477.0763559917114 1583.5943514723572
2718.588242732943 1599.8992816978068
2983.64724028334 1616.5646121560446
3274.5491628777318 1633.5575306320366
3593.8136638046294 1650.8465054831643
3944.20605943766 1668.4015664341746
4328.7612810830615 1686.1945121162694
4750.810162102798 1704.199051147916
5214.00828799969 1722.3908845442295
5722.3676593502205 1740.7477375924411
6280.29144183426 1759.2493491827092
6892.612104349702 1777.8774260670143
7564.633275546291 1796.6155687493642
8302.175681319752 1815.4491747841173
9111.627561154895 1834.3653242531586
10000.0 1853.3526511664957
};

\addplot [line width = 2pt, KITred]
table {%
1.0 1600.0264148280535
1.0974987654930561 1600.0303704114099
1.2045035402587823 1600.0349182996035
1.321941148466029 1600.0401471714335
1.4508287784959397 1600.046158978567
1.5922827933410924 1600.0530709308637
1.7475284000076838 1600.0610177782512
1.9179102616724888 1600.0701544333097
2.1049041445120205 1600.0806589852734
2.3101297000831598 1600.0927361636336
2.535364493970112 1600.1066213180964
2.782559402207125 1600.1225849914892
3.0538555088334154 1600.1409381734097
3.3516026509388426 1600.1620383352597
3.6783797718286344 1600.1862963619526
4.0370172585965545 1600.214184512295
4.430621457583881 1600.246245559086
4.862601580065355 1600.2831032816184
5.3366992312063095 1600.3254745078855
5.857020818056667 1600.374182931646
6.428073117284322 1600.4301749610236
7.054802310718643 1600.4945378908703
7.742636826811272 1600.5685207310457
8.497534359086444 1600.6535580675202
9.326033468832199 1600.7512973830499
10.235310218990262 1600.863630319452
11.233240329780276 1600.9927284243581
12.328467394420665 1601.1410839918285
13.530477745798075 1601.311556678063
14.84968262254465 1601.5074266502313
16.297508346206442 1601.732455107036
17.886495290574352 1601.9909530924917
19.630406500402714 1602.287859607031
21.544346900318846 1602.6288300989468
23.644894126454084 1603.02033648938
25.950242113997373 1603.4697799389498
28.48035868435802 1603.985617594831
31.25715849688237 1604.577504552285
34.30469286314919 1605.256452209875
37.649358067924695 1606.0350040751712
41.32012400115339 1606.927429866269
45.34878508128585 1607.9499384292817
49.770235643321115 1609.12090952542
54.62277217684343 1610.4611439045952
59.94842503189412 1611.994130247888
65.79332246575683 1613.7463265063996
72.20809018385468 1615.747451878169
79.24828983539177 1618.0307841570627
86.97490026177834 1620.6334554954958
95.45484566618342 1623.5967373182475
104.76157527896652 1626.966306953801
114.97569953977369 1630.7924768055495
126.1856883066021 1635.130380638214
138.48863713938732 1640.0401026037252
151.99110829529346 1645.586731096533
166.81005372000593 1651.84033610056
183.07382802953697 1658.8758600552999
200.9233002565048 1666.7729274262788
220.51307399030455 1675.6155835423865
242.01282647943833 1685.4919845792547
265.6087782946687 1696.4940719340925
291.5053062825179 1708.7172748156577
319.92671377973846 1722.260293932046
351.11917342151344 1737.2250247502584
385.3528593710531 1753.716679986627
422.9242874389499 1771.8441670453092
464.1588833612782 1791.7207671237677
509.413801481638 1813.4651494787088
559.0810182512229 1837.2027383614568
613.5907273413176 1863.067433257521
673.4150657750828 1891.2036672444297
739.0722033525783 1921.7687752231282
811.1308307896873 1954.9356347258802
890.2150854450392 1990.8955376290182
977.0099572992257 2029.8612515062666
1072.2672220103243 2072.070234175144
1176.811952434999 2117.7879734971475
1291.549665014884 2167.311435764686
1417.4741629268062 2220.9726190262795
1555.6761439304723 2279.1422214554536
1707.352647470692 2342.233448430748
1873.817422860385 2410.7059945851943
2056.5123083486537 2485.0702481223607
2257.0197196339213 2565.891773833427
2477.0763559917114 2653.7961383806555
2718.588242732943 2749.4741466603696
2983.64724028334 2853.687561736295
3274.5491628777318 2967.275383385643
3593.8136638046294 3091.1607622344854
3944.20605943766 3226.358628287103
4328.7612810830615 3373.9841148396818
4750.810162102798 3535.261861696552
5214.00828799969 3711.5362855690655
5722.3676593502205 3904.2829107361404
6280.29144183426 4115.120859608872
6892.612104349702 4345.8266108423495
7564.633275546291 4598.34914211608
8302.175681319752 4874.826585686657
9111.627561154895 5177.60453733386
10000.0 5509.2561734200835
};

\addplot [line width = 2pt, KITred, dashed]
table {%
1.0 70.35066127356973
1.0974987654930561 70.35169020151424
1.2045035402587823 70.35287321374497
1.321941148466029 70.35423338389523
1.4508287784959397 70.35579724053045
1.5922827933410924 70.3575952843874
1.7475284000076838 70.35966258302031
1.9179102616724888 70.36203945442945
2.1049041445120205 70.36477225297753
2.3101297000831598 70.36791427288225
2.535364493970112 70.37152678685239
2.782559402207125 70.37568024005822
3.0538555088334154 70.38045562262786
3.3516026509388426 70.38594604731756
3.6783797718286344 70.39225856296379
4.0370172585965545 70.39951623887123
4.430621457583881 70.4078605605057
4.862601580065355 70.41745418283955
5.3366992312063095 70.42848409455449
5.857020818056667 70.44116525415649
6.428073117284322 70.45574476805045
7.054802310718643 70.47250669091774
7.742636826811272 70.49177754050481
8.497534359086444 70.5139326323845
9.326033468832199 70.53940335560821
10.235310218990262 70.5686855276797
11.233240329780276 70.60234898723279
12.328467394420665 70.64104860549789
13.530477745798075 70.68553692341389
14.84968262254465 70.73667865048289
16.297508346206442 70.79546729453325
17.886495290574352 70.86304422889768
19.630406500402714 70.94072054555345
21.544346900318846 71.03000208996274
23.644894126454084 71.13261812614795
25.950242113997373 71.25055413937979
28.48035868435802 71.38608934915054
31.25715849688237 71.54183957722955
34.30469286314919 71.72080619483165
37.649358067924695 71.92643195949871
41.32012400115339 72.16266464628545
45.34878508128585 72.43402947928207
49.770235643321115 72.74571147828433
54.62277217684343 73.10364895145472
59.94842503189412 73.51463948804668
65.79332246575683 73.9864599359117
72.20809018385468 74.52800198729969
79.24828983539177 75.14942514504727
86.97490026177834 75.8623290020925
95.45484566618342 76.67994682498822
104.76157527896652 77.61736348012835
114.97569953977369 78.69175855455619
126.1856883066021 79.92267932430929
138.48863713938732 81.33234639462066
151.99110829529346 82.94599427952326
166.81005372000593 84.79225279911364
183.07382802953697 86.9035727326563
200.9233002565048 89.31670168874349
220.51307399030455 92.07321607149558
242.01282647943833 95.22011590307764
265.6087782946687 98.81048989101832
291.5053062825179 102.90425848487367
319.92671377973846 107.5690027841249
351.11917342151344 112.88088681683801
385.3528593710531 118.92567988008553
422.9242874389499 125.79988428165645
464.1588833612782 133.6119720034724
509.413801481638 142.483731666702
559.0810182512229 152.55172494941098
613.5907273413176 163.96884959866185
673.4150657750828 176.90600474547995
739.0722033525783 191.5538537365084
811.1308307896873 208.12468046492796
890.2150854450392 226.85433745358372
977.0099572992257 248.00428782090242
1072.2672220103243 271.8637486869173
1176.811952434999 298.7519503135776
1291.549665014884 329.02053291501625
1417.4741629268062 363.05611108819744
1555.6761439304723 401.28304361464114
1707.352647470692 444.1664534115303
1873.817422860385 492.2155482239816
2056.5123083486537 545.9872969943981
2257.0197196339213 606.0905196880361
2477.0763559917114 673.1904498856512
2718.588242732943 748.0138300442098
2983.64724028334 831.3545994590762
3274.5491628777318 924.0802351562949
3593.8136638046294 1027.1388066882812
3944.20605943766 1141.5668075005722
4328.7612810830615 1268.4978284677356
4750.810162102798 1409.1721435315578
5214.00828799969 1564.947283180227
5722.3676593502205 1737.3096787736633
6280.29144183426 1927.8874693919977
6892.612104349702 2138.4645728899823
7564.633275546291 2370.9961341147427
8302.175681319752 2627.6254757467023
9111.627561154895 2910.7026909461415
10000.0 3222.805031961536
};

\addplot [line width = 2pt, KITred, dotted]
table {%
1.0 1529.6757535544837
1.0974987654930561 1529.6786802098957
1.2045035402587823 1529.6820450858586
1.321941148466029 1529.6859137875383
1.4508287784959397 1529.6903617380367
1.5922827933410924 1529.6954756464763
1.7475284000076838 1529.701355195231
1.9179102616724888 1529.7081149788803
2.1049041445120205 1529.715886732296
2.3101297000831598 1529.7248218907514
2.535364493970112 1529.735094531244
2.782559402207125 1529.746904751431
3.0538555088334154 1529.760482550782
3.3516026509388426 1529.7760922879422
3.6783797718286344 1529.7940377989887
4.0370172585965545 1529.8146682734239
4.430621457583881 1529.8383849985803
4.862601580065355 1529.865649098779
5.3366992312063095 1529.896990413331
5.857020818056667 1529.9330176774895
6.428073117284322 1529.9744301929732
7.054802310718643 1530.0220311999526
7.742636826811272 1530.076743190541
8.497534359086444 1530.1396254351357
9.326033468832199 1530.2118940274418
10.235310218990262 1530.2949447917724
11.233240329780276 1530.3903794371254
12.328467394420665 1530.5000353863306
13.530477745798075 1530.6260197546492
14.84968262254465 1530.7707479997484
16.297508346206442 1530.9369878125028
17.886495290574352 1531.127908863594
19.630406500402714 1531.3471390614775
21.544346900318846 1531.5988280089841
23.644894126454084 1531.8877183632321
25.950242113997373 1532.21922579957
28.48035868435802 1532.5995282456804
31.25715849688237 1533.0356649750554
34.30469286314919 1533.5356460150433
37.649358067924695 1534.1085721156724
41.32012400115339 1534.7647652199835
45.34878508128585 1535.5159089499996
49.770235643321115 1536.3751980471357
54.62277217684343 1537.3574949531405
59.94842503189412 1538.4794907598414
65.79332246575683 1539.759866570488
72.20809018385468 1541.2194498908693
79.24828983539177 1542.8813590120155
86.97490026177834 1544.7711264934032
95.45484566618342 1546.9167904932592
104.76157527896652 1549.3489434736728
114.97569953977369 1552.1007182509934
126.1856883066021 1555.2077013139049
138.48863713938732 1558.7077562091047
151.99110829529346 1562.6407368170096
166.81005372000593 1567.0480833014465
183.07382802953697 1571.9722873226435
200.9233002565048 1577.4562257375353
220.51307399030455 1583.542367470891
242.01282647943833 1590.271868676177
265.6087782946687 1597.6835820430742
291.5053062825179 1605.813016330784
319.92671377973846 1614.691291147921
351.11917342151344 1624.3441379334204
385.3528593710531 1634.7910001065416
422.9242874389499 1646.0442827636527
464.1588833612782 1658.1087951202953
509.413801481638 1670.9814178120068
559.0810182512229 1684.6510134120458
613.5907273413176 1699.0985836588593
673.4150657750828 1714.2976624989496
739.0722033525783 1730.2149214866197
811.1308307896873 1746.8109542609523
890.2150854450392 1764.0412001754346
977.0099572992257 1781.8569636853642
1072.2672220103243 1800.2064854882267
1176.811952434999 1819.03602318357
1291.549665014884 1838.2909028496697
1417.4741629268062 1857.916507938082
1555.6761439304723 1877.8591778408127
1707.352647470692 1898.0669950192175
1873.817422860385 1918.490446361213
2056.5123083486537 1939.0829511279626
2257.0197196339213 1959.801254145391
2477.0763559917114 1980.6056884950042
2718.588242732943 2001.4603166161598
2983.64724028334 2022.3329622772185
3274.5491628777318 2043.1951482293482
3593.8136638046294 2064.021955546204
3944.20605943766 2084.7918207865305
4328.7612810830615 2105.4862863719463
4750.810162102798 2126.089718164994
5214.00828799969 2146.5890023888383
5722.3676593502205 2166.9732319624773
6280.29144183426 2187.233390216874
6892.612104349702 2207.3620379523672
7564.633275546291 2227.353008001337
8302.175681319752 2247.201109939955
9111.627561154895 2266.901846387719
10000.0 2286.4511414585477
};
\nextgroupplot[
    title = coions, 
    title style={yshift=-0.2cm},
    xmode = log,
    ymode = log,
    xmin=0.5, xmax=20000,
    ymin = 0.01, ymax = 20000,
    xtick={1,10,100,1000,10000},
    xticklabels={,,$10^{2}$,,$10^{4}$},
    yticklabel pos=right,
    ytick={0.01,0.1,1,10,100,1000,10000},
    yticklabels={$10^{-2}$,,,,,,$10^{4}$},
    ylabel = {$c_\mathrm{co}^\mathrm{m} \ / \ \frac{\mathrm{mol}}{\mathrm{m}^3}$},
    ylabel style = {yshift = 0.9cm, xshift = -0.05 cm},
    xlabel = {},]
\addplot [line width = 2pt, KITblue]
table {%
1.0 0.02721101983384419
1.0974987654930561 0.031285991738002
1.2045035402587823 0.03597119708605023
1.321941148466029 0.04135801548773683
1.4508287784959397 0.047551508998810894
1.5922827933410924 0.0546724704246549
1.7475284000076838 0.06285977813538413
1.9179102616724888 0.0722731032172821
2.1049041445120205 0.0830960216244626
2.3101297000831598 0.09553959184896425
2.535364493970112 0.10984646764759762
2.782559402207125 0.1262956257210119
3.0538555088334154 0.14520780013042767
3.3516026509388426 0.16695172888399104
3.6783797718286344 0.19195133378308535
4.0370172585965545 0.2206939725802582
4.430621457583881 0.25373992309613336
4.862601580065355 0.2917332825494979
5.3366992312063095 0.33541449240021415
5.857020818056667 0.38563472997271186
6.428073117284322 0.4433724435652781
7.054802310718643 0.5097523482727526
7.742636826811272 0.5860672460482761
8.497534359086444 0.6738030863751792
9.326033468832199 0.7746677441708106
10.235310218990262 0.8906240601477742
11.233240329780276 1.0239277668536781
12.328467394420665 1.1771710121263832
13.530477745798075 1.3533322919484978
14.84968262254465 1.555833717941394
16.297508346206442 1.788606672328654
17.886495290574352 2.056167046451824
19.630406500402714 2.3637014191199217
21.544346900318846 2.7171657093726886
23.644894126454084 3.123398035552979
25.950242113997373 3.5902477294441257
28.48035868435802 4.126722690571661
31.25715849688237 4.743157520677913
34.30469286314919 5.451405149735492
37.649358067924695 6.265054948917743
41.32012400115339 7.199680616752524
45.34878508128585 8.273121413459988
49.770235643321115 9.505800592775799
54.62277217684343 10.921085123379914
59.94842503189412 12.545690980796888
65.79332246575683 14.41013839608758
72.20809018385468 16.54926143299631
79.24828983539177 19.00277608534347
86.97490026177834 21.81591068781174
95.45484566618342 25.040101754712772
104.76157527896652 28.733757336456488
114.97569953977369 32.963088545315514
126.1856883066021 37.80300799093502
138.48863713938732 43.33809144118952
151.99110829529346 49.66359608045715
166.81005372000593 56.88652532767958
183.07382802953697 65.12672643691333
200.9233002565048 74.51800328236202
220.51307399030455 85.20922320881033
242.01282647943833 97.3653926714854
265.6087782946687 111.16868569153448
291.5053062825179 126.81937908435945
319.92671377973846 144.53670573052986
351.11917342151344 164.5595868580807
385.3528593710531 187.147267177654
422.9242874389499 212.57985352649104
464.1588833612782 241.15880564885217
509.413801481638 273.20743808465215
559.0810182512229 309.0715161677765
613.5907273413176 349.12004919066766
673.4150657750828 393.74639709103417
739.0722033525783 443.36981239164874
811.1308307896873 498.4375350124519
890.2150854450392 559.4275440650343
977.0099572992257 626.8520493242205
1072.2672220103243 701.2617787935108
1176.811952434999 783.2510912888863
1291.549665014884 873.4639180798912
1417.4741629268062 972.6005185601553
1555.6761439304723 1081.425023811824
1707.352647470692 1200.7737396195137
1873.817422860385 1331.5641865999296
2056.5123083486537 1474.8048683250179
2257.0197196339213 1631.6057767716559
2477.0763559917114 1803.18966612679
2718.588242732943 1990.9041490844234
2983.64724028334 2196.234692839349
3274.5491628777318 2420.8186140286384
3593.8136638046294 2666.4601923759274
3944.20605943766 2935.1470416352613
4328.7612810830615 3229.0678937948805
4750.810162102798 3550.631968777725
5214.00828799969 3902.4901175712075
5722.3676593502205 4287.557942388157
6280.29144183426 4709.041113653969
6892.612104349702 5170.463120848673
7564.633275546291 5675.695712975798
8302.175681319752 6228.992305099014
9111.627561154895 6835.024650355074
10000.0 7498.923102453584
};
\addplot [line width = 2pt, KITred]
table {%
1.0 0.05282965611179828
1.0974987654930561 0.06074082282031491
1.2045035402587823 0.06983659920739033
1.321941148466029 0.08029434286708335
1.4508287784959397 0.0923179571338098
1.5922827933410924 0.10614186172787342
1.7475284000076838 0.12203555650247246
1.9179102616724888 0.1403088666188462
2.1049041445120205 0.16131797054723018
2.3101297000831598 0.18547232726721283
2.535364493970112 0.21324263619270248
2.782559402207125 0.2451699829783771
3.0538555088334154 0.28187634681932344
3.3516026509388426 0.3240766705190403
3.6783797718286344 0.372592723904448
4.0370172585965545 0.4283690245903741
4.430621457583881 0.4924911181719935
4.862601580065355 0.5662065632364234
5.3366992312063095 0.6509490157709821
5.857020818056667 0.7483658632914397
6.428073117284322 0.8603499220472511
7.054802310718643 0.989075781740605
7.742636826811272 1.1370414620912814
8.497534359086444 1.3071161350402132
9.326033468832199 1.5025947661005656
10.235310218990262 1.7272606389041505
11.233240329780276 1.9854568487169246
12.328467394420665 2.2821679836573905
13.530477745798075 2.6231133561265465
14.84968262254465 3.014853300463552
16.297508346206442 3.4649102140738974
17.886495290574352 3.9819061849854442
19.630406500402714 4.575719214064337
21.544346900318846 5.257660197896152
23.644894126454084 6.040672978764064
25.950242113997373 6.939559877904582
28.48035868435802 7.971235189668384
31.25715849688237 9.155009104577395
34.30469286314919 10.512904419760384
37.649358067924695 12.070008150354159
41.32012400115339 13.85485973255349
45.34878508128585 15.89987685858346
49.770235643321115 18.241819050863878
54.62277217684343 20.922287809220748
59.94842503189412 23.98826049581292
65.79332246575683 27.49265301284281
72.20809018385468 31.49490375638934
79.24828983539177 36.06156831412571
86.97490026177834 41.26691099099195
95.45484566618342 47.19347486491373
104.76157527896652 53.93261419327886
114.97569953977369 61.58495372474668
126.1856883066021 70.26076141524565
138.48863713938732 80.08020523419496
151.99110829529346 91.1734622121307
166.81005372000593 103.6806722157044
183.07382802953697 117.75172011587168
200.9233002565048 133.5458548575004
220.51307399030455 151.2311670860874
242.01282647943833 170.98396915933904
265.6087782946687 192.988143868772
291.5053062825179 217.4345496316177
319.92671377973846 244.52058786432767
351.11917342151344 274.4500495006838
385.3528593710531 307.4333599733679
422.9242874389499 343.6883340907033
464.1588833612782 383.44153424760805
509.413801481638 426.93029895747543
559.0810182512229 474.4054767229606
613.5907273413176 526.1348665150819
673.4150657750828 582.4073344888984
739.0722033525783 643.5375504462866
811.1308307896873 709.8712694517873
890.2150854450392 781.7910752580627
977.0099572992257 859.7225030125571
1072.2672220103243 944.1404683503084
1176.811952434999 1035.5759469943162
1291.549665014884 1134.6228715293917
1417.4741629268062 1241.9452380525788
1555.6761439304723 1358.2844429109268
1707.352647470692 1484.4668968615138
1873.817422860385 1621.4119891704065
2056.5123083486537 1770.1404962447402
2257.0197196339213 1931.7835476668727
2477.0763559917114 2107.5922767613306
2718.588242732943 2298.9482933207596
2983.64724028334 2507.375123472607
3274.5491628777318 2734.550766771304
3593.8136638046294 2982.3215244689877
3944.20605943766 3252.717256574223
4328.7612810830615 3547.9682296793826
4750.810162102798 3870.523723393122
5214.00828799969 4223.072571138153
5722.3676593502205 4608.565821472304
6280.29144183426 5030.241719217766
6892.612104349702 5491.653221684722
7564.633275546291 5996.698284232185
8302.175681319752 6549.653171373338
9111.627561154895 7155.209074667748
10000.0 7818.512346840191
};

\end{groupplot}

\end{tikzpicture}

%% file: 5_Results/Figures/PartitioningNaCl.tex
\begin{tikzpicture}

\begin{groupplot}[
    group style={
        group name=my plots,
        group size=2 by 1,
        horizontal sep= 0pt,
    },
    tick align=inside,
]

\nextgroupplot[
    height = 5cm,
    width = 6cm,
    scale only axis,
    xmode = log,
    ymode = log,
    xmin=8, xmax=7000,
    ymin = 0.1, ymax = 10000,
    xlabel={$c_\mathrm{s}^\mathrm{b} \ / \ \frac{\mathrm{mol}}{\mathrm{m^3}}$},
    xtick={1,10,100,1000,10000},
    xticklabels={,$10^{1}$,,$10^{3}$,},
    yticklabel pos=right,
    ytick={0.001,0.01,0.1,1,10,100,1000,10000},
    yticklabels={,,$10^{-1}$,,,,,$10^{4}$},
    ylabel = {$c_\mathrm{co}^\mathrm{m} \ / \ \frac{\mathrm{mol}}{\mathrm{m}^3}$},
    xlabel style = {yshift = 0.5cm, xshift = -0.6cm},
    ylabel style = {yshift = 0.9cm, xshift = -0.05 cm},
    legend style={
        nodes={anchor=west},
        anchor = south east, 
        at={(0.97,0.04)}, 
        fill opacity=0.9,
        text opacity=1,
        draw opacity=1,
    },
    ]
\addlegendimage{KITblack, only marks, mark=x, mark size=4, mark options={line width=1.5pt}, legend image post style={xshift=-3.5pt}}
\addlegendentry{exp. \cite{Galizia2019}}
\addlegendimage{line legend, color=KITblack, line width = 2pt, legend image post style={scale=0.6}}
\addlegendentry{IOM}
\addlegendimage{line legend, color=KITred, dotted, line width = 2pt, legend image post style={scale=0.6}}
\addlegendentry{D-M}
\addlegendimage{line legend, color=KITblue, dashed, line width = 2pt, legend image post style={scale=0.6}}
\addlegendentry{SDE}
\addlegendimage{line legend, color=KITorange, dashdotted, line width = 2pt, legend image post style={scale=0.6}}
\addlegendentry{low-T*}

\addplot [line width = 2pt, KITorange, dashdotted]
table {%
10.0 0.10992492628614728
11.829087159678172 0.15382257182680617
13.992730303126296 0.21524980878160405
16.552122635755094 0.30120491385424447
19.579650133602893 0.42147753728474097
23.1609387986393 0.5897591435219841
27.397276374907616 0.8251920386110972
32.40847701765738 1.154526965194902
38.33626993542959 1.61512186839906
45.34830784431464 2.2590968854095674
53.64290860343152 3.1590658611359954
63.45466413686417 4.415992373381926
75.0610752763071 6.169859451514054
88.79040017426006 8.613969309168796
105.03093826040269 12.013737881528638
124.24201231450792 16.73068678505628
146.9669592562123 23.251719726752356
173.84849706346063 32.2223029146101
205.64690243427285 44.479296769312484
243.26151330128494 61.07437316960855
287.7561643436108 83.27225169207024
340.38927487552485 112.50125454690715
402.6494400722235 150.23217018876863
476.29753214099406 197.7739845154814
563.416502163544 256.00912874059
666.4702911293562 325.1402759557315
788.3735163105241 404.55528592333127
932.5739038819152 492.8965553559623
1103.149799186051 588.338467270316
1304.925512475327 688.9830571144048
1543.6077623958336 793.2484184711353
1825.947076173611 900.1648221221734
2159.9287113017167 1009.5677777396415
2554.998498467936 1122.2360751312942
3022.3299931224074 1240.0522038957556
3575.1404913954457 1366.2809724455103
4229.064848081138 1506.1023490754144
5002.597669188292 1667.64431180202
5917.616385363117 1864.0451757725982
7000.0 2117.8374324329407
};
\addplot [line width = 2pt, KITblue, dashed]
table {%
10.0 0.0328632485760811
11.829087159678172 0.0459796924896475
13.992730303126296 0.06432986586694343
16.552122635755094 0.09000122795489768
19.579650133602893 0.12591319024558995
23.1609387986393 0.17614824791675424
27.397276374907616 0.24641458994718418
32.40847701765738 0.344692159395315
38.33626993542959 0.4821344805465569
45.34830784431464 0.6743267873553771
53.64290860343152 0.9430398879633561
63.45466413686417 1.318672677648435
75.0610752763071 1.8436494544968856
88.79040017426006 2.5771373961537876
105.03093826040269 3.601582671002889
124.24201231450792 5.031740502837354
146.9669592562123 7.02710204282873
173.84849706346063 9.80890712623696
205.64690243427285 13.683278455458714
243.26151330128494 19.07236698825168
287.7561643436108 26.555781445331068
340.38927487552485 36.924710324083904
402.6494400722235 51.25098144665169
476.29753214099406 70.97231413369767
563.416502163544 97.99288147657546
666.4702911293562 134.794495125701
788.3735163105241 184.54860197118427
932.5739038819152 251.213663265862
1103.149799186051 339.6003408957611
1304.925512475327 455.3913590845873
1543.6077623958336 605.1147888812266
1825.947076173611 796.0922935810825
2159.9287113017167 1036.39257525273
2554.998498467936 1334.819879491814
3022.3299931224074 1700.9429049110463
3575.1404913954457 2145.114885889258
4229.064848081138 2678.396371471999
5002.597669188292 3312.1694519261205
5917.616385363117 4057.039902278494
7000.0 4920.077310831037
};
\addplot [line width = 2pt, KITred, dotted]
table {%
10.0 0.055998681435672465
11.829087159678172 0.0783510484081805
13.992730303126296 0.10962362642356216
16.552122635755094 0.15337479136265866
19.579650133602893 0.2145813007068499
23.1609387986393 0.30020279522181526
27.397276374907616 0.4199702539032777
32.40847701765738 0.5874866451703152
38.33626993542959 0.8217622616371646
45.34830784431464 1.14935428944452
53.64290860343152 1.6073444187792305
63.45466413686417 2.247475212767489
75.0610752763071 3.1418819394240844
88.79040017426006 4.391008362729221
105.03093826040269 6.134487879276526
124.24201231450792 8.56600544120619
146.9669592562123 11.95341834233138
173.84849706346063 16.665666795433964
205.64690243427285 23.20816340761344
243.26151330128494 32.26825555126344
287.7561643436108 44.771751222651595
340.38927487552485 61.95005139551208
402.6494400722235 85.41484675592562
476.29753214099406 117.23368721361916
563.416502163544 159.99599559561386
666.4702911293562 216.85759214805583
788.3735163105241 291.55569197676397
932.5739038819152 388.39736878151075
1103.149799186051 512.2397166943656
1304.925512475327 668.4912507283688
1543.6077623958336 863.1633664621763
1825.947076173611 1102.987351688733
2159.9287113017167 1395.5936141009038
2554.998498467936 1749.744310003858
3022.3299931224074 2175.581838957228
3575.1404913954457 2684.897390397984
4229.064848081138 3291.3712431839203
5002.597669188292 4010.744531049376
5917.616385363117 4860.808503444467
7000.0 5860.933144714023
};
\addplot [line width = 2pt, KITblack]
table {%
9.681052479253339 0.08071986266785557
12.14828107939447 0.12708838740021908
15.24428604227089 0.2000831198598544
19.129311827723896 0.31498183303714516
24.004441401034345 0.49581341747727586
30.12200397823995 0.7803505397434073
37.79863519865146 1.227923969484537
47.43166569903015 1.9316147004922966
59.519686336844714 3.037188307439275
74.68835448696494 4.772284601474458
93.7227771060572 7.490909553298112
117.60814666233829 11.740060201319242
147.58073318397368 18.35675573845965
185.19187169619667 28.603259956583805
232.38825693857507 44.34261390901717
291.61270129361884 68.2410002969658
365.9305710022973 103.95538020659842
459.1884448107114 156.2345714823009
576.2132069756917 230.86451378978285
723.0618793773842 334.4673388959776
907.3351236651032 474.30213189492986
1138.5706398258199 658.293774354817
1428.7368228807334 895.4321722658102
1792.8522286220302 1196.5095516852325
2249.7629109845534 1575.0637163567583
2823.1178648403506 2048.423104665664
3542.5930616363794 2638.8317497788335
4445.427432079221 3374.6920631703874
5578.350295970535 4291.9710652553085
7000.0 5435.740399455195
};
\addplot [white, only marks, mark = x,  mark size=6, mark options={line width=3pt}]
table {%
10 0.306772
30 1.536783
100 10.927972
300 61.524296
1000 0380.295953
4000 2704.261477
5000 3663.48589
}; 
\addplot [KITblack, only marks, mark = x,  mark size=5, mark options={line width=1.5pt}]
table {%
10 0.306772
30 1.536783
100 10.927972
300 61.524296
1000 0380.295953
4000 2704.261477
5000 3663.48589
}; 

\addplot[KITgrey, loosely dotted, line width = 0.7pt]
table {%
1 1
10000 10000
}; 
\node[rotate=27, text=KITlightgrey] at (axis cs:25,17) {\tiny{equal concentrations}};

\nextgroupplot[
    axis background/.style={fill=white},
    font=\footnotesize,
    xshift = -6.3cm,
    yshift=1.55cm,
    height = 2.5cm,
    width = 3cm,
    scale only axis,
    xticklabel pos=upper,
    xlabel={$c_\mathrm{s}^\mathrm{b} \ / \ \frac{\mathrm{mol}}{\mathrm{m^3}}$},
    xlabel style = {yshift = -0.5cm, xshift = -0.1cm},
    xmin=5, xmax=10000,
    xmode=log,
    xtick={1,10,100,1000,10000},
    xticklabels={,$10^{1}$,,,$10^{4}$},
    ylabel={$c_\mathrm{ct}^\mathrm{m}\ / \ \frac{\mathrm{mol}}{\mathrm{m}^3}$},
    ylabel style = {yshift = -0.7cm, xshift = 0.1cm},
    ymin= 2700, ymax=8000,
    ytick={3000,4000,5000,6000,7000,8000},
    yticklabels={3000,,,,,8000},
]
\addplot [line width = 2pt, KITorange, dashdotted]
table {%
10.0 3042.989869536368
11.829087159678172 3043.3492273833263
13.992730303126296 3043.7838997007807
16.552122635755094 3044.311487889453
19.579650133602893 3044.954337638051
23.1609387986393 3045.741011982363
27.397276374907616 3046.708271450967
32.40847701765738 3047.9037446838215
38.33626993542959 3049.3895383395043
45.34830784431464 3051.2471210224644
53.64290860343152 3053.5839255169576
63.45466413686417 3056.5422466806745
75.0610752763071 3060.3111603369553
88.79040017426006 3065.142324259367
105.03093826040269 3071.3705818536955
124.24201231450792 3079.4401412463626
146.9669592562123 3089.9365055389108
173.84849706346063 3103.62288710603
205.64690243427285 3121.4769993616183
243.26151330128494 3144.7193505905707
287.7561643436108 3174.817524058717
340.38927487552485 3213.4442783094596
402.6494400722235 3262.3659086841194
476.29753214099406 3323.2500459190433
563.416502163544 3397.416396805779
666.4702911293562 3485.6036645046624
788.3735163105241 3587.86097404225
932.5739038819152 3703.6525751583895
1103.149799186051 3832.183251336014
1304.925512475327 3972.860218166813
1543.6077623958336 4125.774828943777
1825.947076173611 4292.133073553564
2159.9287113017167 4474.647086117166
2554.998498467936 4677.974679540742
3022.3299931224074 4909.352985051947
3575.1404913954457 5179.6485100780865
4229.064848081138 5505.205104920079
5002.597669188292 5911.242698456587
5917.616385363117 6438.466762666698
7000.0 7156.9532196311875
};
\addplot [line width = 2pt, KITblue, dashed]
table {%
10.0 3042.9128078586587
11.829087159678172 3043.241384503989
13.992730303126296 3043.632979757866
16.552122635755094 3044.1002842035537
19.579650133602893 3044.658773291012
23.1609387986393 3045.3274010867576
27.397276374907616 3046.129494002303
32.40847701765738 3047.0939098780223
38.33626993542959 3048.2565509516517
45.34830784431464 3049.6623509244105
53.64290860343152 3051.367899543785
63.45466413686417 3053.444926984941
75.0610752763071 3055.984950339938
88.79040017426006 3059.105492346352
105.03093826040269 3062.9584266431702
124.24201231450792 3067.7411949641437
146.9669592562123 3073.7118878549877
173.84849706346063 3081.209491317656
205.64690243427285 3090.680981047765
243.26151330128494 3102.717344409213
287.7561643436108 3118.101053811978
340.38927487552485 3137.867734086636
402.6494400722235 3163.3847199420024
476.29753214099406 3196.4483755372594
563.416502163544 3239.4001495417647
666.4702911293562 3295.2578836746316
788.3735163105241 3367.8542900901025
932.5739038819152 3461.9696830682897
1103.149799186051 3583.44512496146
1304.925512475327 3739.2685201369954
1543.6077623958336 3937.641199353868
1825.947076173611 4188.060545012472
2159.9287113017167 4501.471883630254
2554.998498467936 4890.558483901262
3022.3299931224074 5370.243686067239
3575.1404913954457 5958.482423521834
4229.064848081138 6677.499127316664
5002.597669188292 7555.767838580688
5917.616385363117 8631.461489172594
7000.0 9959.193098029284
};

\addplot [line width = 2pt, KITred, dotted]
table {%
10.0 3042.935943291518
11.829087159678172 3043.251791193956
13.992730303126296 3043.6269717329005
16.552122635755094 3044.0737809316756
19.579650133602893 3044.6074632297377
23.1609387986393 3045.2470474993756
27.397276374907616 3046.016456183992
32.40847701765738 3046.9459850182393
38.33626993542959 3048.0742868715006
45.34830784431464 3049.451042561933
53.64290860343152 3051.140568494686
63.45466413686417 3053.226701282416
75.0610752763071 3055.8194197074563
88.79040017426006 3059.0638214440182
105.03093826040269 3063.152271483486
124.24201231450792 3068.340781683425
146.9669592562123 3074.9709542850915
173.84849706346063 3083.499089040216
205.64690243427285 3094.534229971101
243.26151330128494 3108.8868566923966
287.7561643436108 3127.629353844873
340.38927487552485 3152.1679878215227
402.6494400722235 3184.3235938734915
476.29753214099406 3226.4146088741804
563.416502163544 3281.33246263077
666.4702911293562 3352.5979990284604
788.3735163105241 3444.391729865536
932.5739038819152 3561.5621029363565
1103.149799186051 3709.631725601117
1304.925512475327 3894.8335921470516
1543.6077623958336 4124.209895018004
1825.947076173611 4405.794686284952
2159.9287113017167 4748.886165536923
2554.998498467936 5164.414614364341
3022.3299931224074 5665.393476783654
3575.1404913954457 6267.50182214971
4229.064848081138 6989.831333000217
5002.597669188292 7855.917317081565
5917.616385363117 8895.276908184418
7000.0 10145.955495002474
};
\addplot [line width = 2pt, KITblack]
table {%
10.0 3042.9816626774823
11.829087159678172 3043.3377184388983
13.992730303126296 3043.7677537053837
16.552122635755094 3044.2888262279366
19.579650133602893 3044.922514069568
23.1609387986393 3045.6962949183435
27.397276374907616 3046.645392304766
32.40847701765738 3047.8152545074468
38.33626993542959 3049.2648894432
45.34830784431464 3051.0713534201227
53.64290860343152 3053.335785164955
63.45466413686417 3056.1914876752107
75.0610752763071 3059.814680336809
88.79040017426006 3064.4386485235505
105.03093826040269 3070.372056908284
124.24201231450792 3078.0220733040533
146.9669592562123 3087.9225255257256
173.84849706346063 3100.766396690761
205.64690243427285 3117.4404057461643
243.26151330128494 3139.0573165808537
287.7561643436108 3166.97965294745
340.38927487552485 3202.8282021343175
402.6494400722235 3248.472091414808
476.29753214099406 3306.0052623636975
563.416502163544 3377.7245380442646
666.4702911293562 3466.1313960344432
788.3735163105241 3573.9777011115834
932.5739038819152 3704.3655186963997
1103.149799186051 3860.8995749602805
1304.925512475327 4047.8854797823133
1543.6077623958336 4270.570241727385
1825.947076173611 4535.432054836527
2159.9287113017167 4850.541489626284
2554.998498467936 5226.03687137444
3022.3299931224074 5674.788904156182
3575.1404913954457 6213.388350975774
4229.064848081138 6863.709363929949
5002.597669188292 7655.56131828079
5917.616385363117 8631.561668543105
7000.0 9856.9938093026
};
\addplot [white, only marks, mark = x,  mark size=6, mark options={line width=3pt}]
table {%
10 3184.509849
30 3111.004279
100 3184.509849
300 3259.752178
1000 3663.485891
4000 5982.411295
5000 7734.604899
};
\addplot [KITblack, only marks, mark = x,  mark size=5, mark options={line width=1.5pt}]
table {%
10 3184.509849
30 3111.004279
100 3184.509849
300 3259.752178
1000 3663.485891
4000 5982.411295
5000 7734.604899
};

\end{groupplot}

\end{tikzpicture}

%% file: 5_Results/Figures/PartitioningCaCl2.tex
\begin{tikzpicture}

\begin{groupplot}[
    group style={
        group name=my plots,
        group size=2 by 1,
        horizontal sep= 0pt,
    },
    tick align=inside,
]

\nextgroupplot[
    height = 5cm,
    width = 6cm,
    scale only axis,
    xmode = log,
    ymode = log,
    xmin=8, xmax=7000,
    ymin = 2, ymax = 10000,
    xlabel={$c_\mathrm{s}^\mathrm{b} \ / \ \frac{\mathrm{mol}}{\mathrm{m^3}}$},
    xtick={1,10,100,1000,10000},
    xticklabels={,$10^{1}$,,$10^{3}$,},
    yticklabel pos=right,
    ytick={0.01,0.1,1,10,100,1000,10000},
    yticklabels={,,,$10^{1}$,,,$10^{4}$},
    ylabel = {$c_\mathrm{co}^\mathrm{m} \ / \ \frac{\mathrm{mol}}{\mathrm{m}^3}$},
    xlabel style = {yshift = 0.5cm, xshift = -0.6cm},
    ylabel style = {yshift = 0.7cm, xshift = 0.3 cm},
    legend style={
        nodes={anchor=west},
        anchor = south east, 
        at={(0.97,0.04)}, 
        fill opacity=0.9,
        text opacity=1,
        draw opacity=1,
    },
    ]
\addlegendimage{KITblack, only marks, mark=x, mark size=4, mark options={line width=1.5pt}, legend image post style={xshift=-3.5pt}}
\addlegendentry{exp. \cite{Galizia2017}}
\addlegendimage{line legend, color=KITblack, line width = 2pt, legend image post style={scale=0.6}}
\addlegendentry{IOM}
\addlegendimage{line legend, color=KITred, dotted, line width = 2pt, legend image post style={scale=0.6}}
\addlegendentry{D-M}
\addlegendimage{line legend, color=KITblue, dashed, line width = 2pt, legend image post style={scale=0.6}}
\addlegendentry{SDE}
\addlegendimage{line legend, color=KITorange, dashdotted, line width = 2pt, legend image post style={scale=0.6}}
\addlegendentry{low-T*}

\addplot [line width = 2pt, KITblue, dashed]
table {%
10.0 1.595912631664253
11.829087159678172 2.0528769071416186
13.992730303126296 2.6405846907755146
16.552122635755094 3.3963835420457236
19.579650133602893 4.36825370816435
23.1609387986393 5.617813536366659
27.397276374907616 7.2241612136718505
32.40847701765738 9.28877725818241
38.33626993542959 11.941768586592415
45.34830784431464 15.349799072533479
53.64290860343152 19.726130223841615
63.45466413686417 25.34328213416449
75.0610752763071 32.54890809772913
88.79040017426006 41.78558251590569
105.03093826040269 53.61524971047196
124.24201231450792 68.7491338547002
146.9669592562123 88.08386017554767
173.84849706346063 112.74440495073506
205.64690243427285 144.13419259935688
243.26151330128494 183.99220833681028
287.7561643436108 234.45642160451104
340.38927487552485 298.1320714034993
402.6494400722235 378.16314150920914
476.29753214099406 478.30501198408325
563.416502163544 602.9975427798755
666.4702911293562 757.4392886671868
788.3735163105241 947.6661509504307
932.5739038819152 1180.639860190841
1103.149799186051 1464.3539239516329
1304.925512475327 1807.9565446099414
1543.6077623958336 2221.8948249908813
1825.947076173611 2718.0575192523393
2159.9287113017167 3309.880120641771
2554.998498467936 4012.3239346794476
3022.3299931224074 4841.537795224282
3575.1404913954457 5813.754454525023
4229.064848081138 6942.286172792031
5002.597669188292 8229.2523363785
5917.616385363117 9640.389292558262
7000.0 11009.207171106
};
\addplot [line width = 2pt, KITred, dotted]
table {%
10.0 2.9575377460730214
11.829087159678172 3.8030230109194734
13.992730303126296 4.889514245161085
16.552122635755094 6.285267488021727
19.579650133602893 8.077590034645892
23.1609387986393 10.377983110682232
27.397276374907616 13.328565510540011
32.40847701765738 17.11003306991981
38.33626993542959 21.95142198146583
45.34830784431464 28.14193430403615
53.64290860343152 36.04503854994545
63.45466413686417 46.11496382791789
75.0610752763071 58.91555537983899
88.79040017426006 75.14126157770924
105.03093826040269 95.63981730494298
124.24201231450792 121.43605882928374
146.9669592562123 153.75637402505654
173.84849706346063 194.05368929351846
205.64690243427285 244.03368621072008
243.26151330128494 305.68404224759195
287.7561643436108 381.3096309295657
340.38927487552485 473.57742316807156
402.6494400722235 585.5750159480507
476.29753214099406 720.8862629353313
563.416502163544 883.6866835016085
666.4702911293562 1078.8606119537947
788.3735163105241 1312.14134192026
932.5739038819152 1590.2787315898725
1103.149799186051 1921.2298439170982
1304.925512475327 2314.385502389644
1543.6077623958336 2780.8284137710693
1825.947076173611 3333.629303791966
2159.9287113017167 3988.1813452643987
2554.998498467936 4762.567616824059
3022.3299931224074 5677.940639771946
3575.1404913954457 6758.852243877079
4229.064848081138 8033.357405801192
5002.597669188292 9532.349538851375
5917.616385363117 11286.162767073603
7000.0 13308.94492119128
};
\addplot [line width = 2pt, KITorange, dashdotted]
table {%
10.0 2.802979009428675
11.829087159678172 3.6050185562534347
13.992730303126296 4.6360367163373075
16.552122635755094 5.96104822246137
19.579650133602893 7.663283882835234
23.1609387986393 9.849130391478415
27.397276374907616 12.654300216985847
32.40847701765738 16.25146448934088
38.33626993542959 20.859574474332582
45.34830784431464 26.75504497681702
53.64290860343152 34.28484149717232
63.45466413686417 43.88125079396176
75.0610752763071 56.07765016055845
88.79040017426006 71.52383490850409
105.03093826040269 90.99832540014916
124.24201231450792 115.4135064344303
146.9669592562123 145.80753964449815
173.84849706346063 183.31511056407774
205.64690243427285 229.10809521726324
243.26151330128494 284.2986664791047
287.7561643436108 349.8032050710526
340.38927487552485 426.177278906656
402.6494400722235 513.4494284611765
476.29753214099406 610.9997684057525
563.416502163544 717.5387875988023
666.4702911293562 831.23122536752
788.3735163105241 949.9756399881672
932.5739038819152 1071.8032835032602
1103.149799186051 1195.3243455710774
1304.925512475327 1320.1488169988806
1543.6077623958336 1447.2498098157791
1825.947076173611 1579.307702858401
2159.9287113017167 1721.161532176164
2554.998498467936 1880.6131682897692
3022.3299931224074 2070.0558433323704
3575.1404913954457 2309.9607217052585
4229.064848081138 2636.880753351815
5002.597669188292 3124.080509465626
5917.616385363117 3945.483067710982
7000.0 5644.895634592636
};
\addplot [line width = 2pt, KITblack]
table {%
10.0 3.4218531705477635
11.829087159678172 4.397741743274919
13.992730303126296 5.650289774689122
16.552122635755094 7.256873334760443
19.579650133602893 9.315845242170141
23.1609387986393 11.951812477608259
27.397276374907616 15.321995512974569
32.40847701765738 19.62374746787126
38.33626993542959 25.103228431018977
45.34830784431464 32.065095970196055
53.64290860343152 40.88288297123301
63.45466413686417 52.00950396768596
75.0610752763071 65.9871137341348
88.79040017426006 83.45543512825917
105.03093826040269 105.15782733462193
124.24201231450792 131.94493559024863
146.9669592562123 164.77678264750793
173.84849706346063 204.72558083947928
205.64690243427285 252.98283085406482
243.26151330128494 310.8750450550527
287.7561643436108 379.89215586368306
340.38927487552485 461.73139837981034
402.6494400722235 558.35767127982
476.29753214099406 672.0798416076259
563.416502163544 805.6417957833996
666.4702911293562 962.3274810780321
788.3735163105241 1146.0805693986383
932.5739038819152 1361.6413796754148
1103.149799186051 1614.7060734651384
1304.925512475327 1912.115947818943
1543.6077623958336 2262.0883973200116
1825.947076173611 2674.5070903096976
2159.9287113017167 3161.2999502491325
2554.998498467936 3736.956340999397
3022.3299931224074 4419.286648609661
3575.1404913954457 5230.6576262205135
4229.064848081138 6200.3069987330555
5002.597669188292 7369.574841739039
5917.616385363117 8807.025326231995
7000.0 10670.606664228337
};

\addplot [white, only marks, mark = x,  mark size=6, mark options={line width=3pt}]
table {%
10 4.13063125727462
30 16.2741746834076
100 73.8895739176099
300 258.661032107662
1000 1153.9925732231
2000 3023.38053812949
4000 5942.21044448895
6000 10361.6518810075
}; 
\addplot [KITblack, only marks, mark = x,  mark size=5, mark options={line width=1.5pt}]
table {%
10 4.13063125727462
30 16.2741746834076
100 73.8895739176099
300 258.661032107662
1000 1093.9925732231
2000 3023.38053812949
4000 6042.21044448895
6000 10361.6518810075
}; 

\addplot[KITgrey, loosely dotted, line width = 0.7pt]
table {%
1 2
10000 20000
}; 
\node[rotate=33.5, text=KITlightgrey] at (axis cs:25,70) {\tiny{equal concentrations}};

\nextgroupplot[
    axis background/.style={fill=white},
    font=\footnotesize,
    xshift = -6.3cm,
    yshift=1.55cm,
    height = 2.5cm,
    width = 3cm,
    scale only axis,
    xticklabel pos=upper,
    xlabel={$c_\mathrm{s}^\mathrm{b} \ / \ \frac{\mathrm{mol}}{\mathrm{m^3}}$},
    xlabel style = {yshift = -0.5cm, xshift = -0.1cm},
    xmin=5, xmax=10000,
    xmode=log,
    xtick={1,10,100,1000,10000},
    xticklabels={,$10^{1}$,,,$10^{4}$},
    ylabel={$c_\mathrm{ct}^\mathrm{m}\ / \ \frac{\mathrm{mol}}{\mathrm{m}^3}$},
    ylabel style = {yshift = -0.7cm},
    ymin= 1000, ymax=8000,
    ytick={1000,2000,3000,4000,5000,6000,7000,8000},
    yticklabels={1000,,,,,,,8000},
]
\addplot [line width = 2pt, KITorange, dashdotted]
table {%
10.0 1572.5188591656442
11.829087159678172 1573.6208329689623
13.992730303126296 1575.0067939937942
16.552122635755094 1576.7518777684636
19.579650133602893 1578.9513108251967
23.1609387986393 1581.725702993975
27.397276374907616 1585.2276380975768
32.40847701765738 1589.649810028372
38.33626993542959 1595.2349470713386
45.34830784431464 1602.2877191594953
53.64290860343152 1611.188696428708
63.45466413686417 1622.4101719651771
75.0610752763071 1636.5332060075139
88.79040017426006 1654.2645045025654
105.03093826040269 1676.4506219762998
124.24201231450792 1704.0854305761156
146.9669592562123 1738.30491276522
173.84849706346063 1780.3614943502112
205.64690243427285 1831.5692109533281
243.26151330128494 1893.2125116179693
287.7561643436108 1966.4174555726904
340.38927487552485 2051.996113741948
402.6494400722235 2150.292692316044
476.29753214099406 2261.078517195102
563.416502163544 2383.552925659363
666.4702911293562 2516.497447696751
788.3735163105241 2658.5977471129095
932.5739038819152 2808.9030173838323
1103.149799186051 2967.3607606183828
1304.925512475327 3135.3707217032916
1543.6077623958336 3316.354503370453
1825.947076173611 3516.43118284724
2159.9287113017167 3745.4241666199728
2554.998498467936 4018.6440982995973
3022.3299931224074 4360.350698975374
3575.1404913954457 4810.9463239902525
4229.064848081138 5443.253420774467
5002.597669188292 6404.24339349494
5917.616385363117 8044.182439662813
7000.0 11460.177253580789
};
\addplot [line width = 2pt, KITblue, dashed]
table {%
10.0 1785.7563327479525
11.829087159678172 1785.9021005775614
13.992730303126296 1786.0793899166765
16.552122635755094 1786.2959041222282
19.579650133602893 1786.5615274888823
23.1609387986393 1786.8890309923518
27.397276374907616 1787.295028528002
32.40847701765738 1787.8012770274338
38.33626993542959 1788.4364491355043
45.34830784431464 1789.2385554894367
53.64290860343152 1790.2582594963958
63.45466413686417 1791.563416402288
75.0610752763071 1793.2452869149104
88.79040017426006 1795.4270305634006
105.03093826040269 1798.275280837545
124.24201231450792 1802.0158432916423
146.9669592562123 1806.9548270107316
173.84849706346063 1813.5067812764973
205.64690243427285 1822.2315787825685
243.26151330128494 1833.8817067621674
287.7561643436108 1849.4610443998051
340.38927487552485 1870.2947852107009
402.6494400722235 1898.107616564104
476.29753214099406 1935.1036732123407
563.416502163544 1984.0381218101395
666.4702911293562 2048.2688400800716
788.3735163105241 2131.7807135435078
932.5739038819152 2239.1863489126063
1103.149799186051 2375.7226153144316
1304.925512475327 2547.2743154715395
1543.6077623958336 2760.4564983929995
1825.947076173611 3022.7751205681734
2159.9287113017167 3342.869507320242
2554.998498467936 3730.8390850773612
3022.3299931224074 4198.636312437418
3575.1404913954457 4760.565015126994
4229.064848081138 5433.90205822395
5002.597669188292 6239.735035674922
5917.616385363117 7204.189082279173
7000.0 8360.443419547324
};
\addplot [line width = 2pt, KITred, dotted]
table {%
10.0 1571.1946490292517
11.829087159678172 1571.881455292528
13.992730303126296 1572.731719752155
16.552122635755094 1573.786634020264
19.579650133602893 1575.098093695237
23.1609387986393 1576.7314704995438
27.397276374907616 1578.7690603271355
32.40847701765738 1581.314341269885
38.33626993542959 1584.4971836330824
45.34830784431464 1588.480150819701
53.64290860343152 1593.4660093495718
63.45466413686417 1599.7065217404897
75.0610752763071 1607.5125240346722
88.79040017426006 1617.265195184018
105.03093826040269 1629.4283285047147
124.24201231450792 1644.5613577581091
146.9669592562123 1663.3329343406235
173.84849706346063 1686.5350635225118
205.64690243427285 1715.0982206309861
243.26151330128494 1750.1084420604097
287.7561643436108 1792.8279887210374
340.38927487552485 1844.7216278375874
402.6494400722235 1907.4907353746366
476.29753214099406 1983.1172882401788
563.416502163544 2073.9195527350557
666.4702911293562 2182.62112294656
788.3735163105241 2312.4349231649358
932.5739038819152 2467.1658871165037
1103.149799186051 2651.3323786354867
1304.925512475327 2870.316403061841
1543.6077623958336 3130.546349873978
1825.947076173611 3439.725572419685
2159.9287113017167 3807.12495114202
2554.998498467936 4243.970812101357
3022.3299931224074 4763.986454491238
3575.1404913954457 5384.2076197413
4229.064848081138 6126.3544541044785
5002.597669188292 7019.538342763288
5917.616385363117 8106.959837459279
7000.0 9469.119610086727
};

\addplot [line width = 2pt, KITblack]
table {%
10.0 1571.2928186349861
11.829087159678172 1572.0423864321651
13.992730303126296 1572.9743659069918
16.552122635755094 1574.1344665351335
19.579650133602893 1575.579892428388
23.1609387986393 1577.3820965414538
27.397276374907616 1579.6300833338967
32.40847701765738 1582.4342995950692
38.33626993542959 1585.931113386276
45.34830784431464 1590.287816817017
53.64290860343152 1595.707998410399
63.45466413686417 1602.4370237177789
75.0610752763071 1610.7672648412681
88.79040017426006 1621.0426912988703
105.03093826040269 1633.6625060503127
124.24201231450792 1649.0838720569277
146.9669592562123 1667.824270540642
173.84849706346063 1690.464801302528
205.64690243427285 1717.6564334086913
243.26151330128494 1750.1316476831018
287.7561643436108 1788.723859632252
340.38927487552485 1834.3964910972886
402.6494400722235 1888.2828468385635
476.29753214099406 1951.7374772327823
563.416502163544 2026.3998664785784
666.4702911293562 2114.272325039455
788.3735163105241 2217.816020032205
932.5739038819152 2340.0723527562272
1103.149799186051 2484.8219413538063
1304.925512475327 2656.801624434729
1543.6077623958336 2862.0140831174717
1825.947076173611 3108.1911721062343
2159.9287113017167 3405.5249649431125
2554.998498467936 3767.8935779027433
3022.3299931224074 4215.069760399545
3575.1404913954457 4777.060972345831
4229.064848081138 5503.611820459653
5002.597669188292 6488.16254691177
5917.616385363117 7941.485176531251
7000.0 10501.16271854243
};
\addplot [white, only marks, mark = x,  mark size=6, mark options={line width=3pt}]
table {%
10 1460.9686069190782
30 1506.2621132107
100 1568.8452621285028
300 1617.4832016531925
1000 2086.109095285926
2000 3009.213757626229
4000 5430.085024385972
6000 6862.211327693474
};
\addplot [KITblack, only marks, mark = x,  mark size=5, mark options={line width=1.5pt}]
table {%
10 1460.9686069190782
30 1506.2621132107
100 1568.8452621285028
300 1617.4832016531925
1000 2086.109095285926
2000 3009.213757626229
4000 5430.085024385972
6000 6862.211327693474
};

\end{groupplot}

\end{tikzpicture}

%% file: 5_Results/Figures/Donnan_potentials.tex
\begin{tikzpicture}

\begin{axis}[
scale only axis,
name = DonnanPotential,
tick align=inside,
xlabel={$c_\mathrm{s}^\mathrm{b} \ / \ \frac{\mathrm{mol}}{\mathrm{m^3}}$},
xmin=0.7, xmax=1600,
xmode=log,
xtick={0.1,1,10,100,1000,10000},
xticklabels={$10^{-1}$,$10^{0}$,$10^{1}$,,$10^{3}$,$10^{4}$},
xlabel style = {yshift = 0.4cm},
ylabel={$\Phi_\mathrm{D} \ / \ \mathrm{V}$},
ylabel style = {yshift = -0.7cm, xshift = -0cm},
ymin= -0.22, ymax=0.02,
ytick={-0.2,-0.15, -0.1,-0.05,0},
yticklabels={$-0.2$,,,,$0$},
height = 4cm,
width = 6.5cm,
legend style={nodes={anchor=west}, anchor = south east, at={(0.97,0.05)}}
]
\addlegendimage{KITgrey, only marks, mark=x, mark size=4, mark options={line width=1.5pt}, legend image post style={xshift=-3.5pt}}
\addlegendentry{exp. \cite{Gokturk2022}}
\addlegendimage{line legend, color=KITgrey, line width = 2pt, legend image post style={scale=0.6}}
\addlegendentry{IOM}
\addlegendimage{line legend, color=KITgrey, dotted, line width = 2pt, legend image post style={scale=0.6}}
\addlegendentry{D-M}
\addlegendimage{line legend, color=KITgrey, dashed, line width = 2pt, legend image post style={scale=0.6}}
\addlegendentry{SDE}
\addlegendimage{line legend, color=KITgrey, dashdotted, line width = 2pt, legend image post style={scale=0.6}}
\addlegendentry{low-T*}

\addplot [line width = 2pt, KITblack, dashdotted]
table {%
1.0 -0.1528692528953514
1.2548512783530967 -0.14700007756604144
1.5746517307844012 -0.1411308317803688
1.9759537373357223 -0.13526150546472265
2.47952807326231 -0.1293920905074555
3.111438972445601 -0.12352258386113481
3.904393172091009 -0.11765299293006473
4.899432763191505 -0.11178334468331699
6.148059466095905 -0.10591370079192483
7.714900280421304 -0.10004418244609846
9.681052479253339 -0.09417501064887983
12.14828107939447 -0.0883065711490209
15.24428604227089 -0.08243951845016853
19.129311827723896 -0.07657494155294858
24.004441401034345 -0.07071462679303295
30.12200397823995 -0.06486147253396156
37.79863519865146 -0.059020139476418015
47.43166569903015 -0.053198062228583136
59.519686336844714 -0.047407004580948854
74.68835448696494 -0.04166540872822755
93.7227771060572 -0.03600184589911398
117.60814666233829 -0.03045985820355619
147.58073318397368 -0.02510423894947057
185.19187169619667 -0.020028053059704388
232.38825693857507 -0.015358107075917424
291.61270129361884 -0.011254170392901075
365.9305710022973 -0.007895494586449957
459.1884448107114 -0.0054507525538158704
576.2132069756917 -0.004037444575138724
723.0618793773842 -0.003688875053361019
907.3351236651032 -0.004347008670530299
1138.5706398258199 -0.005883017702506786
1428.7368228807334 -0.00813008970054711
1792.8522286220302 -0.010911057436922876
2249.7629109845534 -0.014052829789849917
2823.1178648403506 -0.017388087585416327
3542.5930616363794 -0.020747105857023038
4445.427432079221 -0.023940355139218192
5578.350295970535 -0.026727115145411158
7000.0 -0.028753359896796938
};
\addplot [line width = 2pt, KITblack, dashed]
table {%
1.0 -0.20733432758963244
1.2548512783530967 -0.20146585735172182
1.5746517307844012 -0.19559748315500572
1.9759537373357223 -0.18972922977503895
2.47952807326231 -0.18386112847334488
3.111438972445601 -0.17799321874928603
3.904393172091009 -0.17212555059459916
4.899432763191505 -0.16625818741200285
6.148059466095905 -0.1603912098175199
7.714900280421304 -0.15452472063242428
9.681052479253339 -0.14865885149476835
12.14828107939447 -0.14279377169131877
15.24428604227089 -0.13692970011665795
19.129311827723896 -0.13106692158907804
24.004441401034345 -0.12520580944310025
30.12200397823995 -0.11934685712475962
37.79863519865146 -0.11349072316574858
47.43166569903015 -0.10763829466246076
59.519686336844714 -0.10179078131611823
74.68835448696494 -0.0959498496854146
93.7227771060572 -0.09011782183532957
117.60814666233829 -0.08429796879924498
147.58073318397368 -0.07849493914342849
185.19187169619667 -0.07271538951850263
232.38825693857507 -0.06696889163495968
291.61270129361884 -0.06126921186597246
365.9305710022973 -0.05563603732475326
459.1884448107114 -0.05009714070306516
576.2132069756917 -0.04469074443005636
723.0618793773842 -0.03946740658774748
907.3351236651032 -0.034490113783680915
1138.5706398258199 -0.029830850372059618
1428.7368228807334 -0.025562472387402994
1792.8522286220302 -0.02174704806657274
2249.7629109845534 -0.01842479383937401
2823.1178648403506 -0.015608897607948372
3542.5930616363794 -0.01328888548192355
4445.427432079221 -0.011441490569507782
5578.350295970535 -0.010047109349044672
7000.0 -0.009115050742004742
};
\addplot [line width = 2pt, KITblack, dotted]
table {%
1.0 -0.19355977323999093
1.2548512783530967 -0.18769118165756005
1.5746517307844012 -0.18182265616606935
1.9759537373357223 -0.17595421446819665
2.47952807326231 -0.17008587927179997
3.111438972445601 -0.16421767984898147
3.904393172091009 -0.15834965415520635
4.899432763191505 -0.15248185174453147
6.148059466095905 -0.14661433783058878
7.714900280421304 -0.1407471990159982
9.681052479253339 -0.13488055147777905
12.14828107939447 -0.12901455280342733
15.24428604227089 -0.12314941929963176
19.129311827723896 -0.11728545156357996
24.004441401034345 -0.11142307260036748
30.12200397823995 -0.10556288507016016
37.79863519865146 -0.09970575777544707
47.43166569903015 -0.09385295686242667
59.519686336844714 -0.08800634525609793
74.68835448696494 -0.08216868661923421
93.7227771060572 -0.07634409982956608
117.60814666233829 -0.07053875241326278
147.58073318397368 -0.06476187097677123
185.19187169619667 -0.05902719822825569
232.38825693857507 -0.053354987608211585
291.61270129361884 -0.04777452353057026
365.9305710022973 -0.04232685485960999
459.1884448107114 -0.03706684333553313
576.2132069756917 -0.0320628291989598
723.0618793773842 -0.027391769724544584
907.3351236651032 -0.02312875326731266
1138.5706398258199 -0.019332971130107773
1428.7368228807334 -0.016035714458707263
1792.8522286220302 -0.0132360144911207
2249.7629109845534 -0.010905208902290823
2823.1178648403506 -0.008996968440613757
3542.5930616363794 -0.00745804051123824
4445.427432079221 -0.006237013443745113
5578.350295970535 -0.005290826370698917
7000.0 -0.004591349239533151
};
\addplot [line width = 2pt, KITblack]
table {%
1.0 -0.1764437172245635
1.2548512783530967 -0.1705751455179906
1.5746517307844012 -0.16470664739763805
1.9759537373357223 -0.15883824387235224
2.47952807326231 -0.1529699626005848
3.111438972445601 -0.14710184032894383
3.904393172091009 -0.14123392637910642
4.899432763191505 -0.1353662876949518
6.148059466095905 -0.12949901623212098
7.714900280421304 -0.12363223989101985
9.681052479253339 -0.11776613884411245
12.14828107939447 -0.11190097011721666
15.24428604227089 -0.10603710484985084
19.129311827723896 -0.10017508508327046
24.004441401034345 -0.09431571066124592
30.12200397823995 -0.08846017356129386
37.79863519865146 -0.08261025849499168
47.43166569903015 -0.07676866278954769
59.519686336844714 -0.07093947113184226
74.68835448696494 -0.06512888093870442
93.7227771060572 -0.059346281408178485
117.60814666233829 -0.05360582015812739
147.58073318397368 -0.047928582188738224
185.19187169619667 -0.042345405330289555
232.38825693857507 -0.036900054549051625
291.61270129361884 -0.03165184554599682
365.9305710022973 -0.02667586344055358
459.1884448107114 -0.02205820446955689
576.2132069756917 -0.017884483107402507
723.0618793773842 -0.014223294250378353
907.3351236651032 -0.011110848305463565
1138.5706398258199 -0.008544082455457907
1428.7368228807334 -0.006484917484828796
1792.8522286220302 -0.00487204806891531
2249.7629109845534 -0.003634194099281283
2823.1178648403506 -0.00270056421145025
3542.5930616363794 -0.0020072680726536548
4445.427432079221 -0.001500322997229424
5578.350295970535 -0.0011365355860534876
7000.0 -0.0008836137385108433
};
\addplot [line width = 2pt, KITred, dashdotted]
table {%
1.0 -0.08054322147565739
1.2548512783530967 -0.07760870470240343
1.5746517307844012 -0.07467425334165818
1.9759537373357223 -0.07173992032410656
2.47952807326231 -0.06880578676205362
3.111438972445601 -0.06587197507368503
3.904393172091009 -0.062938667848122
4.899432763191505 -0.06000613486839623
6.148059466095905 -0.0570747717012502
7.714900280421304 -0.05414515464035152
9.681052479253339 -0.05121811869834689
12.14828107939447 -0.04829486796267687
15.24428604227089 -0.04537713118022259
19.129311827723896 -0.04246738015872833
24.004441401034345 -0.03956913468155759
30.12200397823995 -0.036687385181927094
37.79863519865146 -0.033829173040946174
47.43166569903015 -0.0310043767213594
59.519686336844714 -0.0282267567360102
74.68835448696494 -0.025515306791272105
93.7227771060572 -0.02289592947792916
117.60814666233829 -0.020403381237503296
147.58073318397368 -0.018083282953794057
185.19187169619667 -0.015993739262125665
232.38825693857507 -0.014205750392134004
291.61270129361884 -0.01280122349815173
365.9305710022973 -0.011867249904956646
459.1884448107114 -0.011485828785017548
576.2132069756917 -0.011719717568752138
723.0618793773842 -0.012597267230606321
907.3351236651032 -0.014100633986055761
1138.5706398258199 -0.016160991267906016
1428.7368228807334 -0.018660977086450772
1792.8522286220302 -0.021440353041821632
2249.7629109845534 -0.02429780808060612
2823.1178648403506 -0.026979729426143885
3542.5930616363794 -0.02914119282565163
4445.427432079221 -0.030242235041805263
5578.350295970535 -0.02924933602162435
7000.0 -0.023481516701993624
};
\addplot [line width = 2pt, KITred, dashed]
table {%
1.0 -0.09510597133585715
1.2548512783530967 -0.09217197698078343
1.5746517307844012 -0.0892381044892337
1.9759537373357223 -0.08630439005375606
2.47952807326231 -0.08337088125112412
3.111438972445601 -0.08043764062805366
3.904393172091009 -0.07750475088414106
4.899432763191505 -0.07457232152012432
6.148059466095905 -0.07164049806419419
7.714900280421304 -0.06870947449955433
9.681052479253339 -0.06577951007587779
12.14828107939447 -0.06285095238728182
15.24428604227089 -0.05992426845546727
19.129311827723896 -0.05700008730416085
24.004441401034345 -0.05407925767731424
30.12200397823995 -0.05116292631775008
37.79863519865146 -0.04825264377905512
47.43166569903015 -0.045350506031997063
59.519686336844714 -0.04245934298264904
74.68835448696494 -0.039582964313799096
93.7227771060572 -0.036726474133395186
117.60814666233829 -0.03389665953411561
147.58073318397368 -0.03110244614432707
185.19187169619667 -0.02835538847926548
232.38825693857507 -0.025670120621370276
291.61270129361884 -0.02306463352148695
365.9305710022973 -0.02056018224640605
459.1884448107114 -0.01818059628011138
576.2132069756917 -0.01595082891619987
723.0618793773842 -0.013894786844015976
907.3351236651032 -0.012032812804310074
1138.5706398258199 -0.010379499002461046
1428.7368228807334 -0.008942600386488713
1792.8522286220302 -0.007723618941805867
2249.7629109845534 -0.00672034016557566
2823.1178648403506 -0.005931740304097457
3542.5930616363794 -0.005367089253472852
4445.427432079221 -0.005066636127209787
5578.350295970535 -0.005166544456823606
7000.0 -0.006212891966211085
};
\addplot [line width = 2pt, KITred, dotted]
table {%
1.0 -0.07912745194077904
1.2548512783530967 -0.07619379775061916
1.5746517307844012 -0.0732604205927251
1.9759537373357223 -0.07032742394501146
2.47952807326231 -0.06739495092892961
3.111438972445601 -0.06446319973220303
3.904393172091009 -0.061532445063410435
4.899432763191505 -0.05860306797088164
6.148059466095905 -0.055675597225652226
7.714900280421304 -0.052750766616885846
9.681052479253339 -0.04982959399223214
12.14828107939447 -0.046913489714619803
15.24428604227089 -0.04400440432817966
19.129311827723896 -0.04110502780604115
24.004441401034345 -0.03821905095332768
30.12200397823995 -0.03535150854720842
37.79863519865146 -0.032509201897195195
47.43166569903015 -0.02970120006605295
59.519686336844714 -0.02693937893770307
74.68835448696494 -0.024238909322412724
93.7227771060572 -0.021618532666234504
117.60814666233829 -0.019100387657787356
147.58073318397368 -0.016709120395238062
185.19187169619667 -0.014470099275133019
232.38825693857507 -0.012406818596933173
291.61270129361884 -0.010537960687872057
365.9305710022973 -0.00887489050880829
459.1884448107114 -0.007420332861896591
576.2132069756917 -0.0061685752376003464
723.0618793773842 -0.005106983203035696
907.3351236651032 -0.004218240109096279
1138.5706398258199 -0.0034826840856257927
1428.7368228807334 -0.0028802844321406187
1792.8522286220302 -0.0023921738332423497
2249.7629109845534 -0.0020016804828647066
2823.1178648403506 -0.001695192658314431
3542.5930616363794 -0.0014632814726404104
4445.427432079221 -0.0013033088367306448
5578.350295970535 -0.0012281160469399717
7000.0 -0.0013086533048155956
};

\addplot [line width = 2pt, KITred]
table {%
1.0 -0.07853230578754448
1.2548512783530967 -0.07559887987742026
1.5746517307844012 -0.07266583453209813
1.9759537373357223 -0.06973331796275016
2.47952807326231 -0.06680153681783452
3.111438972445601 -0.06387077934973523
3.904393172091009 -0.060941447741092246
4.899432763191505 -0.058014103656179414
6.148059466095905 -0.05508952883932475
7.714900280421304 -0.0521688149969894
9.681052479253339 -0.04925348157638307
12.14828107939447 -0.046345640577522385
15.24428604227089 -0.04344821978024817
19.129311827723896 -0.040565262034101095
24.004441401034345 -0.03770231887185635
30.12200397823995 -0.0348669536733616
37.79863519865146 -0.0320693577812659
47.43166569903015 -0.029323055084896025
59.519686336844714 -0.02664561714848211
74.68835448696494 -0.024059226423372843
93.7227771060572 -0.021590808080641493
117.60814666233829 -0.019271356434568596
147.58073318397368 -0.017134069472083912
185.19187169619667 -0.01521111942131983
232.38825693857507 -0.013529366995616673
291.61270129361884 -0.01210594110821551
365.9305710022973 -0.010944982384879258
459.1884448107114 -0.010036637463121924
576.2132069756917 -0.009358602033528332
723.0618793773842 -0.008879598225704957
907.3351236651032 -0.00856364067993332
1138.5706398258199 -0.00837396680129548
1428.7368228807334 -0.008275887931522475
1792.8522286220302 -0.008238249615459284
2249.7629109845534 -0.00823343481294766
2823.1178648403506 -0.008235736892574176
3542.5930616363794 -0.008217111859019143
4445.427432079221 -0.008136199933706754
5578.350295970535 -0.00790196942495138
7000.0 -0.007187497746283914
};


\addplot [ white, only marks, mark = x,  mark size=6, mark options={line width=3pt}, error bars/.cd, y dir=both, y explicit, error bar style={line width=3pt},error mark options={line width=6pt, mark size=1pt}]
table[x index=0, y index=1, y error index=2,] {%
1 -0.08 0.01
10 -0.066 0.01
30 -0.043 0.02
100 -0.035 0.02
300 -0.02 0.03
1000 -0.005 0.005
};

\addplot [ KITred, only marks, mark = x,  mark size=5, mark options={line width=1.5pt}, error bars/.cd, y dir=both, y explicit, error bar style={line width=1.5pt},error mark options={line width=5pt, mark size=0.5pt}]
table[x index=0, y index=1, y error index=2,] {%
1 -0.08 0.01
10 -0.066 0.01
30 -0.043 0.02
100 -0.035 0.02
300 -0.02 0.03
1000 -0.005 0.005
};

\addplot [ white, only marks, mark = x,  mark size=6, mark options={line width=3pt}, error bars/.cd, y dir=both, y explicit, error bar style={line width=3pt},error mark options={line width=6pt, mark size=1pt}]
table[x index=0, y index=1, y error index=2,] {%
1 -0.182 0.02
10 -0.113 0.015
30 -0.087 0.02
100 -0.065 0.02
300 -0.058 0.015
1000 -0.026 0.02
};
\addplot [ KITblack, only marks, mark = x,  mark size=5, mark options={line width=1.5pt}, error bars/.cd, y dir=both, y explicit, error bar style={line width=1.5pt},error mark options={line width=5pt, mark size=0.5pt}]
table[x index=0, y index=1, y error index=2,] {%
1 -0.182 0.02
10 -0.113 0.015
30 -0.087 0.02
100 -0.065 0.02
300 -0.058 0.015
1000 -0.026 0.02
};
\end{axis}

\begin{axis}[
    scale only axis,
    hide axis,
    height = 4cm,
    width = 6.5cm,
    xmin=0, xmax=1,
    ymin=0, ymax=1,
    at={(DonnanPotential.south west)},
    legend pos=north west,
    legend style= {nodes={anchor=west}, at={(0.03,0.95)}}
  ]
\addlegendimage{only marks, mark=square*, color=KITblack}
\addlegendentry{$\ \mathrm{NaCl}$}
\addlegendimage{only marks, mark=square*, color=KITred}
\addlegendentry{$\ \mathrm{MgCl_2}$}
\end{axis}

\end{tikzpicture}

%% file: 5_Results/Figures/Permeability.tex
\begin{tikzpicture}

\begin{axis}[
scale only axis,
name = permeability,
tick align=inside,
xlabel={$c_\mathrm{s}^\mathrm{b} \ / \ \frac{\mathrm{mol}}{\mathrm{m^3}}$},
xmin=7, xmax=3000,
xmode=log,
xtick={1,10,100,1000,10000},
xticklabels={$10^{0}$,$10^{1}$,,$10^{3}$,$10^{4}$},
xlabel style = {yshift = 0.4cm},
ylabel={$P_\mathrm{s} \ / \ \frac{\mathrm{m^2}}{\mathrm{s}}$},
ylabel style = {yshift = -0.7cm, xshift = 0.1cm},
ymin= 7e-13, ymax=1e-10,
ytick={1e-12, 1e-11, 1e-10},
ymode = log,
yticklabels={$10^{-12}$,,$10^{-10}$},
height = 4cm,
width = 6.5cm,
legend style={nodes={anchor=west}, anchor = south east, at={(0.95,0.05)}}
]

\addlegendimage{KITgrey, only marks, mark=x, mark size=4, mark options={line width=1.5pt}, legend image post style={xshift=-3.5pt}}
\addlegendentry{exp. \cite{Kamcev2017}}
\addlegendimage{line legend, color=KITgrey, line width = 2pt, legend image post style={scale=0.6}}
\addlegendentry{IOM}
\addlegendimage{line legend, color=KITgrey, dotted, line width = 2pt, legend image post style={scale=0.6}}
\addlegendentry{D-M}
\addlegendimage{line legend, color=KITgrey, dashed, line width = 2pt, legend image post style={scale=0.6}}
\addlegendentry{SDE}
\addlegendimage{line legend, color=KITgrey, dashdotted, line width = 2pt, legend image post style={scale=0.6}}
\addlegendentry{low-T*}

\addplot [line width = 2pt, KITblack, dashdotted]
table {%
10.0 1.2455875210009252e-12
11.829087159678172 1.4733235446100142e-12
13.992730303126296 1.7426406863043124e-12
16.552122635755094 2.0610910013812085e-12
19.579650133602893 2.4375708284182295e-12
23.1609387986393 2.8825416631183665e-12
27.397276374907616 3.40827485070636e-12
32.40847701765738 4.02911363773878e-12
38.33626993542959 4.761738042795112e-12
45.34830784431464 5.625404217919193e-12
53.64290860343152 6.642107062394885e-12
63.45466413686417 7.836577731622345e-12
75.0610752763071 9.235969236917709e-12
88.79040017426006 1.0868994648790168e-11
105.03093826040269 1.2764154920483368e-11
124.24201231450792 1.4946525528406146e-11
146.9669592562123 1.7432384985683838e-11
173.84849706346063 2.0220843040666446e-11
205.64690243427285 2.3281754801725684e-11
243.26151330128494 2.6539959626164064e-11
287.7561643436108 2.985780092492948e-11
340.38927487552485 3.302139049710684e-11
402.6494400722235 3.574063314768534e-11
476.29753214099406 3.7675756641604754e-11
563.416502163544 3.849845712027535e-11
666.4702911293562 3.797935329548742e-11
788.3735163105241 3.607045473713011e-11
932.5739038819152 3.294033842785351e-11
1103.149799186051 2.89378575168931e-11
1304.925512475327 2.4499474181328672e-11
1543.6077623958336 2.0044359858359116e-11
1825.947076173611 1.5898739375425702e-11
2159.9287113017167 1.2265809233177004e-11
2554.998498467936 9.233493944344318e-12
3022.3299931224074 6.802346297912286e-12
3575.1404913954457 4.917956139377018e-12
4229.064848081138 3.498779452531855e-12
5002.597669188292 2.456110778338922e-12
5917.616385363117 1.7063812196314342e-12
7000.0 1.1774254441158798e-12
};
\addplot [line width = 2pt, KITblack, dashed]
table {%
10.0 4.325533567133773e-13
11.829087159678172 5.116150170523257e-13
13.992730303126296 6.051145537437307e-13
16.552122635755094 7.156830071170709e-13
19.579650133602893 8.464284221207902e-13
23.1609387986393 1.0010211877302183e-12
27.397276374907616 1.1837940813139624e-12
32.40847701765738 1.3998592594423219e-12
38.33626993542959 1.6552446001226783e-12
45.34830784431464 1.9570518868828464e-12
53.64290860343152 2.313639171269238e-12
63.45466413686417 2.7348292442431e-12
75.0610752763071 3.232145015939588e-12
88.79040017426006 3.819070620894704e-12
105.03093826040269 4.511333442494698e-12
124.24201231450792 5.3271960967566276e-12
146.9669592562123 6.287737443761264e-12
173.84849706346063 7.417085858337825e-12
205.64690243427285 8.742543825543995e-12
243.26151330128494 1.0294507490146361e-11
287.7561643436108 1.2106035696142366e-11
340.38927487552485 1.4211861006455786e-11
402.6494400722235 1.664656872607482e-11
476.29753214099406 1.9441624218363038e-11
563.416502163544 2.2620956798603905e-11
666.4702911293562 2.619499697553871e-11
788.3735163105241 3.015351551576472e-11
932.5739038819152 3.445837171812115e-11
1103.149799186051 3.903818944732228e-11
1304.925512475327 4.378756474461231e-11
1543.6077623958336 4.857297602821223e-11
1825.947076173611 5.324573029511817e-11
2159.9287113017167 5.765959052934904e-11
2554.998498467936 6.168864415870751e-11
3022.3299931224074 6.524077505595344e-11
3575.1404913954457 6.826379247196742e-11
4229.064848081138 7.07436404412303e-11
5002.597669188292 7.269581633641538e-11
5917.616385363117 7.415137903816115e-11
7000.0 7.513702444784088e-11
};
\addplot [line width = 2pt, KITblack]
table {%
10.0 8.546080093132e-13
13.372338725601036 1.1425045051371781e-12
17.881944299220915 1.5271539254913748e-12
23.912341624151257 2.0407900507570208e-12
31.976393192043943 2.7260277434735694e-12
42.75991609870146 3.6387368452036176e-12
57.1800081950117 4.851078195006825e-12
76.46304379163396 6.4537465010342315e-12
102.24897215721943 8.55501406758861e-12
136.7307890030889 1.1271366119160087e-11
182.8410424767988 1.4700376059329787e-11
244.50123529417635 1.8864357243510112e-11
326.95533371816055 2.3625273343346796e-11
437.21574706211686 2.8614359479134034e-11
584.6597065881333 3.327207685766815e-11
781.8267635707035 3.704959007556915e-11
1045.4852307207843 3.964711346845454e-11
1398.058263781148 4.109292734905953e-11
1869.5308661407196 4.163902394387207e-11
2500.0 4.1604413088515155e-11
};
\addplot [line width = 2pt, KITblack, dotted]
table {%
10.0 7.370540346751133e-13
11.52087968577631 8.490869315127015e-13
13.273066873413322 9.781345919777999e-13
15.2917406509858 1.126775238037734e-12
17.61743042261021 1.297975230584785e-12
20.296829627142756 1.4951463479870631e-12
23.383733213701174 1.722210960816779e-12
26.94011769593426 1.983675921831215e-12
31.037385469551186 2.284715908103365e-12
35.7577983755761 2.63126680891299e-12
41.19612929132598 3.0301293556005832e-12
47.46156490850518 3.4890826985459896e-12
54.67989790095513 4.0170065272155694e-12
62.99605249474363 4.624008835949526e-12
72.57699414707902 5.321553959883148e-12
83.61508175237887 6.122581787192918e-12
96.33192967855071 7.0416035804533456e-12
110.98285717252473 8.094751961364225e-12
127.86201446683529 9.299751646820188e-12
147.30828850533985 1.0675762937591653e-11
169.71210685876466 1.224303182608702e-11
195.52327643394406 1.4022260295838081e-11
225.26001435642507 1.603359237453994e-11
259.51935234166183 1.829510484387229e-11
298.9891234458878 2.0820712288473063e-11
344.4617718575792 2.3617467248431037e-11
396.8502629920495 2.6682382472180304e-11
457.2064133200091 2.999913349570929e-11
526.7420079425143 3.353528400544161e-11
606.8531298949932 3.7240910784052954e-11
699.1481896457002 4.1049522107183755e-11
805.4802175436434 4.4881807504824263e-11
927.9840675593239 4.8652027101992916e-11
1069.119279266829 5.227599533101553e-11
1231.719458617701 5.567904737346038e-11
1419.049168936407 5.880239677968791e-11
1634.8694743517199 6.160685522864509e-11
1883.513451595453 6.407365843395523e-11
2169.9731862372464 6.620277066988983e-11
2500.0 6.800937632053919e-11
};
\addplot [line width = 2pt, KITred, dashdotted]
table {%
10.0 9.04821976839688e-12
11.829087159678172 9.840670987510138e-12
13.992730303126296 1.0702143456427435e-11
16.552122635755094 1.1638439160688678e-11
19.579650133602893 1.2655746759383069e-11
23.1609387986393 1.3760619888853837e-11
27.397276374907616 1.4959929253527905e-11
32.40847701765738 1.6260775839598247e-11
38.33626993542959 1.7670347149619856e-11
45.34830784431464 1.9195690705503377e-11
53.64290860343152 2.0843368400070845e-11
63.45466413686417 2.2618940620397622e-11
75.0610752763071 2.4526209413495734e-11
88.79040017426006 2.6566124697029424e-11
105.03093826040269 2.8735227435998138e-11
124.24201231450792 3.10234732689051e-11
146.9669592562123 3.341126172780874e-11
173.84849706346063 3.586551780467205e-11
205.64690243427285 3.833478771711201e-11
243.26151330128494 4.074360725204523e-11
287.7561643436108 4.2986993271346804e-11
340.38927487552485 4.492688002697146e-11
402.6494400722235 4.6393573475526297e-11
476.29753214099406 4.719630128090059e-11
563.416502163544 4.714654876967816e-11
666.4702911293562 4.609462563475011e-11
788.3735163105241 4.3973269359145005e-11
932.5739038819152 4.083431203276779e-11
1103.149799186051 3.6860996505140074e-11
1304.925512475327 3.234475350536445e-11
1543.6077623958336 2.7630411554581598e-11
1825.947076173611 2.3049095476899752e-11
2159.9287113017167 1.886301297717909e-11
2554.998498467936 1.5238210291921862e-11
3022.3299931224074 1.2247210585401258e-11
3575.1404913954457 9.892946234532681e-12
4229.064848081138 8.143709117861641e-12
5002.597669188292 6.976916750087518e-12
5917.616385363117 6.4564464684703825e-12
7000.0 7.007701467529723e-12
};
\addplot [line width = 2pt, KITred]
table {%
10.0 1.0679340851018265e-11
13.372338725601036 1.1736653347994271e-11
17.881944299220915 1.2977449780668782e-11
23.912341624151257 1.4416073668305858e-11
31.976393192043943 1.6058547625173383e-11
42.75991609870146 1.7895713041282873e-11
57.1800081950117 1.9893897632418197e-11
76.46304379163396 2.1984662807182333e-11
102.24897215721943 2.4058281765146146e-11
136.7307890030889 2.5968914852500652e-11
182.8410424767988 2.7558684587265994e-11
244.50123529417635 2.869826317407234e-11
326.95533371816055 2.932688331811157e-11
437.21574706211686 2.946983596894764e-11
584.6597065881333 2.922501769783148e-11
781.8267635707035 2.872851625470086e-11
1045.4852307207843 2.8117375827736154e-11
1398.058263781148 2.7503525490707816e-11
1869.5308661407196 2.696212465598901e-11
2500.0 2.6530986696024847e-11
};
\addplot [line width = 2pt, KITred, dotted]
table {%
10.0 1.1323130740208386e-11
11.52087968577631 1.1829803688269533e-11
13.273066873413322 1.2375236750452502e-11
15.2917406509858 1.2960185970858208e-11
17.61743042261021 1.3584869501729421e-11
20.296829627142756 1.424882218873402e-11
23.383733213701174 1.4950741236801574e-11
26.94011769593426 1.568833521301297e-11
31.037385469551186 1.645819629044696e-11
35.7577983755761 1.725572231989894e-11
41.19612929132598 1.8075120918678228e-11
47.46156490850518 1.8909528388890704e-11
54.67989790095513 1.9751268280725766e-11
62.99605249474363 2.0592255872643887e-11
72.57699414707902 2.1424525442776663e-11
83.61508175237887 2.2240822829678516e-11
96.33192967855071 2.3035176458824874e-11
110.98285717252473 2.3803347035498964e-11
127.86201446683529 2.454306868706932e-11
147.30828850533985 2.5254031939917256e-11
169.71210685876466 2.5937610678305802e-11
195.52327643394406 2.6596386393060923e-11
225.26001435642507 2.7233557330767785e-11
259.51935234166183 2.7852330886447083e-11
298.9891234458878 2.8455385688948754e-11
344.4617718575792 2.9044467010877064e-11
396.8502629920495 2.9620158751131486e-11
457.2064133200091 3.018186739728353e-11
526.7420079425143 3.072803368759356e-11
606.8531298949932 3.1256497115400876e-11
699.1481896457002 3.176482322931389e-11
805.4802175436434 3.225046090042305e-11
927.9840675593239 3.271079450161995e-11
1069.119279266829 3.3143202728240766e-11
1231.719458617701 3.354513448789554e-11
1419.049168936407 3.391414609053863e-11
1634.8694743517199 3.424783895380521e-11
1883.513451595453 3.4543644689815944e-11
2169.9731862372464 3.479840039096176e-11
2500.0 3.5007631357596945e-11
};
\addplot [white, only marks, mark = x,  mark size=6, mark options={line width=3pt}]
table {%
10 1.229654966016453e-11
30 1.5669186826535668e-11
100 2.0938686423873617e-11
300 2.4147320900183272e-11
1000 2.6453423218552183e-11
};
\addplot [KITred, only marks, mark = x,  mark size=5, mark options={line width=1.5pt}]
table {%
10 1.229654966016453e-11
30 1.5669186826535668e-11
100 2.0938686423873617e-11
300 2.4147320900183272e-11
1000 2.6453423218552183e-11
};
\addplot [white, only marks, mark = x,  mark size=6, mark options={line width=3pt}]
table {%
10 2.233638507312483e-12
30 4.061612072208954e-12
100 8.858666991206468e-12
300 1.8814515292095255e-11
1000 3.457616133680615e-11
};
\addplot [KITblack, only marks, mark = x,  mark size=5, mark options={line width=1.5pt}]
table {%
10 2.233638507312483e-12
30 4.061612072208954e-12
100 8.858666991206468e-12
300 1.8814515292095255e-11
1000 3.457616133680615e-11
};

\end{axis}
\begin{axis}[
    scale only axis,
    hide axis,
    height = 4cm,
    width = 6.5cm,
    xmin=0, xmax=1,
    ymin=0, ymax=1,
    at={(permeability.south west)},
    legend pos=north west,
    legend style= {nodes={anchor=west}, at={(0.03,0.95)}}
  ]
\addlegendimage{only marks, mark=square*, color=KITblack}
\addlegendentry{$\ \mathrm{NaCl}$}
\addlegendimage{only marks, mark=square*, color=KITred}
\addlegendentry{$\ \mathrm{MgCl_2}$}
\end{axis}

\end{tikzpicture}

%% file: 5_Results/Figures/Mixture_qeff.tex
\begin{tikzpicture}
\def\qeffmin{-1000}
\def\qeffmax{1000}

\def\ccomin{1}
\def\ccomax{10000}

\pgfmathsetmacro{\ccominlog}{log10(\ccomin)}
\pgfmathsetmacro{\ccomaxlog}{log10(\ccomax)}

\pgfplotsset{
  colormap={custom_qeff}{
    color(\qeffmin) = (black!50!blue)
    color(0.6*\qeffmin) = (blue)
    color(0)    = (white)
    color(0.6*\qeffmax)  = (red) 
    color(\qeffmax)  = (red!70!black)
  },
}

\begin{groupplot}[
    group style={
        group size=2 by 1,
        horizontal sep= 4pt,
    },
    tick align=inside,
    width = 3.4cm,
    height = 3cm,
    scale only axis,
    xmin = 5, xmax = 5000,
    ymin = 5, ymax = 5000,
    xmode = log,
    ymode = log,
    axis on top,
    view={0}{90},  
    xtick = {1,10,100,1000,10000},
    ytick = {1,10,100,1000,10000},
    yticklabels = {,$10^{1}$,, $10^{3}$,},
    tick style={color=black!85},
    point meta min=\qeffmin,
    point meta max=\qeffmax,
    colormap name=custom_qeff,
]

\nextgroupplot[
    xticklabels = {,$10^{1}$,$10^{2}$,,},
    xlabel = $c_\mathrm{NaCl}^\mathrm{b}$,
    ylabel = $c_\mathrm{MgCl_2}^\mathrm{b}$,
    xlabel style = {yshift = 0.5cm, xshift = 1.75cm},
    ylabel style = {yshift = -0.6cm, xshift = -0.2cm},
    legend style={anchor = south west, at={(0,1)}, font=\small},
]

\addlegendimage{empty legend}
\addlegendentry{IOM}
\addplot3[surf,shader=interp,draw=none,point meta=\thisrow{z},]
    table {5_Results/Figures/Mixture_qeff_IOM.tex};
    
\nextgroupplot[
    ylabel = $c_\mathrm{MgCl_2}^\mathrm{b}$,
    xticklabels = {,,$10^{2}$,$10^{3}$,,},
    yticklabel pos=right,
    ylabel style = {yshift = 0.75cm, xshift = -0.2cm},
    legend style={anchor = south east, at={(1,1)}, font=\small},
    colorbar horizontal,
    colorbar style={
        at={(-0.02,1.085)},
        anchor=south,
        title=$q_\mathrm{eff}$,
        title style={at={(axis description cs:0.5,0.7)},anchor=south},
        xtick={-1000,-500,0,500,1000},
        xticklabels={$-1000$,$-500$,,$500$,$1000$},
        xticklabel pos=upper,
        width = 4.5cm,
        height = 0.3cm,
    },
]

\addlegendimage{empty legend}
\addlegendentry{D-M}


\path [color of colormap={(-307.48 - \qeffmin) / (\qeffmax - \qeffmin) * 1000},draw=none,fill=., ] (axis cs:5,5) rectangle (axis cs:5000,5000);
    
\end{groupplot}
\end{tikzpicture}

%% file: 5_Results/Figures/Mixture_partitioning.tex
\begin{tikzpicture}
\def\ccomin{1}
\def\ccomax{10000}

\pgfmathsetmacro{\ccominlog}{log10(\ccomin)}
\pgfmathsetmacro{\ccomaxlog}{log10(\ccomax)}

\def\cdiffmin{0.75}
\def\cdiffmax{1.25}
\pgfmathsetmacro{\cbluepos}{1 + 0.6*(\cdiffmin-1)}
\pgfmathsetmacro{\credpos}{1 + 0.6*(\cdiffmax-1)}

\pgfplotsset{
  colormap={custom_c_co}{
    rgb(0)    = (0.002, 0.000, 0.014)
    rgb(0.25)    = (0.103, 0.063, 0.258)
    rgb(0.5) = (0.421, 0.113, 0.505)
    rgb(0.75) = (0.858, 0.280, 0.415)
    rgb(1) = (0.987, 0.991, 0.750)
  }
}

\pgfplotsset{
  colormap={custom_c_co_diff}{
    color(\cdiffmin) = (black!50!blue)
    color(\cbluepos) = (blue)
    color(1)    = (white)
    color(\credpos)  = (red) 
    color(\cdiffmax)  = (black!50!red)
  }
}

\pgfplotsset{
  colormap={custom_c_co_diff}{
    rgb255(0.5)  = ( 64,   0,  75) 
    rgb255(0.6)  = (118,  42, 131)
    rgb255(0.7)  = (153, 112, 171)
    rgb255(0.8)  = (194, 165, 207)
    rgb255(0.9)  = (231, 212, 232)
    rgb255(1)  = (255,255,255) 
    rgb255(1.1)  = (217, 240, 211)
    rgb255(1.2)  = (166, 219, 160)
    rgb255(1.3)  = (102, 194, 164)
    rgb255(1.4)  = ( 27, 120,  55)
    rgb255(1.5) = (  0,  68,  27)
  },
}

\begin{groupplot}[
    group style={
        group size=2 by 1,
        horizontal sep= 4pt,
    },
    tick align=inside,
    width = 3.4cm,
    height = 3cm,
    scale only axis,
    xmin = 5, xmax = 5000,
    ymin = 5, ymax = 5000,
    xmode = log,
    ymode = log,
    axis on top,
    view={0}{90},  
    xtick = {1,10,100,1000,10000},
    yticklabels = {,$10^{1}$,, $10^{3}$,},
    tick style={color=black!85},
    point meta min=\ccominlog,
    point meta max=\ccomaxlog,
    colormap name=custom_c_co,
]

\nextgroupplot[
    xticklabels = {,$10^{1}$,$10^{2}$,,},
    xlabel = $c_\mathrm{NaCl}^\mathrm{b}$,
    ylabel = $c_\mathrm{MgCl_2}^\mathrm{b}$,
    xlabel style = {yshift = 0.5cm, xshift = 1.75cm},
    ylabel style = {yshift = -0.6cm, xshift = -0.2cm},
    legend style={anchor = south west, at={(0,1)}, font=\small},
    colorbar horizontal,
    colorbar style={
        at={(1,1.1)},
        anchor=south east,
        title=$c_\mathrm{co}^\mathrm{m} \ / \ \mathrm{\frac{mol}{m^3}}$,
        title style={at={(axis description cs:0.5,0.5)},anchor=south},
        xtick={0,4},
        xticklabels = {$10^{0}$, $10^{4}$ },
        xticklabel pos=upper,
        width = 2.2cm,
        height = 0.3cm,
        font = \small,
    },
]

\addlegendimage{empty legend}
\addlegendentry{IOM}

\addplot3[surf,shader=interp,draw=none,point meta={log10(\thisrow{z})}]
    table {5_Results/Figures/Mixture_c_co_IOM.tex};
    
 \nextgroupplot[
    colormap name=custom_c_co_diff,
    point meta min=\cdiffmin,
    point meta max=\cdiffmax,
    ylabel = $c_\mathrm{MgCl_2}^\mathrm{b}$,
    xticklabels = {,,$10^{2}$,$10^{3}$,,},
    yticklabel pos=right,
    ylabel style = {yshift = 0.75cm, xshift = -0.2cm},
    colorbar horizontal,
    colorbar style={
        at={(1,1.1)},
        anchor=south east,
        xtick={0.75,1.25},
        xticklabels = {$0.75$,$1.25$},
        xticklabel pos=upper,
        title={$\eta_\mathrm{\,D\text{-M}}^\mathrm{\,IOM}$},
        title style={at={(axis description cs:0.5,0.5)},anchor=south},
        width = 2.3cm,
        height = 0.3cm,
        font = \small,
    },
    legend style={anchor = south west, at={(0,1)}, font=\small},
]

\addlegendimage{empty legend}
\addlegendentry{\raisebox{1pt}{$\frac{\text{IOM}}{\text{D-M}}$}}

\addplot3[surf,shader=interp,draw=none,point meta=\thisrow{z},]
    table {5_Results/Figures/Mixture_c_co_IOM_DM_diff.tex};
    
\end{groupplot}
\end{tikzpicture}

%% file: 5_Results/Figures/Mixture_permeability.tex
\begin{tikzpicture}
\def\permeamin{-10e-10}
\def\permeamax{2e-10}

\pgfmathdeclarefunction{permeanorm}{1}{%
  \pgfmathparse{(#1<0) ? (#1/abs(\permeamin)) : (#1/\permeamax)}%
}


\pgfplotsset{
  colormap={custom_permea}{
    color(-1) = (black!50!blue)
    color(-0.5) = (blue)
    color(0)    = (white)
    color(0.5)  = (red) 
    color(1)  = (black!50!red)
  },
}

\pgfplotsset{
  colormap={custom_permea_diff}{
    rgb255(0.5)  = ( 64,   0,  75) 
    rgb255(0.6)  = (118,  42, 131)
    rgb255(0.7)  = (153, 112, 171)
    rgb255(0.8)  = (194, 165, 207)
    rgb255(0.9)  = (231, 212, 232)
    rgb255(1)  = (255,255,255) 
    rgb255(1.2)  = (217, 240, 211)
    rgb255(1.4)  = (166, 219, 160)
    rgb255(1.6)  = (102, 194, 164)
    rgb255(1.8)  = ( 27, 120,  55)
    rgb255(2) = (  0,  68,  27)
  },
}
\begin{groupplot}[
    group style={
        group size=2 by 2,
        horizontal sep= 4pt,
        vertical sep= 4pt,
    },
    tick align=inside,
    width = 3.4cm,
    height = 3cm,
    scale only axis,
    xmin = 5, xmax = 5000,
    ymin = 5, ymax = 5000,
    xmode = log,
    ymode = log,
    view={0}{90},  
    xtick = {1,10,100,1000,10000},
    ytick = {1,10,100,1000,10000},
    tick style={color=black!85},
]

\nextgroupplot[
    name = topleft,
    colormap name=custom_permea,
    point meta min=-1,
    point meta max=1,
    xticklabels = {},
    yticklabels = {,,$10^{2}$, $10^{3}$,},
    ylabel = $c_\mathrm{MgCl_2}^\mathrm{b}$,
    ylabel style = {yshift = -0.6cm, xshift = -1.6cm},
    colorbar horizontal,
    colorbar style={
        name = topleftcolorbar,
        at={(0,1.05)},
        anchor=south west,
        title=$P_i^\mathrm{\,IOM} \ / \ \mathrm{\frac{m^2}{s}}$,
        title style={at={(axis description cs:1.04,1.4)},anchor=south},
        xtick={-1,0},
        xticklabels = {$-10^{-9}$,$\hspace{2pt} 0$},
        xticklabel pos=upper,
        width = 1.35cm,
        height = 0.3cm,
        xmin=-1,
        xmax=0,
    },
]

\addplot3[surf,shader=interp,draw=none,point meta={permeanorm(\thisrow{z})},]
    table {5_Results/Figures/Mixture_P_Na_IOM.tex};

\nextgroupplot[
    name = topright,
    colormap name=custom_permea_diff,
    point meta min=0.5,
    point meta max=2,
    ylabel = $c_\mathrm{MgCl_2}^\mathrm{b}$,
    xticklabels = {},
    yticklabels = {,,$10^{2}$, $10^{3}$,},
    yticklabel pos=right,
    ylabel style = {yshift = 0.75cm, xshift = -1.6cm},
    colorbar horizontal,
    colorbar style={
        title = $P_i^\mathrm{\,IOM} \ / \ P_i^\mathrm{\,D\text{-}M}$,
        title style={at={(0.5,1.4)},anchor=south},
        at={(1,1.05)},
        anchor=south east,
        xtick={0.5,1,1.5,2},
        xticklabels = {$0.5$,$1$,$1.5$,$2$},
        xticklabel pos=upper,
        width = 2.8cm,
        height = 0.3cm,
    },
]

\addplot3[surf,shader=interp,draw=none,point meta=\thisrow{z},]
    table {5_Results/Figures/Mixture_P_Na_IOM_DM_diff.tex};

\coordinate (toprowmiddlelabel) at (rel axis cs:-0.02,1);
\coordinate (rightmiddlelabel) at (rel axis cs:1,-0.02);

\nextgroupplot[
    colormap name=custom_permea,
    point meta min=-1,
    point meta max=1,
    xticklabels = {,$10^{1}$,$10^{2}$,,},
    yticklabels = {,$10^{1}$, $10^{2}$,,},
    xlabel = $c_\mathrm{NaCl}^\mathrm{b}$,
    xlabel style = {yshift = 0.5cm, xshift = 1.75cm},
    ylabel style = {yshift = -0.6cm, xshift = -0.2cm},
    colorbar horizontal,
    colorbar style={
        at={(topleftcolorbar.south east)},
        anchor=south west,
        yshift = 0.3cm,
        xshift = 4pt,
        xtick={1},
        xticklabels = {$10^{-10}$},
        xticklabel pos=upper,
        width = 1.35cm,
        height = 0.3cm,
        xmin=0,
        xmax=1,
    },
    legend style={anchor = south west, at={(1,1.1)}, font=\small},
]

\addplot3[surf,shader=interp,draw=none,point meta={permeanorm(\thisrow{z})},]
    table {5_Results/Figures/Mixture_P_Mg_IOM.tex};
  
\nextgroupplot[
    colormap name=custom_permea_diff,
    point meta min=0.5,
    point meta max=2,
    xticklabels = {,,$10^{2}$,$10^{3}$,,},
    yticklabels = {,$10^{1}$, $10^{2}$,,},
    yticklabel pos=right,
    ylabel style = {yshift = 0.65cm, xshift = -0.2cm},
]

\addplot3[surf,shader=interp,draw=none,point meta=\thisrow{z},]
    table {5_Results/Figures/Mixture_P_Mg_IOM_DM_diff.tex};

\coordinate (bottomrowmiddlelabel) at (rel axis cs:-0.02,1);
    
\end{groupplot}
\node[
    anchor=north,
    fill=white,
    draw=black,
    inner sep=4pt,
    font=\small
] at (toprowmiddlelabel) {$\mathrm{Na^+}$};
\node[
    anchor=north,
    fill=white,
    draw=black,
    inner sep=4pt,
    font=\small
] at (bottomrowmiddlelabel) {$\mathrm{Mg^{2+}}$};
\node[
    anchor=west,
    fill=white,
    yshift = -2pt,
    draw=black,
    inner sep=4pt,
    inner ysep=6pt,
    font=\small
] at (topleft.south west) {IOM};
\node[
    anchor=east,
    fill=white,
    yshift = -2pt,
    draw=black,
    inner sep = 4pt,
    font=\small
] at (topright.south east) {$\frac{\text{IOM}}{\text{D-M}}$};
\end{tikzpicture}

%% file: 6_Conclusion/0_main.tex
\section{Conclusion and outlook}
Through a consistent derivation from fundamental thermodynamics, combining mass-action site occupation with mean-field electrostatic contributions, we obtain a model with broad structural applicability for ion exchange membranes. 
The main benefit of this model is its internal consistency: It links static and kinetic properties on the microscale through a thermodynamic framework and couples them coherently to mesoscale porosity-tortuosity effects, yielding a unified description of macroscopic partitioning and transport.\par
In the low-salt limit, the IOM replicates the saturation of the effective polymer charge predicted by Manning theory. For salt mixtures and elevated concentrations, the IOM remains structurally applicable, whereas classical Donnan–Manning approaches require empirical corrections. Also, the IOM recovers correlations similar to these empirical corrections, which supports the consistency of the unified approach. However, systematic experimental data for mixed electrolytes, especially mixtures with counterions of different valence, are still needed to validate these predictions directly.\par
Overall, the broad structural applicability of this model makes it a promising basis for theory-driven membrane optimization, supporting the targeted development of ion-exchange membranes for various technological applications. In particular, the IOM could be integrated into larger models, such as battery models to analyze the membrane's influence on rate capability, crossover, and cell degradation.

%% file: 7_Appendix/0_main.tex
\setcounter{figure}{0}
\renewcommand{\thefigure}{S\arabic{figure}}
\setcounter{section}{0}
\renewcommand{\thesection}{S}
\setcounter{equation}{0}
\renewcommand{\thesubsection}{\thesection\arabic{subsection}}
\renewcommand{\theequation}{S\arabic{equation}}
\setcounter{table}{0}
\renewcommand{\thetable}{S\arabic{table}}
\section*{Supporting Information}
\label{C:Appendix}
\subsection{SDE: hindrance factors}
\label{C:SDE_hindrance_factors}
The dependence of hindrance factors $k$ (see eq. \ref{E:Nernst_Planck_SDE}) in cylindrical and slit pores on the particle to pore size ratio $\lambda = r_i/r_p$ is well known \cite{Dechadilok2006, Deen1987}. However, many existing fits are not valid within the entire relevant parameter range $\lambda \in \left[ 0,\ 1\right]$. To obtain useful fits for the whole range, the available data \cite{Dechadilok2006, Deen1987, Dufresne2001, Brenner1977, Ennis1996, Ganatos1980, Happel1983, Higdon1995, Mavrovouniotis1988, Nitsche1994, Pawar1993, Weinbaum1981} can be refitted with meaningful conditions for the extreme values enforced as shown in table \ref{T:Extreme_value_conditions_k_refit}.
\begin{table}[h]
    \centering
    \resizebox{0.9\linewidth}{!}{
    \begin{tabular}{|c|c|c|c|c|}
        \hline \rule{0pt}{1.3em}
        & $\left.k^\mathrm{d}\right|_{\lambda = 0}$ 
        & $\left.k^\mathrm{d}\right|_{\lambda = 1}$
        & $\left.\frac{\partial k^\mathrm{d}}{\partial \lambda}\right|_{\lambda = 0}$
        & $\left.\frac{\partial k^\mathrm{d}}{\partial \lambda}\right|_{\lambda = 1}$
        \\[0.6em]\hline \rule{0pt}{1.3em}
        cylindrical & $1$ & $0$ & $\sim$ & $0$
        \\[0.3em]\hline \rule{0pt}{1.3em}
        slit & $1$ & $0$ & $\sim$ & $\sim$
        \\[0.3em]\hline\hline \rule{0pt}{1.3em}
        & $\left.k^\mathrm{c}\right|_{\lambda = 0}$ 
        & $\left.k^\mathrm{c}\right|_{\lambda = 1}$
        & $\left.\frac{\partial k^\mathrm{c}}{\partial \lambda}\right|_{\lambda = 0}$
        & $\left.\frac{\partial k^\mathrm{c}}{\partial \lambda}\right|_{\lambda = 1}$
        \\[0.6em]\hline \rule{0pt}{1.3em}
        cylindrical & $1$ & $0$ & $\sim$ & $\sim$
        \\[0.3em]\hline \rule{0pt}{1.3em}
        slit & $1$ & $0$ & $1$ & $\sim$
        \\[0.3em]\hline
    \end{tabular}}
    \caption{Conditions for the fit of hindrance factors.}
    \label{T:Extreme_value_conditions_k_refit}
\end{table}\par
To match all conditions for the extreme values, the data has to be fitted with third order polynomials at least. Under appropriate weighting, the following fits are achieved:
\begin{equation}
\begin{split}
    k^\mathrm{d}_\mathrm{cyl} = & 1 - 2.52\lambda + 2.04\lambda^2 - 0.52\lambda^3 \\
    k^\mathrm{c}_\mathrm{cyl} = & 1 + 1.59\lambda - 1.96\lambda^2 + 0.37\lambda^3 \\
    k^\mathrm{d}_\mathrm{slit} = & 1 - 1.83\lambda + 2.63\lambda^2 - 1.80\lambda^3 \\
    k^\mathrm{c}_\mathrm{slit} = & 1 + 1.00\lambda - 0.47\lambda^2 - 1.53\lambda^3 \\
\end{split}
\end{equation}

\subsection{Steric exclusion factor}
\label{C:Steric_exclusion_Giddings}
\begin{figure}[h]
    \centering
    \def\svgwidth{10cm}
    \resizebox{0.9\linewidth}{!}{
    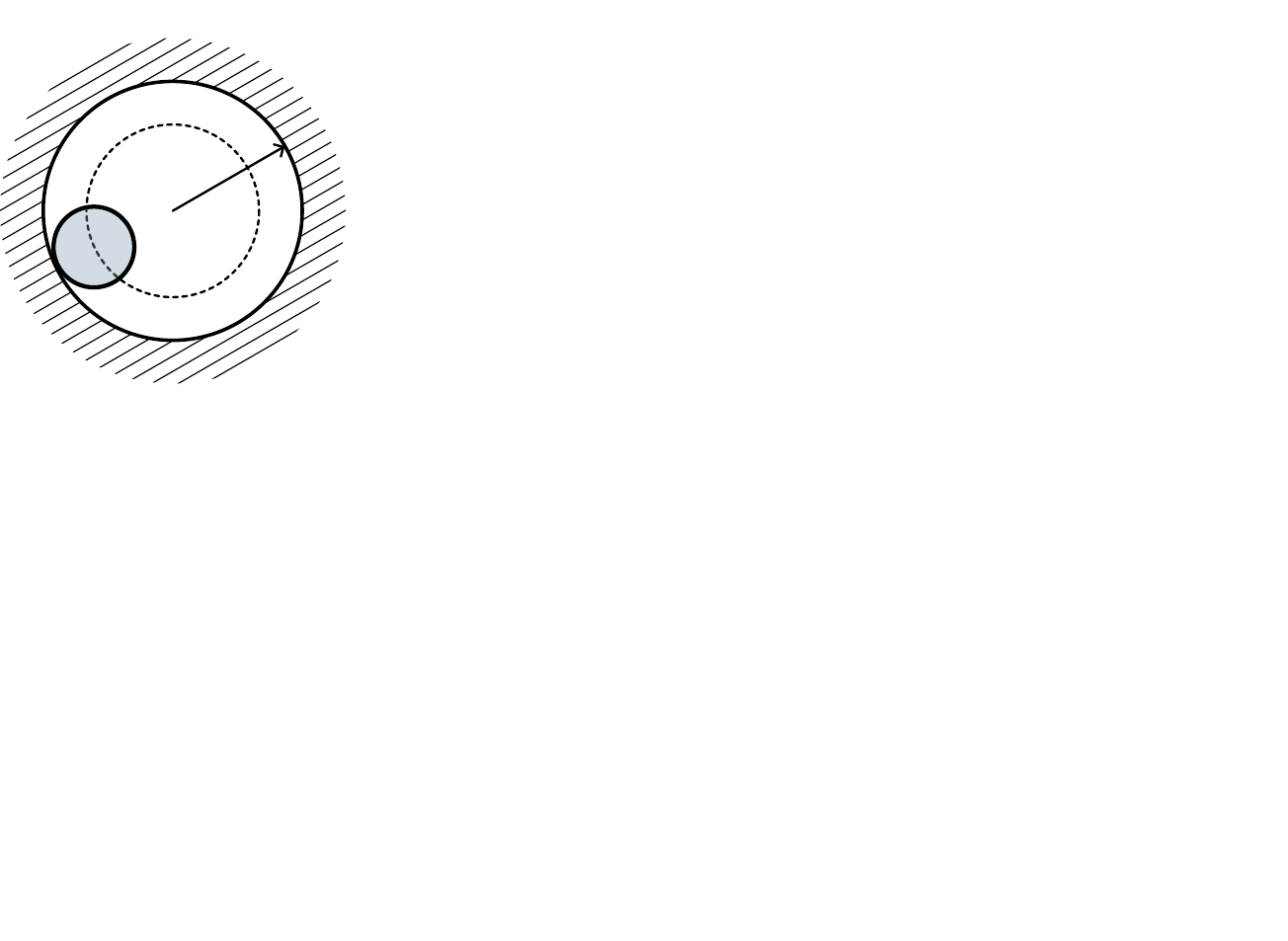}
    \vspace{0.3cm}
    \caption{Steric exclusion factors for various pore geometries\cite{Giddings1968}. Here, $w$ denotes width, $h$ height, $A$ area, $V$ volume and $l_\mathrm{f}$ fibre length. The index $\mathrm{p}$ refers to the pore, $i$ to a species. $\overline{r}$ is the mean radius, $\overline{A}$ the mean projected area.}
    \label{F:Giddings_results_overview}
\end{figure}

Analytical expressions for the steric exclusion factor (see eq. \ref{E:Steric_exclusion}) for several pore and particle geometries have been derived by Giddings et al. \cite{Giddings1968}. The results are summarized in figure \ref{F:Giddings_results_overview}.
\par

\subsection{Derivation of Manning condensation}
\label{C:Manning_condensation_derivation}
The idea of Manning condensation is, that the effective polyion line charge density is limited to a critical value; if the bare polyion line charge density $\beta$ is supercritical, mobile ions condense closely to the polyion and reduce $\beta$ to its critical value. \par To motivate this, the polyion is idealized as a homogeneous line charge density $\beta = z_\mathrm{X} e \rho_\mathrm{X}$, where $\rho_X$ is the number of fixed charges with valence $z_\mathrm{X}$ per unit length. The amount of mobile ions within distance $r_0$ from the line charge is described by a Boltzmann distribution with a Coulomb interaction energy $U_\mathrm{C}(r)$ \cite{Manning1969}:
\begin{equation}
\begin{split}
    &N(r_0) = 2\pi \int_b^{r_0} n(r)\, r \, \mathrm{d} r\\
    &\text{with}\quad
    n(r) = n_0 \exp\left(- \frac{U_\mathrm{C}(r)}{k_\mathrm{B} T}\right)\\
    &\text{and} \quad
    U_\mathrm{C}(r) = - \frac{\beta z_i e }{2\pi \varepsilon_0 \varepsilon_\mathrm{r}} \ln\left(\frac{r}{b}\right)
\end{split}
\end{equation}
Here, $b$ is the radial distance at closest approach. Rewriting yields:
\begin{equation}
\begin{split}
    N(r_0) = 2 \pi \int_0^{r_0} r^{1 + 2 z_i z_\mathrm{X} \xi} 
    \quad \text{with} \quad
\xi = \frac{ e^2 \rho_X}{4 \pi k_\mathrm{B} T \varepsilon_0 \varepsilon_\mathrm{r} }
\end{split}
\end{equation}
Manning noted, that this integral diverges in the lower limit if the exponent $1 + 2 z_i z_\mathrm{X} \xi$ is larger than $-1$, indicating thermodynamic instability. This implies, that the polymer charge is reduced to a critical value by counterion condensation \cite{Manning1969, Kamcev2015} and therefore motivates a critical value $\xi_\mathrm{crit}$ of the Manning parameter $\xi$:
\begin{equation}
    \xi_\mathrm{crit} = (-z_i z_\mathrm{X})^{-1}
\end{equation}

\subsection{Manning hindrance factor}
The hindrance factor $k^\mathrm{e}$ raised by inhomogeneous electric potential for the Donnan-Manning model can be calculated as follows \cite{Kamcev2017,Wang2023}:
\begin{equation}
    k^\mathrm{e}_i = 1 - \frac{{z_i}^2}{3} A 
\end{equation}
Here, $f^\mathrm{u}_i$ represents the fraction of uncondensed ions of species $i$.
The function $A$ is defined as \cite{Wang2023}
\begin{equation}
\resizebox{0.89\linewidth}{!}{$
    A = 
    \begin{cases}
        \displaystyle \sum_{ m_1,m_2= - \infty}^{ \infty} 
        \left( \frac{\left| z_\mathrm{X} \right| c_\mathrm{X}}{\pi \xi^{-1} \left| c_\mathrm{X} \right| \left({m_1}^2 + {m_2}^2\right) + \sum_i {z_i}^2 c_i^m}\right)^2
        & ,\ \forall \xi \leq \xi_\mathrm{crit}\\
        \displaystyle \sum_{ m_1,m_2= - \infty}^{ \infty} 
        \left( \frac{\xi_\mathrm{crit} \left| z_\mathrm{X} \right| c_\mathrm{X} }{\pi \left| z_\mathrm{X} \right|c_\mathrm{X} \left({m_1}^2 + {m_2}^2\right) + \xi \sum_i {z_i}^2 c_i^m f_{\mathrm{u},i}}\right)^2
        & ,\ \forall \xi > \xi_\mathrm{crit}
    \end{cases}
$}
\label{E:Manning_hindrance_factor}
\end{equation}

\subsection{Derivation of the free energy}
\label{C:Derivation_free_energy}
Sections \ref{C:Approach} and \ref{C:Thermodynamics} introduced various assumptions that lead to the free energy density in equation \ref{E:Free_energy_membrane}, which are elaborated in further detail in the following. 

\subsubsection{Uncondensed entropy}
\label{C:Uncondensed_entropy}
Generally, the ideal mixing contribution to the free energy density contains contributions not only from the ionic species $i\in\{1,\ldots,N\}$, but also from the neutral solvent $i=0$ \cite{Atkins2023, Crothers2020}:
\begin{equation}
    g^{\mathrm{u},\mathrm{id}} =
    RT\sum_{i=0}^{N} c_i^\mathrm{u} \left(\ln\left(v_i c_i^\mathrm{u}\right)-1\right)
\label{E:Ideal_mixing_full}
\end{equation}
Here, $v$ denotes partial molar volumes. Subtraction of $1$ within the sum comes from a convenient choice of reference to avoid additional constants on differentiation.\\
Incompressibility imposes:
\begin{equation}
    v_0 c_0^\mathrm{u} + \sum_{i=1}^{N} v_i c_i^\mathrm{u} = 1
\end{equation}
Further, assuming that the solvent is sufficiently abundant $\sum_{i=1}^{N} v_i c_i^\mathrm{u} \ll v_0 c_0^\mathrm{u}$ leads
$c_0^\mathrm{u} v_0 \approx 1$. Therefore, equation \ref{E:Ideal_mixing_full} becomes:
\begin{equation}
    g^{\mathrm{u},\mathrm{id}} = -\frac{RT}{v_0}
    + RT\sum_{i=1}^{N} c_i^\mathrm{u} \left(\ln\left(v_i c_i^\mathrm{u}\right)-1\right)
\end{equation}
Introduction of a standard concentration $c^\circ$ gives:
\begin{equation}
    \ln\left(v_i c_i^\mathrm{u}\right)
    =
    \ln\left(\frac{c_i^\mathrm{u}}{c^\circ}\right)
    +
    \ln\left(v_i c^\circ\right)
\end{equation}
Now, both the constant contribution $- RT / v_0$ and the linear contribution $RT\sum_{i=1}^{N} c_i^\mathrm{u}\ln\left(v_i c^\circ\right)$ can be absorbed into the reference. Thus, the ideal mixing contribution to the free energy density can be written as:
\begin{impeq}
\begin{equation}
    g^{\mathrm{u},\mathrm{id}} =
    RT\sum_{i=1}^{N}c_i^\mathrm{u} \left( \ln\frac{c_i^\mathrm{u}}{c^\circ} - 1 \right)
\end{equation}
\end{impeq}

\subsubsection{Electric potential contribution}
\label{C:Electric_potential_contribution}
Generally, condensed ions may experience state-dependent local electric potentials $\Phi^\mathrm{c}_\alpha$ that differ from the electric potential in the uncondensed state $\Phi^\mathrm{u}$. Both contribute to the free energy density through the respective concentrations \cite{Curran1965, Bockris1998}:
\begin{equation}
    g^{\mathrm{m},\mathrm{el}} =
    F\sum_{i=1}^{N} c_i^\mathrm{u} z_i \Phi^\mathrm{u}
    + F c_\mathrm{X}\sum_\alpha \theta_\alpha  \sum_{i=1}^{N}\nu_{i\alpha}z_i \Phi^\mathrm{c}_\alpha 
\end{equation}
To simplify this, we introduce a state-specific potential difference $\Delta\Phi^\mathrm{c-u}_\alpha = \Phi^\mathrm{c}_\alpha - \Phi^\mathrm{u}$. Then, by application of $c_i^\mathrm{m} = c_i^\mathrm{c} + c_i^\mathrm{u}$ from equation \ref{E:Total_concentration} and $c_i^\mathrm{c} = c_\mathrm{X}\sum_\alpha \nu_{i\alpha}\theta_\alpha$ from equation \ref{E:Condensed_concentrations} this can be rewritten to:
\begin{equation}
    g^{\mathrm{m},\mathrm{el}} =
    F\sum_{i=1}^{N} c_i^\mathrm{m} z_i \Phi^\mathrm{u}
    + F c_\mathrm{X}\sum_\alpha \theta_\alpha \left( \sum_{i=1}^{N}\nu_{i\alpha}z_i \right) \Delta\Phi^{\mathrm{c-u}}_\alpha 
\end{equation}
If the offsets $\Delta\Phi^{\mathrm{c-u}}_\alpha$ are treated as independent of the local composition, the second term can be absorbed into the effective standard formation free energies of the occupation states:
\begin{equation}
    \Delta \hat{G}_\alpha^\circ =
    \Delta \Tilde{G}_\alpha^\circ
    + F \left( \sum_{i=1}^{N}\nu_{i\alpha}z_i \right) \Delta\Phi^{\mathrm{c-u}}_\alpha
    \label{E:G_absorption_potential}
\end{equation}
Therefore, the electric contribution to the free energy is:
\begin{impeq}
\begin{equation}
    g^{\mathrm{m},\mathrm{el}} =
    F\sum_{i=1}^{N} c_i^\mathrm{m} z_i \Phi^\mathrm{u}
\end{equation}
\end{impeq}

\subsubsection{Excess contribution}
\label{C:Excess_free_energy_contribution}
Similar to the electric potential contribution, we absorb differences of the excess contributions into the effective standard formation free energies of the occupation states. Decomposing the excess chemical potential contributions into the uncondensed contributions $\mu_i^{\mathrm{u},\mathrm{ex}}$, and the condensed contributions $\mu_{\alpha}^{\mathrm{c},\mathrm{ex}}$, the excess contribution to the free energy density is:
\begin{equation}
    g^{\mathrm{m,ex}}
    =
    \sum_{i=1}^{N} c_i^\mathrm{u}\mu_i^{\mathrm{u,ex}}
    + c_\mathrm{X}\sum_\alpha \theta_\alpha \mu_{\alpha}^{\mathrm{c,ex}}
\end{equation}
Analogously to the electric potential, we introduce the state-specific difference $\Delta\mu_{\alpha}^{\mathrm{c-u},\mathrm{ex}} = \mu_{\alpha}^{\mathrm{c},\mathrm{ex}} - \sum_i \nu_{i\alpha}\mu_i^{\mathrm{u},\mathrm{ex}}$. Then, by application of $c_i^\mathrm{m} = c_i^\mathrm{c} + c_i^\mathrm{u}$ from equation \ref{E:Total_concentration} and $c_i^\mathrm{c} = c_\mathrm{X}\sum_\alpha \nu_{i\alpha}\theta_\alpha$ from equation \ref{E:Condensed_concentrations} this can be rewritten to:
\begin{equation}
    g^{\mathrm{m,ex}}
    = \sum_{i=1}^{N} c_i^\mathrm{m}\mu_i^{\mathrm{u},\mathrm{ex}}
    + c_\mathrm{X}\sum_\alpha \theta_\alpha \Delta\mu_{\alpha}^{\mathrm{c-u},\mathrm{ex}}
\end{equation}
If the offsets $\Delta\mu_{\alpha}^{\mathrm{c-u},\mathrm{ex}}$ are treated as independent of the local composition, the second term can once again be absorbed into the effective standard formation free energies of the occupation states:
\begin{equation}
    \Delta G_\alpha^\circ =
    \Delta \hat{G}_\alpha^\circ + \Delta\mu_{\alpha}^{\mathrm{c-u},\mathrm{ex}}
    \label{E:G_absorption_excess}
\end{equation}
Therefore, the excess contribution to the free energy is:
\begin{impeq}
\begin{equation}
    g^{\mathrm{m},\mathrm{ex}} =
    \sum_{i=1}^{N} c_i^\mathrm{m}\mu_i^{\mathrm{u},\mathrm{ex}}
\end{equation}
\end{impeq}

\paragraph{The formation free energy of the occupation states} $\Delta G_\alpha^\circ$ with the absorbed contributions therefore is:
\begin{equation}
\begin{split}
    \Delta G_\alpha^\circ = & \
    \Delta \Tilde{G}_\alpha^\circ
    + F \left( \sum_{i=1}^{N}\nu_{i\alpha}z_i \right) \left( \Phi^{\mathrm{c}}_\alpha - \Phi^{\mathrm{u}} \right) \\&
    + \mu_\alpha^{\mathrm{c},\mathrm{ex}} - \sum_{i=1}^{N}\nu_{i\alpha} \mu_i^{\mathrm{u},\mathrm{ex}}
\end{split}
\label{E:Effective_formation_free_energy}
\end{equation}

\subsection{Derivation of thermodynamically consistent transport equations}
\label{C:Derivation_transport_equations}
The starting point for the derivation of thermodynamically consistent transport equations for the regarded membranes is the fundamental relation for the internal energy $e$ of a polarizable system \cite{Groot1984, Henjes1993, Latz2011, Stamm2017, Schammer2021}. 
\begin{equation}
\begin{split}
    &\mathrm{d} e
    =
    T \mathrm{d} s
    + \mathbf{E} \cdot \mathrm{d} \mathbf{D}
    + \mathbf{H} \cdot \mathrm{d} \mathbf{B}
    +  \sum_i \mu_i \mathrm{d} c_i
    \label{E:FundamentalRelationInternalEnergy} \\
\end{split}
\end{equation}
Here, the energy density $e$, absolute temperature $T$, entropy density  $s$, electric field $\mathbf{E}$, electric displacement $\mathbf{D}$, magnetic induction $\mathbf{B}$, magnetic field $\mathbf{H}$, chemical potential $\mu$ and concentration $c$ are all volume-averaged and therefore include both the condensed and uncondensed state. The index $i \in \left\{ X,0,1, ...,N \right\}$ represents all species with $X$ being the fixed charge species, $0$ the neutral solvent, $1$ a charged mobile species and $2,...,N$ all other mobile species. \par
Further, assuming that the local volumes are additive and partial molar volumes $v_i$ are constant, the volume filling constraint reads \cite{Lorenz2022,Kilchert2023}:
\begin{equation}
    \sum_i v_i \mathrm{d} c_i =  0
    \label{E:Continuity}
\end{equation}
The fixed charge concentration is constant in space and time:
\begin{equation}
    \mathrm{d} c_\mathrm{X}
    = 0
    \label{E:FixedChargeIsConstant}
\end{equation}
Using equations \ref{E:Continuity}-\ref{E:FixedChargeIsConstant}, species $0$ and $\mathrm{X}$ are eliminated from relation \ref{E:FundamentalRelationInternalEnergy}:
\begin{equation} 
\begin{split}
    &\mathrm{d} e 
    =
    T \mathrm{d} s 
    + \mathbf{E} \cdot \mathrm{d} \mathbf{D}
    + \mathbf{H} \cdot \mathrm{d} \mathbf{B}
    + \sum_{i\neq 0,X} \Tilde{\mu}_i \mathrm{d} c_i
    \\
    &\text{with} \quad
    \Tilde{\mu}_i = \mu_i - \frac{v_i}{v_0}\mu_0 
\end{split}
\label{E:FundamentalRelationInternalEnergySimplified01}
\end{equation}
To further simplify this, we first use electromagnetic energy balance obtained from the Maxwell equations, also known as Poynting's theorem \cite{Jackson2009, Landau1984, Latz2011, Schammer2021}:
\begin{equation}
    \mathbf{E} \cdot \frac{\partial \mathbf{D}}{\partial t} + \mathbf{H} \cdot \frac{\partial \mathbf{B}}{\partial t}  
    =
    - \nabla \cdot \left( \mathbf{E} \times \mathbf{H} \right) - \mathbf{j} \cdot \mathbf{E}
    \label{E:Maxwell}
\end{equation}
Here, $\mathbf{j}$ denotes the electric current density.\\
Next, we write local entropy balance as \cite{Groot1984, Bedeaux2016, Latz2011, Schammer2021}:
\begin{equation}
    \frac{\partial s}{\partial t}
    = -\nabla \cdot \left(\frac{\mathbf{q}}{T}\right) + \sigma_\mathrm{s}
    \label{E:Local_entropy_balance}
\end{equation}
Here, $\mathbf{q}$ denotes the heat flux and $\sigma_\mathrm{s}$ the entropy production rate density. The second law of thermodynamics requires $\sigma_\mathrm{s} \geq 0$ \cite{Bedeaux2016}. \\
With this, equation \ref{E:FundamentalRelationInternalEnergySimplified01} becomes:
\begin{equation}
\begin{split}
    \frac{\partial e}{\partial t}
    = &\ 
    - T \nabla \cdot \left(\frac{\mathbf{q}}{T}\right) + T \sigma_\mathrm{s}
    - \nabla \cdot \left( \mathbf{E} \times \mathbf{H} \right) \\ &\ 
    - \mathbf{j} \cdot \mathbf{E}
    + \sum_{i\neq 0,X} \Tilde{\mu}_i \frac{\partial c_i}{\partial t}
\end{split}
\label{E:FundamentalRelationInternalEnergySimplified02}
\end{equation}\par
So far, we deliberately avoided to distinguish between the condensed state $\mathrm{c}$ and the uncondensed state $\mathrm{u}$ of the species by pointing out that all variables are volume averaged and therefore include both states. To elaborate on how transport through condensed and uncondensed states contribute to the total transport, this distinction has to be made. Therefore, we decompose the averaged quantities into state-resolved quantities for the states $\mathbb{S} \in \{\mathrm{c},\, \mathrm{u}\}$ \cite{Curran1965,Maugin1994,Groot1984}:
\begin{equation}
    c_i = \sum_\mathbb{S} c_i^\mathbb{S} 
    \qquad \text{and} \qquad 
    \mathbf{j} = \sum_\mathbb{S} \mathbf{j}^\mathbb{S} 
\end{equation}
Since these states can be subject to different local electric potentials, state transfer is associated with local electrical work \cite{Curran1965}. Therefore, the material and electrical work terms become:
\begin{equation}
\begin{split}
    &- \mathbf{j} \cdot \mathbf{E}
    + \sum_{i\neq 0,X} \Tilde{\mu}_i \frac{\partial c_i}{\partial t}
    \\ & \quad =
    \sum_{\mathbb{S}\in\{\mathrm{c,u}\}} \left[
    -\mathbf{j}^\mathbb{S}\cdot\mathbf{E}^\mathbb{S}
    + \sum_{i\neq 0,X} \left(
    \Tilde{\mu}_i^\mathbb{S} \frac{\partial c_i^\mathbb{S}}{\partial t}
    +F z_i\Phi^\mathbb{S}\sigma_i^\mathbb{S} \right) \right]
\end{split}
\label{E:Material_electrical}
\end{equation}
Here, $\sigma_i^\mathbb{S}$ denotes the source of species $i$ into state $\mathbb{S}$. \\
These source terms also appear in the species continuity of each state \cite{Curran1965, Groot1984}:
\begin{equation}
\begin{gathered}
    \frac{\partial c_i^\mathbb{S}}{\partial t} 
    = 
    - \nabla \cdot \mathbf{N}_i^\mathbb{S} + \sigma_i^\mathbb{S} 
    \quad \text{with} \quad \sigma_i^\mathrm{c} = - \sigma_i^\mathrm{u} \equiv \sigma_i\\
\end{gathered}
\label{E:FluxDivergenceStateResolved}
\end{equation}
For ionic transport, the electric current is given by the charge fluxes \cite{Curran1965,Newman2004}. With $\mathbf{E}^\mathbb{S}=-\nabla\Phi^\mathbb{S}$ we have:
\begin{equation}
    \mathbf{j}^\mathbb{S} \cdot \mathbf{E}^\mathbb{S} =
    - F\sum_{i\neq 0,X} z_i\mathbf{N}_i^\mathbb{S} \cdot \nabla \Phi^\mathbb{S}
\end{equation}
Then, by application of the product rule and with the electrochemical potentials $\overline{\mu}_i^\mathbb{S} = \Tilde{\mu}_i^\mathbb{S} + Fz_i\Phi^\mathbb{S}$, the material and electrical work terms (equation \ref{E:Material_electrical}) become:
\begin{equation}
    \sum_{\mathbb{S}\in\{\mathrm{c,u}\}}
    \sum_{i\neq 0,X}
    \left[
        -\nabla\cdot
        \left(
            \Tilde{\mu}_i^\mathbb{S}\mathbf{N}_i^\mathbb{S}
        \right)
        +
        \mathbf{N}_i^\mathbb{S}\cdot
        \nabla\overline{\mu}_i^\mathbb{S}
        +
        \overline{\mu}_i^\mathbb{S}\sigma_i^\mathbb{S}
    \right] 
\label{E:Material_electrical2}
\end{equation}
In local equilibrium $\overline{\mu}_i^\mathrm{c} = \overline{\mu}_i^\mathrm{u} \equiv \overline{\mu}_i^\mathrm{m}$ (compare equation \ref{E:Electrochemical_potential_membrane}), the source contributions vanish:
\begin{equation}
    \sum_{\mathbb{S}} \overline{\mu}_i^\mathbb{S}\sigma_i^\mathbb{S} = 0
\end{equation}
Since local equilibrium also implies
$\nabla\overline{\mu}_i^\mathbb{S}
= \nabla\overline{\mu}_i^\mathrm{m}$, introduction of the total species flux $\mathbf{N}_i^\mathrm{m} = \sum_{\mathbb{S}} \mathbf{N}_i^\mathbb{S}$ further simplifies equation \ref{E:Material_electrical2} to:
\begin{equation}
    \sum_{i\neq 0,X} \mathbf{N}_i^\mathrm{m} \cdot\nabla\overline{\mu}_i^\mathrm{m} 
    - \sum_{i\neq 0,X} \sum_{\mathbb{S}\in\{\mathrm{c,u}\}} \nabla\cdot \left( \Tilde{\mu}_i^\mathbb{S}\mathbf{N}_i^\mathbb{S} \right)
\end{equation}
Therefore, equation \ref{E:FundamentalRelationInternalEnergySimplified02} becomes:
\begin{equation}
\begin{split}
    \frac{\partial e}{\partial t}
    = &\
    -\nabla \cdot \left( \mathbf{q} + \mathbf{E} \times \mathbf{H}  +  \sum_\mathbb{S} \sum_{i\neq 0,X} \Tilde{\mu}_i^\mathbb{S} \mathbf{N}_i^\mathbb{S} \right)\\&\
    + T \sigma_\mathrm{s} + \frac{\mathbf{q} \cdot \nabla T}{T}
    + \sum_{i\neq 0,X} \mathbf{N}_i^\mathrm{m} \cdot \nabla \overline{\mu}_i^\mathrm{m} 
\end{split}
\end{equation}\par
Since the time derivative of the energy has to fulfill the continuity relation $\frac{\partial e}{\partial t} = -\nabla \cdot J_e$, the entropy production $\sigma_\mathrm{s}$ has to be \cite{Latz2011, Schammer2021}:
\begin{equation}
    T \sigma_\mathrm{s} =  
    - \frac{\mathbf{q} \cdot \nabla T}{T}
    - \sum_{i\neq 0,X} \mathbf{N}_i^\mathrm{m} \cdot \nabla \overline{\mu}_i^\mathrm{m}
    \label{E:Entropy_production_full}
\end{equation}
Assumption of a homogeneous temperature distribution throughout the membrane ($\nabla T = 0$) reduces equation \ref{E:Entropy_production_full} to:
\begin{equation}
    T \sigma_\mathrm{s} = 
    - \sum_{i\neq 0,X} \mathbf{N}_i^\mathrm{m} \cdot \nabla \overline{\mu}_i^\mathrm{m}
    \label{E:Entropy_production_simplified}
\end{equation}
Now, the transport equations need to be constructed such that the entropy production $\sigma_\mathrm{s}$ is non-negative. This is achieved by an Onsager approach, assuming that the fluxes are linearly coupled to the forces \cite{Onsager1931, Onsager1931a, Latz2011, Schammer2021}:
\begin{equation}
\begin{split}
    &\mathbf{N}_i^\mathrm{m}
    =
    -\sum_{j\neq 0,X}
    \mathcal{L}_{ij}
    \nabla\overline{\mu}_j^\mathrm{m} \\
    \textrm{with} \quad &
    \mathcal{L}_{ij} =  \mathcal{L}_{ji} 
    \quad \textrm{and} \quad 
    \mathcal{L}_{ij} =  \sum_\mathbb{S} \mathcal{L}_{ij}^\mathbb{S}
\end{split}
\label{E:Onsager_flux}
\end{equation}
Here, non-negative entropy production $\sigma_\mathrm{s}$ requires the Onsager matrix $\boldsymbol{\mathcal{L}}$ to be symmetric and positive semi-definite \cite{Onsager1931, Onsager1931a}.\par 
Using the electrochemical potential from equation \ref{E:Electrochemical_potential_membrane} we can write equation \ref{E:Onsager_flux} as:
\begin{equation}
    \mathbf{N}_i^\mathrm{m}
    =
    -\sum_{j\neq 0,X}\mathcal{L}_{ij}
    \left(
        \frac{RT}{c_j^\mathrm{u}} \nabla c_j^\mathrm{u}
        + z_j F \nabla \Phi^\mathrm{u}
        +\nabla \mu_j^\mathrm{u,ex}
    \right)
\label{E:Effective_flux_onsager_form}
\end{equation}
Next, we replace the state-specific Onsager coefficients $\mathcal{L}_{ij}^\mathbb{S}$ with physically motivated, state-specific Onsager diffusion coefficients $\mathcal{D}_{ij}^\mathbb{S}$ \cite{Latz2011}:
\begin{equation}
    \mathcal{D}_{ij}^\mathbb{S} =
    \mathcal{L}_{ij}^\mathbb{S} \frac{RT}{c_j^\mathbb{S}}
\end{equation} 
Since the state contributions add in the Onsager mobilities $\mathcal{L}_{ij}=\sum_{\mathbb{S}}\mathcal{L}_{ij}^{\mathbb{S}}$ in equation \ref{E:Effective_flux_onsager_form} and
the electrochemical potential is expressed through $c_j^\mathrm{u}$, the effective Onsager diffusion coefficient $\mathcal{D}_{ij}^\mathrm{m}$ is:
\begin{equation}
    \mathcal{D}_{ij}^\mathrm{m} =
    \frac{1}{c_j^\mathrm{u}}
    \sum_{\mathbb{S} \in \{\mathrm{c,u}\}} \mathcal{D}_{ij}^\mathbb{S} c_j^\mathbb{S}
    \label{E:Diffusion_coefficients_state_vs_effective}
\end{equation} 
This reduction to effective transport coefficients is consistent with continuum treatments of rapidly equilibrating electrolyte species \cite{Curran1965, Clark2017}. However, the effective transport coefficients remain dependent on the concentrations of all ions through the state equilibria (see equation \ref{E:Equilibrium_equation_system}).\par
Lastly, we neglect off-diagonal coefficients ($\mathcal{L}_{ij} = 0$ for all $i \neq j$) and rewrite the effective diagonal Onsager diffusion coefficients as $D_i^\mathrm{m} = \mathcal{D}_{ii}^\mathrm{m}$. Then, by assumption of a spatially constant $\mu_j^\mathrm{u,ex}$, we recover a typical Nernst-Planck-type equation \cite{Newman2004}:
\begin{impeq}
\begin{equation}
    \mathbf{N}_i^\mathrm{m} =
    -D_i^\mathrm{m}\nabla c_i^\mathrm{u}
    - \frac{F}{RT} D_i^\mathrm{m} z_i c_i^\mathrm{u} \nabla\Phi^\mathrm{u}
\label{E:Flux_equations_appendix}
\end{equation}
\end{impeq}
For this form, it is also apparent that the entropy production is non-negative for $D_i^\mathrm{m} \geq 0$:
\begin{equation}
    T \sigma_\mathrm{s} = 
    \sum_{i\neq 0,X} \frac{D_i^\mathrm{m} c_i^\mathrm{u}}{RT}
    \left( \nabla\overline{\mu}_i^\mathrm{m} \right)^2
    \geq 0
\end{equation} \par
The flux equations (eq. \ref{E:Flux_equations_appendix}) require a closure relation for the electric potential gradient $\nabla\Phi^\mathrm{u}$, which we get from the current density with a prescribed current $\mathbf{j} = \mathbf{j}^\mathrm{app}$:
\begin{impeq}
\begin{equation}
    \mathbf{j}
    =
    F\sum_{i\neq 0,X}
    z_i\mathbf{N}_i^\mathrm{m}
\end{equation}
\end{impeq}
Equivalently, this relation could be used to eliminate one species from the fluxes, but the presented version is more convenient in the context of membrane transport.\par
If necessary, the solvent flux $N_0$ can be determined by:
\begin{equation}
    \mathbf{N}_\mathrm{0}^\mathrm{m} = - \frac{1}{v_0} \sum_{i \neq 0,X} v_i \mathbf{N}_\mathrm{i}^\mathrm{m}
\end{equation}
Note that the off-diagonal diffusion coefficients neglected in this derivation can be relevant for certain effects, e.g. water drag \cite{Zawodzinski1993}. We still use this simplification for simplicity, since it complicates parameterization otherwise.

\subsection{Limiting behavior of the IOM}
\label{C:Appendix_limiting_IOM}
This section derives the limiting behavior of the IOM needed for the comparison to Manning condensation in section \ref{C:Analytical_low_salt}.
\subsubsection{The limiting value for the residual charge}
\label{C:Appendix_limiting_Z}
In the single-salt, single-occupation case, the residual-charge equation \ref{E:IOM_Residual_charge} becomes:
\begin{equation}
    Z - z_\mathrm{X} + \left(Z - q_\mathrm{ct}  \right)\chi_\mathrm{ct} + \left(Z - q_\mathrm{co}  \right)\chi_\mathrm{co}= 0
    \label{E:IOM_Residual_charge_reduced}
\end{equation}
Here, the index $\mathrm{ct}$ denotes counterions and $\mathrm{co}$ denotes coions.
By introduction of strictly positive coefficients $A_i = K_i c_i^\mathrm{u} / c^\circ > 0$ and $B_i = \lvert z_i/z_\mathrm{X} \rvert \, n_\mathrm{n} U > 0$ for $i \in \{ \mathrm{ct}, \, \mathrm{co}\}$ we can write $\chi_\mathrm{i}$ (equation \ref{E:Definition_of_chi_and_zeta}) compactly as:
\begin{equation}
\begin{split}
    \chi_\mathrm{ct} = &\,
    \frac{\theta_\mathrm{ct}}{\theta_\mathpzc{0}} = 
    A_\mathrm{ct} \exp\left(B_\mathrm{ct} z_\mathrm{X}Z\right) \\
    \chi_\mathrm{co} = &\, 
    \frac{\theta_\mathrm{co}}{\theta_\mathpzc{0}} =
    A_\mathrm{co} \exp\left(-B_\mathrm{co} z_\mathrm{X}Z\right)
\end{split}
\label{E:chi_simplified}
\end{equation}
Here, $B_\mathrm{ct}$ and $B_\mathrm{co}$ depend linearly on the interaction parameter $U$, yielding:
\begin{equation}
    \lim_{U \to \infty} B_i = \infty
\end{equation}
Consequently, if we allow $\lim_{U \to \infty} z_\mathrm{X} Z \neq 0$ at finite nonzero prefactors $A_i$, either $\chi_\mathrm{ct}$ or $\chi_\mathrm{co}$ will diverge, while the other vanishes, depending on the sign of $z_\mathrm{X} Z$. \par
\paragraph{The positive branch} $\lim_{U \to \infty} z_\mathrm{X} Z > 0$ yields $\chi_\mathrm{ct} \to \infty$ and $\chi_\mathrm{co} \to 0$ in the limit. Through normalization (equation \ref{E:Theta_normalization}), this gives:
\begin{equation}
    \theta_\mathrm{ct} \to 1,
    \quad
    \theta_\mathrm{co} \to 0
    \quad \text{and} \quad
    \theta_\mathpzc{0} \to 0
    \quad \text{for} \quad U \to \infty
\end{equation}
By definition of $Z$, and since counterions satisfy $z_\mathrm{X}z_\mathrm{ct}<0$, this results in:
\begin{equation}
\begin{split}
    \lim_{U \to \infty} z_\mathrm{X} Z = &\,
    z_\mathrm{X}\left(z_\mathrm{X} + z_\mathrm{ct}\right)\\
    \le &\, 0
    \quad \text{for} \quad \lvert z_\mathrm{X} \rvert \le \lvert z_\mathrm{ct} \rvert
\end{split}
\end{equation}
This contradicts the assumption of $\lim_{U \to \infty} z_\mathrm{X} Z > 0$ and therefore invalidates this branch.
\paragraph{The negative branch} $\lim_{U \to \infty} z_\mathrm{X} Z < 0$ yields $\chi_\mathrm{ct} \to 0$ and $\chi_\mathrm{co} \to \infty$ in the limit. Through normalization (equation \ref{E:Theta_normalization}), this gives:
\begin{equation}
    \theta_\mathrm{ct} \to 0,
    \quad
    \theta_\mathrm{co} \to 1
    \quad \text{and} \quad
    \theta_\mathpzc{0} \to 0
    \quad \text{for} \quad U \to \infty
\end{equation}
By definition of $Z$, and since coions satisfy $z_\mathrm{X}z_\mathrm{co}>0$, this results in:
\begin{equation}
    \lim_{U \to \infty} z_\mathrm{X} Z = 
    z_\mathrm{X}\left(z_\mathrm{X} + z_\mathrm{co}\right)
    > 0
\end{equation}
This contradicts the assumption of $\lim_{U \to \infty} z_\mathrm{X} Z < 0$ and therefore invalidates this branch too.
\paragraph{The only self-consistent solution} remaining therefore is $\lim_{U \to \infty} z_\mathrm{X} Z = 0$. Since $z_\mathrm{X} \neq 0 $ we obtain:
\begin{impeq}
\begin{equation}
    \lim_{U \to \infty} Z = 0
    \label{E:Limit_of_Z_appendix}
\end{equation}
\end{impeq}
Note, that if we assume scaling $U\propto 1/L\propto c_\mathrm{X}^{1/3}$, the limit $U\to\infty$ is equivalent to the large fixed-charge-density limit $c_\mathrm{X}\to\infty$.

\subsubsection{The limiting value of the occupation fractions}
\label{C:Appendix_limiting_theta}
As a consequence of the limiting behavior $Z\to0$, the exponential factors in the relative occupation weights $\chi_i=\theta_i/\theta_\mathpzc{0}$ (equation \ref{E:chi_simplified}) converge to unity in the $U\to\infty$ limit. This leaves:
\begin{equation}
    \lim_{U\to\infty} \chi_\mathrm{ct} =
    K_\mathrm{ct} \frac{c_\mathrm{ct}^\mathrm{u}}{c^\circ} 
    \quad \text{and} \quad
    \lim_{U\to\infty} \chi_\mathrm{co} = 
    K_\mathrm{co} \frac{c_\mathrm{co}^\mathrm{u}}{c^\circ}
    \label{E:Limits_of_chi}
\end{equation}
Therefore, the limit for the ratio of the occupation fractions of co- and counterions in the $U\to \infty$ limit is:
\begin{equation}
    \lim_{U\to\infty}
    \frac{\theta_\mathrm{co}}{\theta_\mathrm{ct}}
    =
    \frac{K_\mathrm{co}c_\mathrm{co}^\mathrm{u}}
         {K_\mathrm{ct}c_\mathrm{ct}^\mathrm{u}}
\end{equation}
Together with the definition of $Z$ (equation \ref{E:Definition_of_abbreviation_Z}) 
\begin{equation}
    Z =
    z_\mathrm{X}
    + z_\mathrm{ct}\theta_\mathrm{ct}
    + z_\mathrm{co}\theta_\mathrm{co}
\end{equation}
and $\lim_{U\to\infty}Z=0$, this gives:
\begin{equation}
\begin{aligned}
\begin{aligned}
    \lim_{U\to\infty}\theta_\mathrm{ct}
    &=
    -\frac{z_\mathrm{X}}{z_\mathrm{ct}+A z_\mathrm{co}}
    \\
    \lim_{U\to\infty}\theta_\mathrm{co}
    &=
    -\frac{A z_\mathrm{X}}{z_\mathrm{ct}+A z_\mathrm{co}}
\end{aligned}
\qquad \ \text{with} \ \
    A =
    \frac{K_\mathrm{co}c_\mathrm{co}^\mathrm{u}}
         {K_\mathrm{ct}c_\mathrm{ct}^\mathrm{u}}
\end{aligned}
\label{E:Limits_of_theta_full}
\end{equation}
In the low salt limit $A \ll 1$ holds, yielding:
\begin{impeq}
\begin{equation}
    \lim_{U\to\infty}\theta_\mathrm{ct} \approx
    -\frac{z_\mathrm{X}}{z_\mathrm{ct}}
    \quad \text{and} \quad
    \lim_{U\to\infty}\theta_\mathrm{co} \approx
    0
    \label{E:Limit_of_theta_appendix}
\end{equation}
\end{impeq}
Again, all limits can be written similarly for $c_\mathrm{X}$ if a proportionality $U \propto c_\mathrm{X}$ is assumed.

\subsubsection{The limiting behavior of the effective membrane charge}
\label{C:Appendix_limiting_q_eff}
Expressed in IOM variables, the effective membrane charge $q_\mathrm{eff}$ is:
\begin{equation}
    q_\mathrm{eff} = c_\mathrm{X} Z .
\end{equation}
With $Z \to 0$ for $c_\mathrm{X} \to \infty$,  $q_\mathrm{eff}$ can still converge to a finite value. To obtain this value, we use the definition of the occupation weights $\chi_\alpha$ (equation \ref{E:Definition_of_chi_and_zeta}) and write:
\begin{equation}
    \chi_\mathrm{ct}^\mathrm{lim}
    =
    \frac{\theta_\mathrm{ct}^\mathrm{lim}}{\theta_\mathpzc{0}^\mathrm{lim}}
\end{equation}
This requires that the limiting occupation fractions of the counterion state, $\theta_\mathrm{ct}^\mathrm{lim}$, and of the empty state, $\theta_\mathpzc{0}^\mathrm{lim}$, exist and that the denominator limit is not zero. By equation \ref{E:Limits_of_theta_full}, this condition corresponds to:
\begin{equation}
    z_\mathrm{X}+z_\mathrm{ct}
    +A\left(z_\mathrm{X}+z_\mathrm{co}\right)
    \neq 0
\end{equation}
Given that this condition holds, we can then use equation \ref{E:Limits_of_chi} and obtain:
\begin{equation}
    K_\mathrm{ct} \frac{c_\mathrm{ct}^\mathrm{u,lim}}{c^\circ} =
    -\frac{z_\mathrm{X}}
    {z_\mathrm{X}+z_\mathrm{ct} + A\left(z_\mathrm{X}+z_\mathrm{co}\right)}
    \label{E:Condition_limit_q_eff}
\end{equation}
Next, we aim to eliminate $c_\mathrm{ct}^\mathrm{u}$ through charge neutrality (equation \ref{E:IOM_charge_neutrality}) which for the considered case reads:
\begin{equation}
    q_\mathrm{eff}
    + z_\mathrm{ct} c_\mathrm{ct}^\mathrm{u}
    + z_\mathrm{co} c_\mathrm{co}^\mathrm{u}
    = 0
    \label{E:qeff_charge_neutrality_low_salt}
\end{equation}
Solving equation \ref{E:qeff_charge_neutrality_low_salt} for $q_\mathrm{eff}$ and substituting $c_\mathrm{ct}^\mathrm{u}$ through equation \ref{E:Condition_limit_q_eff} then gives:
\begin{equation}
    \lim_{U\to\infty} q_\mathrm{eff} = 
    \frac{1}{K_\mathrm{ct}} 
    \frac{z_\mathrm{X} z_\mathrm{ct} c^\circ }{z_\mathrm{X} + z_\mathrm{ct} + A \left(z_\mathrm{X} + z_\mathrm{co}\right)}
    - z_\mathrm{co} c^\mathrm{u,lim}_\mathrm{co}
\label{E:Limit_of_q_eff_general_appendix}
\end{equation}
This is only meaningful, if the counterion has higher magnitude charge than the fixed site $\lvert z_\mathrm{X}\rvert < \lvert z_\mathrm{ct}\rvert$. Therefore, the conditions for a counterion-dominated low-salt limiting solution are:
\begin{equation}
\begin{split}
    &\lvert A \left(z_\mathrm{X} + z_\mathrm{co}\right) \rvert
    \ll
    \lvert z_\mathrm{X} + z_\mathrm{ct} \rvert \\
    \text{and} \qquad &
    \lvert z_\mathrm{co} c^\mathrm{u}_\mathrm{co} \rvert
    \ll 
    \frac{1}{K_\mathrm{ct}} 
    \frac{\lvert z_\mathrm{X} z_\mathrm{ct} c^\circ \rvert }{\lvert z_\mathrm{X} + z_\mathrm{ct} \rvert}\\
    \text{for} \qquad &
    \lvert z_\mathrm{X}\rvert < \lvert z_\mathrm{ct}\rvert
\end{split}
\end{equation}
Under these conditions, the low salt limiting value of $q_\mathrm{eff}$ (equation \ref{E:Limit_of_q_eff_general_appendix}) reduces to:
\begin{impeq}
\begin{equation}
    \lim_{U\to\infty} q_\mathrm{eff} \approx
    \frac{1}{K_\mathrm{ct}} 
    \frac{z_\mathrm{X} z_\mathrm{ct} c^\circ }{z_\mathrm{X} + z_\mathrm{ct}}
    \label{E:Limit_of_q_eff_appendix}
\end{equation}
\end{impeq}

\paragraph{For same magnitude charge} of fixed sites and counterions $\lvert z_\mathrm{X}\rvert = \lvert z_\mathrm{ct}\rvert$, expression \ref{E:Limit_of_q_eff_appendix} becomes singular as a consequence of the single occupation condition. Relaxing this assumption by allowing for higher occupation states can regularize the limit.
For example, if a two-counterion occupation state is introduced and its association constant is approximated from penalized independent association $K_{2\cdot \mathrm{ct}} \approx \eta^2 K_\mathrm{ct}^2$ with a penalty $0 < \eta \le 1$, an analogous limiting derivation gives:
\begin{impeq}
\begin{equation}
    \lim_{U\to\infty} q_\mathrm{eff} \approx
    \frac{z_\mathrm{X} c^\circ}{\sqrt{K_{2\cdot \mathrm{ct}}}} 
    \approx \frac{z_\mathrm{X} c^\circ}{K_\mathrm{ct}\eta} 
    \label{E:Limit_of_q_eff_appendix_multi}
\end{equation}
\end{impeq}
For simplicity, we assume that the penalty is negligible $\eta \approx 1$.\par
Again, all limits can be written analogously for $c_\mathrm{X}$ if a proportionality $U \propto c_\mathrm{X}$ is assumed.

%% file: 7_Appendix/Figures/Giddings_results_overview_paper.pdf_tex
\begingroup%
  \makeatletter%
  \providecommand\color[2][]{%
    \errmessage{(Inkscape) Color is used for the text in Inkscape, but the package 'color.sty' is not loaded}%
    \renewcommand\color[2][]{}%
  }%
  \providecommand\transparent[1]{%
    \errmessage{(Inkscape) Transparency is used (non-zero) for the text in Inkscape, but the package 'transparent.sty' is not loaded}%
    \renewcommand\transparent[1]{}%
  }%
  \providecommand\rotatebox[2]{#2}%
  \newcommand*\fsize{\dimexpr\f@size pt\relax}%
  \newcommand*\lineheight[1]{\fontsize{\fsize}{#1\fsize}\selectfont}%
  \ifx\svgwidth\undefined%
    \setlength{\unitlength}{625.79422598bp}%
    \ifx\svgscale\undefined%
      \relax%
    \else%
      \setlength{\unitlength}{\unitlength * \real{\svgscale}}%
    \fi%
  \else%
    \setlength{\unitlength}{\svgwidth}%
  \fi%
  \global\let\svgwidth\undefined%
  \global\let\svgscale\undefined%
  \makeatother%
  \begin{picture}(1,0.73041953)%
    \lineheight{1}%
    \setlength\tabcolsep{0pt}%
    \put(0,0){\includegraphics[width=\unitlength,page=1]{Giddings_results_overview_paper.pdf}}%
    \put(0.17046271,0.59804024){\color[rgb]{0,0,0}\makebox(0,0)[rt]{\lineheight{1.25}\smash{\begin{tabular}[t]{r}$r_\mathrm{p}$\end{tabular}}}}%
    \put(0.10988344,0.54417749){\color[rgb]{0,0,0}\makebox(0,0)[lt]{\lineheight{1.25}\smash{\begin{tabular}[t]{l}$r_i$\end{tabular}}}}%
    \put(0,0){\includegraphics[width=\unitlength,page=2]{Giddings_results_overview_paper.pdf}}%
    \put(0.13430495,0.72310928){\color[rgb]{0,0,0}\makebox(0,0)[t]{\lineheight{1.25}\smash{\begin{tabular}[t]{c}\textbf{Cylindrical}\end{tabular}}}}%
    \put(0,0){\includegraphics[width=\unitlength,page=3]{Giddings_results_overview_paper.pdf}}%
    \put(0.49996951,0.49665332){\color[rgb]{0,0,0}\makebox(0,0)[t]{\lineheight{1.25}\smash{\begin{tabular}[t]{c}\colorbox{white}{$w_\mathrm{p}$}\end{tabular}}}}%
    \put(0.559987,0.56438219){\color[rgb]{0,0,0}\makebox(0,0)[rt]{\lineheight{1.25}\smash{\begin{tabular}[t]{r}\colorbox{white}{$h_\mathrm{p}$}\end{tabular}}}}%
    \put(0.50512004,0.72313502){\color[rgb]{0,0,0}\makebox(0,0)[t]{\lineheight{1.25}\smash{\begin{tabular}[t]{c}\textbf{Rectangular}\end{tabular}}}}%
    \put(0.86454353,0.28262387){\color[rgb]{0,0,0}\makebox(0,0)[t]{\lineheight{1.25}\smash{\begin{tabular}[t]{c}\textbf{Random thin fibers}\end{tabular}}}}%
    \put(0.49869451,0.28350636){\color[rgb]{0,0,0}\makebox(0,0)[t]{\lineheight{1.25}\smash{\begin{tabular}[t]{c}\textbf{Random thin infinite planes}\end{tabular}}}}%
    \put(0.13411769,0.37375552){\color[rgb]{0,0,0}\makebox(0,0)[t]{\lineheight{1.25}\smash{\begin{tabular}[t]{c}$\displaystyle{S^\mathrm{st} = \left(1-\frac{r_i}{r_\mathrm{p}}\right)^2}$\end{tabular}}}}%
    \put(0.86582137,0.37425044){\color[rgb]{0,0,0}\makebox(0,0)[t]{\lineheight{1.25}\smash{\begin{tabular}[t]{c}$\displaystyle{S^{st} = \left(1-\frac{r_i}{r_\mathrm{p}}\right)^3}$\end{tabular}}}}%
    \put(0.49996955,0.37375552){\color[rgb]{0,0,0}\makebox(0,0)[t]{\lineheight{1.25}\smash{\begin{tabular}[t]{c}$\displaystyle{S^\mathrm{st} = \frac{\left(h_\mathrm{p}-2r_i\right)\left(w_\mathrm{p}-2r_i\right)}{h_\mathrm{p}w_\mathrm{p}}}$\end{tabular}}}}%
    \put(0.86599458,0.72310928){\color[rgb]{0,0,0}\makebox(0,0)[t]{\lineheight{1.25}\smash{\begin{tabular}[t]{c}\textbf{Spherical}\end{tabular}}}}%
    \put(0,0){\includegraphics[width=\unitlength,page=4]{Giddings_results_overview_paper.pdf}}%
    \put(0.90216649,0.59804026){\color[rgb]{0,0,0}\makebox(0,0)[rt]{\lineheight{1.25}\smash{\begin{tabular}[t]{r}$r_\mathrm{p}$\end{tabular}}}}%
    \put(0,0){\includegraphics[width=\unitlength,page=5]{Giddings_results_overview_paper.pdf}}%
    \put(0.13411822,0.0019803){\color[rgb]{0,0,0}\makebox(0,0)[t]{\lineheight{1.25}\smash{\begin{tabular}[t]{c}$\displaystyle{S^\mathrm{st} = 1-2\frac{r_i}{h_\mathrm{p}}}$\end{tabular}}}}%
    \put(0.49894722,0.0019803){\color[rgb]{0,0,0}\makebox(0,0)[t]{\lineheight{1.25}\smash{\begin{tabular}[t]{c}$\displaystyle{S^\mathrm{st}= \exp{\left(\overline{r}_i \frac{A_\mathrm{p}}{V_\mathrm{p}}\right)}}$\end{tabular}}}}%
    \put(0.86479858,0.0019803){\color[rgb]{0,0,0}\makebox(0,0)[t]{\lineheight{1.25}\smash{\begin{tabular}[t]{c}$\displaystyle{S^\mathrm{st} = \exp{\left(\overline{A}_\mathrm{p} \frac{l_\mathrm{f}}{V_\mathrm{p}}\right)}}$\end{tabular}}}}%
    \put(0.13367881,0.28352976){\color[rgb]{0,0,0}\makebox(0,0)[t]{\lineheight{1.25}\smash{\begin{tabular}[t]{c}\textbf{Parallel planes}\\\end{tabular}}}}%
    \put(0,0){\includegraphics[width=\unitlength,page=6]{Giddings_results_overview_paper.pdf}}%
    \put(0.55522392,0.23631356){\color[rgb]{0,0,0}\makebox(0,0)[rt]{\lineheight{1.25}\smash{\begin{tabular}[t]{r}$\overline{r}_\mathrm{p}$\end{tabular}}}}%
    \put(0,0){\includegraphics[width=\unitlength,page=7]{Giddings_results_overview_paper.pdf}}%
    \put(0.22990303,0.14723452){\color[rgb]{0,0,0}\makebox(0,0)[rt]{\lineheight{1.25}\smash{\begin{tabular}[t]{r}$h_\mathrm{p}$\end{tabular}}}}%
  \end{picture}%
\endgroup%